\documentclass{article}
\usepackage{arxiv}

\usepackage[utf8]{inputenc} 
\usepackage[T1]{fontenc}    
\usepackage{hyperref}       
\usepackage{url}            
\usepackage{booktabs}       
\usepackage{amsfonts}       
\usepackage{amsmath}        
\usepackage{bm}             
\usepackage{nicefrac}       
\usepackage{microtype}      
\usepackage{lipsum}		
\usepackage{graphicx}
\usepackage{gensymb}
\usepackage{natbib}
\usepackage{doi}
\usepackage{authblk}
\usepackage{setspace}
\title{Machine Learned Particle Detector Simulations}

\author[1]{D.~Darulis}
\author[1]{R.~Tyson}
\author[1]{D.~G.~Ireland}
\author[1,*]{D.~I.~Glazier}
\author[1]{B.~McKinnon}
\author[1]{P.~Pauli}
\affil[1]{SUPA, School of Physics and Astronomy, University of Glasgow, Glasgow G12 8QQ, United Kingdom}
\affil[*]{Corresponding author: {Derek.Glazier@glasgow.ac.uk}}

\renewcommand{\headeright}{}

\makeatletter
\edef\orig@output{\the\output}
\output{\setbox\@cclv\vbox{\unvbox\@cclv\vspace{0pt plus 20pt}}\orig@output}
\makeatother

\begin{document}
\maketitle

\begin{abstract}

The use of machine learning algorithms is an attractive way to produce very fast 
detector simulations for scattering reactions that can otherwise be computationally expensive. Here we develop 
a factorised approach where we deal with each particle produced in a reaction individually: first determine if it
was detected (acceptance) and second determine its reconstructed variables such as four momentum (reconstruction). For the acceptance we propose using a probability classification density ratio technique to determine the probability the particle was detected as a function of many variables. Neural Network and Boosted Decision Tree classifiers were tested for this purpose
and we found using a combination of both, through a reweighting stage, provided the most reliable results. For reconstruction a simple method of synthetic data generation, based on nearest neighbour or decision trees was developed. Using a toy parameterised detector we demonstrate that such a method can reliably and accurately reproduce kinematic distributions from a physics reaction. The relatively simple algorithms allow for small training overheads whilst producing reliable results. Possible applications for such fast simulated data include Toy-MC studies of parameter extraction, preprocessing expensive simulations or generating templates for background distributions shapes.

\end{abstract}


\section{Introduction} \label{sec_intro}

The use of simulated data samples is key to many analyses in hadron physics experiments.
With increased luminosities and detector complexity comes a corresponding increased  
overhead in generating these samples. In addition, more sophisticated analyses requiring, for example toy Monte-Carlo tests of parameter extraction, require larger volumes of independent
simulated data to mock up realistic datasets.
The computational expense for simulated samples comes from: 1) tracking a particle through a 
detector geometry using Monte-Carlo techniques to simulate a wide variety of processes in 
different materials; 2) reconstructing the particle's four-momentum from the hits in the
detectors, which may incorporate tracking through magnetic fields and other expensive operations.

We propose a method to largely remove these two overheads by training machine learning algorithms
to directly produce the output variables from a set of generated, or truth, events.
First we check each generated particle in an event to see if it was successfully reconstructed
(acceptance), 
second we distort the particles four-momentum to account for detector and reconstruction algorithm
effects (reconstruction).

The two components can be handled independently with suitable parameterisations dependent on
the initial particle variables such as four-momentum. 
Acceptance requires the detection probability for each particle species as a function of its variables. In one dimension one may just use the ratio of accepted to generated histograms
for this, but when the acceptance depends on multiple variables this is no longer viable.

Reconstruction requires sampling from resolution functions which are also likely to depend on 
the particles initial four-momentum. Often these can be approximated by Gaussian distributions
but there may also be tails or asymmetries on these distributions.

In the current work we prioritise speed and simplicity, over more sophisticated
machine learning techniques, for the algorithms we propose to use.

\subsection{Machine learning approaches}

Particle physics has long been familiar with various machine learning based methods. 
This was originally led by the ROOT TMVA package (\cite{Hocker:2007ht}) and more recently many different approaches in the Python ecosystem have proven very successful at a number of diverse tasks.
Probability density estimation has been handled with a number of methods, such as
neural density estimation techniques: deep generative neural networks (\cite{Liue2101344118}), normalising flows (\cite{papamakarios2018masked}) and Graph Neural Networks (\cite{DiBello2021}).
Another relevant approach is to use probabilistic classification for density estimation (\cite{sugiyama_suzuki_kanamori_2012}) and due to its simplicity and speed this is the technique applied here. Indeed, this method has already been proposed as part of simulation chains, but more focussed on reweighting to improve agreement between simulated and real data (\cite{PhysRevD.101.091901}).

Encoding reconstruction resolution effects is akin to the machine learning problem of 
producing synthetic data, i.e. data generated from another data set, 
which has been tackled in many different ways. 
Generative Adversarial Networks (GANs, \cite{goodfellow2014generative}) and autoencoders are commonly used for this. GANs have already been investigated as means for producing high level reaction specific data in \cite{hashemi2019lhc}. The "Deep neural networks using the Classification for Tuning and Reweighting" protocol of \cite{Diefenbacher_2020}, enhances the use of GANs through a reweighting scheme to produce better predictions from the generative model. While a different approach,
using Gaussian resolutions with mean and sigma parameters trained via neural networks to be dependent on input variables such as momentum, was proposed in \cite{Chen2021}. Using GANs to produce full pipeline simulated datasets, including the event generator phase, was explored in \cite{Alanazi:2020jod}. In this case the GAN not only encodes resolution effects of the detector but also generates the four-momentum of the particles in a single step.

\subsection{Proposed scheme}

The scheme adopted in this work splits the simulation into two distinct parts: (I) deciding
if a particle was successfully detected and its momentum reconstructed (acceptance);
(II) applying distortions to its reconstructed momentum components (resolution). The detection of each particle in the generated event must be simulated independently, with separate 
models for each particle species arising from the generator.

In general, there are any number of algorithms that could be used to perform these two tasks.
Here we focus on two machine learning methods which are relatively straightforward to apply
and can be trained and applied with comparatively small CPU overheads. We note that other
methods may be substituted for either or both stages if found to improve performance.

\begin{figure}[hbt!]
 \begin{center}
   \raisebox{0.5mm}{\includegraphics[trim = 2 2 2 2, clip, width=0.95\linewidth]{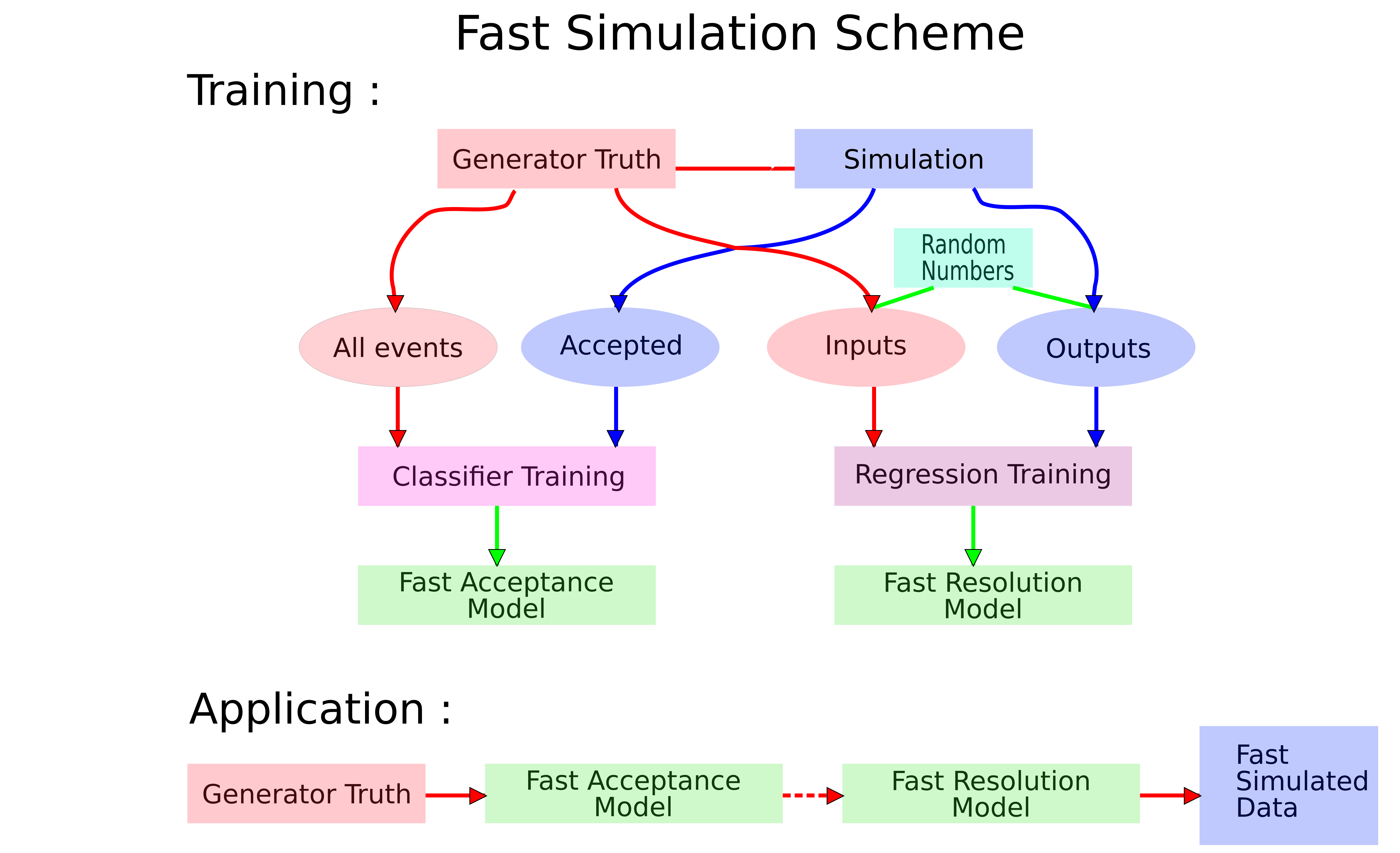}}
  \caption[]{ \label{fig:intro_scheme} Diagram of training and application stages of the fast
  simulation. In training, truth and accepted simulation events are passed as two different classes to the Fast Acceptance classifier. While truth events are also used with uniform random numbers as inputs to the Fast Resolution regressor and the difference between truth and reconstructed events are given as the output. Once the two models are trained they can be used directly on truth events to produce fast simulated 
  events. Note, Particle Truth represents simulations of individual particles, while Generator
  Truth would use complete events of particular reactions from an event generator with acceptance and resolution models trained for each particle species.}
 \end{center}
\end{figure}

In both cases the main requirement is a large amount of full simulated and reconstructed data, for each particle species under consideration. The algorithms can then be trained to reproduce similar detector and reconstruction effects.
Once the routines are trained, events from any full reaction generator may then be passed
for fast simulation. The scheme is illustrated in Fig. \ref{fig:intro_scheme}.

Here we adopt a Density Ratio method for Fast Acceptance and a Decision Tree based method 
for the Fast Resolution Model. Further details of these are given in Sections \ref{sec_accept} and \ref{sec_recon} respectively.

\section{Toy Detector}\label{sec_toy}

To illustrate the effectiveness of this fast simulation we used a toy detector setup
designed to incorporate features and correlations seen in full detector simulations.
We considered detection and reconstruction of a particle's three-momentum. The acceptance
and resolutions were functions of the momentum magnitude ($\bm{p}$), polar ($\bm{\theta}$) and azimuthal ($\bm{\phi}$) angles.

The true momentum components were randomly sampled from uniform distributions: $0<\bm{p}<10$ (GeV), $0<\cos{\bm{\theta}}<1$ and $-180\degree<\bm{\phi}<180\degree$.

For the probability of accepting a particle with a given three-momentum we used,

    \[ A(\bm{p},\bm{\theta},\bm{\phi}) =   0.9\times C(N(\bm{p};2,0.5))\times A_{a}(\bm{\theta},\bm{\phi}) \]
    \[\mbox{where, } A_{a}(\bm{\theta},\bm{\phi}) = \begin{cases}   
                0 & \mbox{if } \bm{\theta} <5\degree \mbox{ or, } \bm{\theta}>175\degree, \\
                0 & \mbox{if } |\bm{\phi} +\frac{\bm{\theta}-90\degree}{5}| < 10\degree, \\
                0 & \mbox{if } |\bm{\phi} +\frac{\bm{\theta}-90\degree}{5}| > 170\degree, \\
                0.95 & \mbox{otherwise. }\end{cases} \]

We use $N(\bm{x};\mu,\sigma)$ to be a general normal distribution as a function of
$\bm{x}$ with mean $\mu$ and standard deviation $\sigma$; 
and $C(N)$ to be the cumulative distribution function of $N$.

\begin{figure}[hbt!]
 \begin{center}
   \raisebox{0.5mm}{\includegraphics[trim = 2 2 2 2, clip, width=0.95\linewidth]{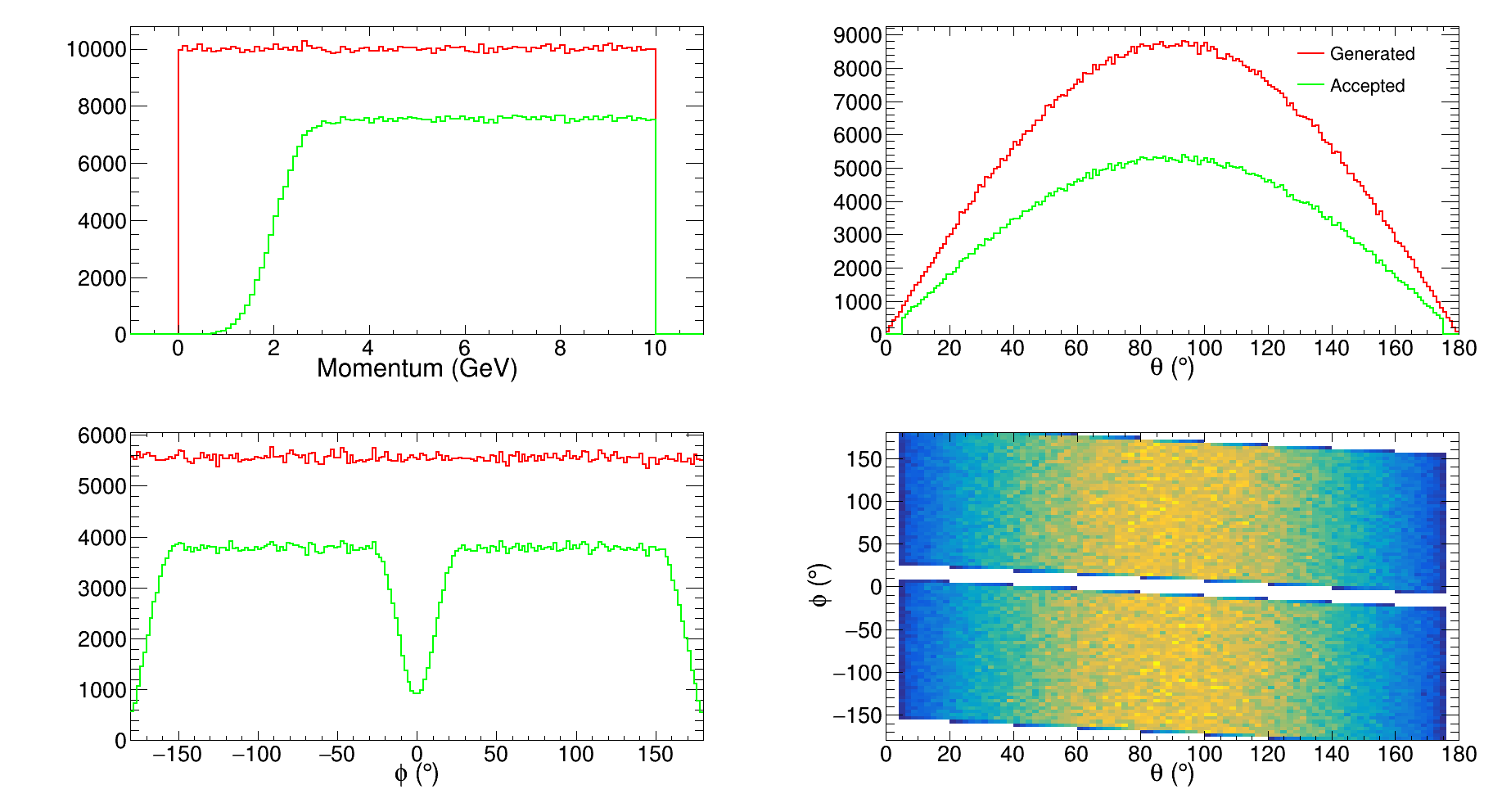}}
  \caption[]{ \label{fig:toy_1DDist} Generated (red) and accepted (green) distributions of the three momentum variables for 1 million generated events. Bottom-right shows the 2D correlation between $\bm{\theta}$ and $\bm{\phi}.$}
 \end{center}
\end{figure}

The resulting generated and accepted distributions from this model are illustrated in Fig. \ref{fig:toy_1DDist}.

The resolution function for $\bm{p}$ which returns the distortion $\delta\bm{p}$ was based on a "Crystal Ball" shape (i.e. normal distribution with a tail) with an additional linear dependence on $\bm{p}$,
    
    \[  CB(\delta\bm{p}) = \begin{cases}  
        N(\delta\bm{p};\mu_{p},\sigma_{p}) 
        & \mbox{ for } \frac{\delta\bm{p}-\mu_{p}}{\sigma_{p}} > \alpha, \\
        S\times\left( T - \frac{\delta\bm{p}-\mu_{p}}{\sigma_{p}}\right)^{-n} 
        & \mbox{ for }  \frac{\delta\bm{p}-\mu_{p}}{\sigma_{p}} \leq \alpha.
                                    \end{cases} \]
                                    
where 
\[  S = \left( \frac{n}{|\alpha|} \right) ^{n} \exp{\frac{-\alpha^{2}}{2}}  
        \mbox{ and, } T = \frac{n}{|\alpha|} - |\alpha|, \]
        
and we used
\[  \mu_{p} = 0, \medspace{} \sigma_{p}=0.05,\medspace{} \alpha=0.5,\medspace{} \mbox{and, } n=10. \]

Once $\delta\bm{p}$ has been sampled from $CB(\delta\bm{p})$ an additional linear $\bm{p}$ dependence was applied to get the final reconstruction distortion,
\[ \delta\bm{p}\prime = 0.5(\bm{p}+1)\times\delta\bm{p} \]

The resolutions on $\theta$ and $\phi$ were correlated and generated from a two-dimensional
normal distribution,  with means of 0 and covariance matrix $\bm{\Sigma}$,

\[ R(\delta\bm{\theta},\delta\bm{\phi})
    =   N(\delta\bm{\theta},\delta\bm{\phi};\mu=0,\bm{\Sigma}) 
    \mbox{ with, } \bm{\Sigma} = \left( \begin{matrix} 1.0 & -0.5 \\ -0.5 & 1.0 \end{matrix} \right) \]
    
The $\bm{\phi}$ distortion was given an additional $\bm{\theta}$ dependence typical of spherical coordinate systems,

\[  \delta\bm{\phi}\prime  = \frac{\delta\bm{\phi}}{\sin{\theta}} \]

The truth values of the three-momentum were then summed with their distortions to give the 
reconstructed variables for the event.

\begin{figure}[hbt!]
 \begin{center}
   \raisebox{0.5mm}{\includegraphics[trim = 2 2 2 2, clip, width=0.33\linewidth]{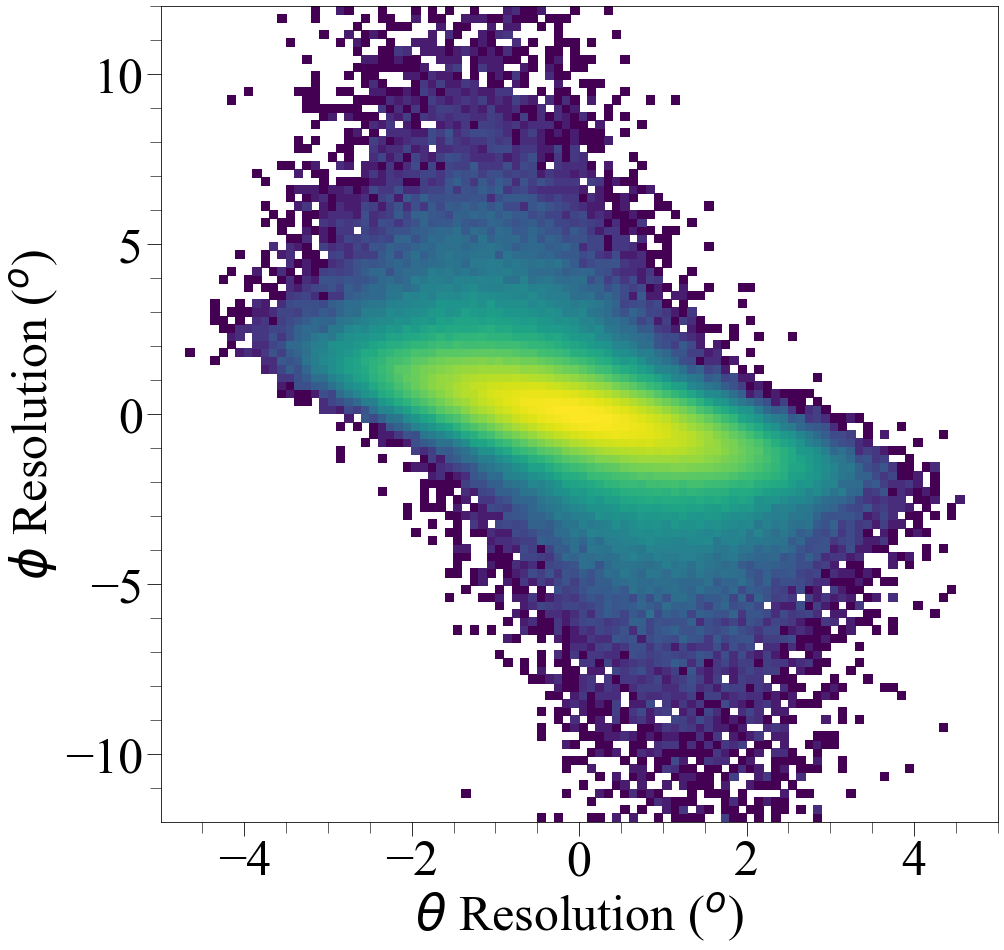}
   \includegraphics[trim = 2 2 2 2, clip, width=0.33\linewidth]{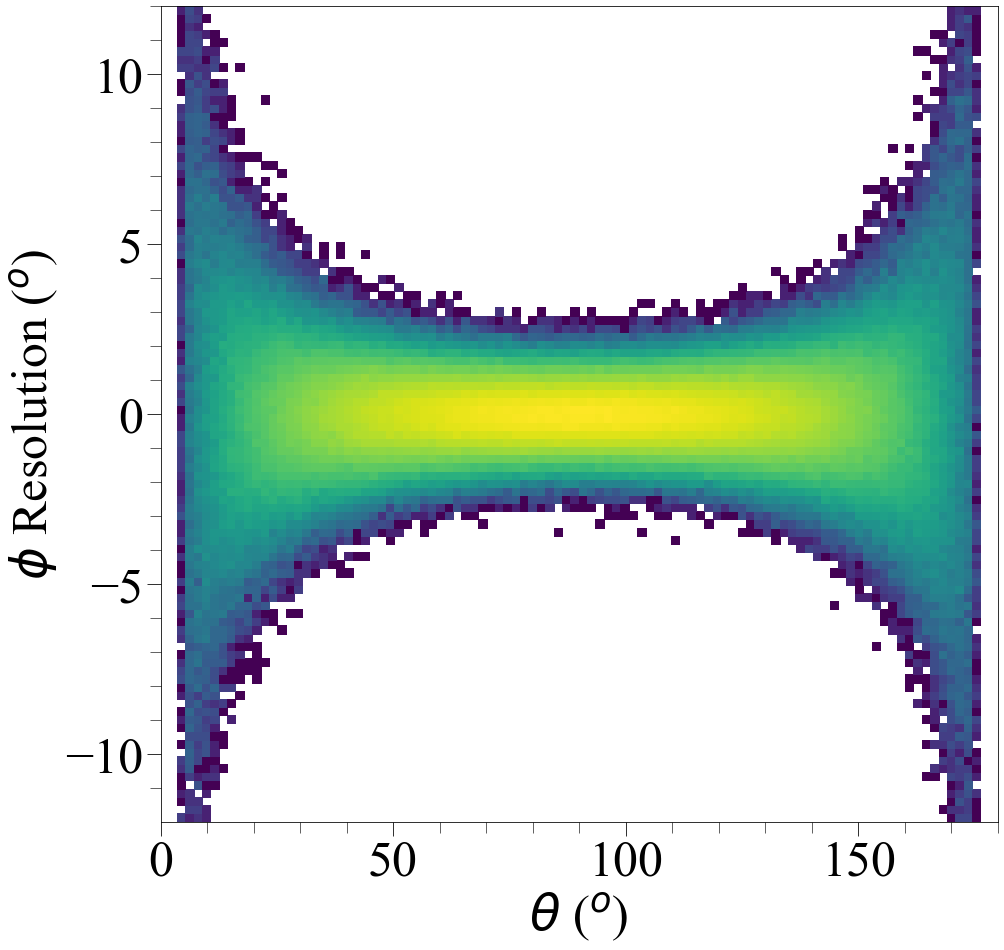}
   \includegraphics[trim = 2 2 2 2, clip, width=0.33\linewidth]{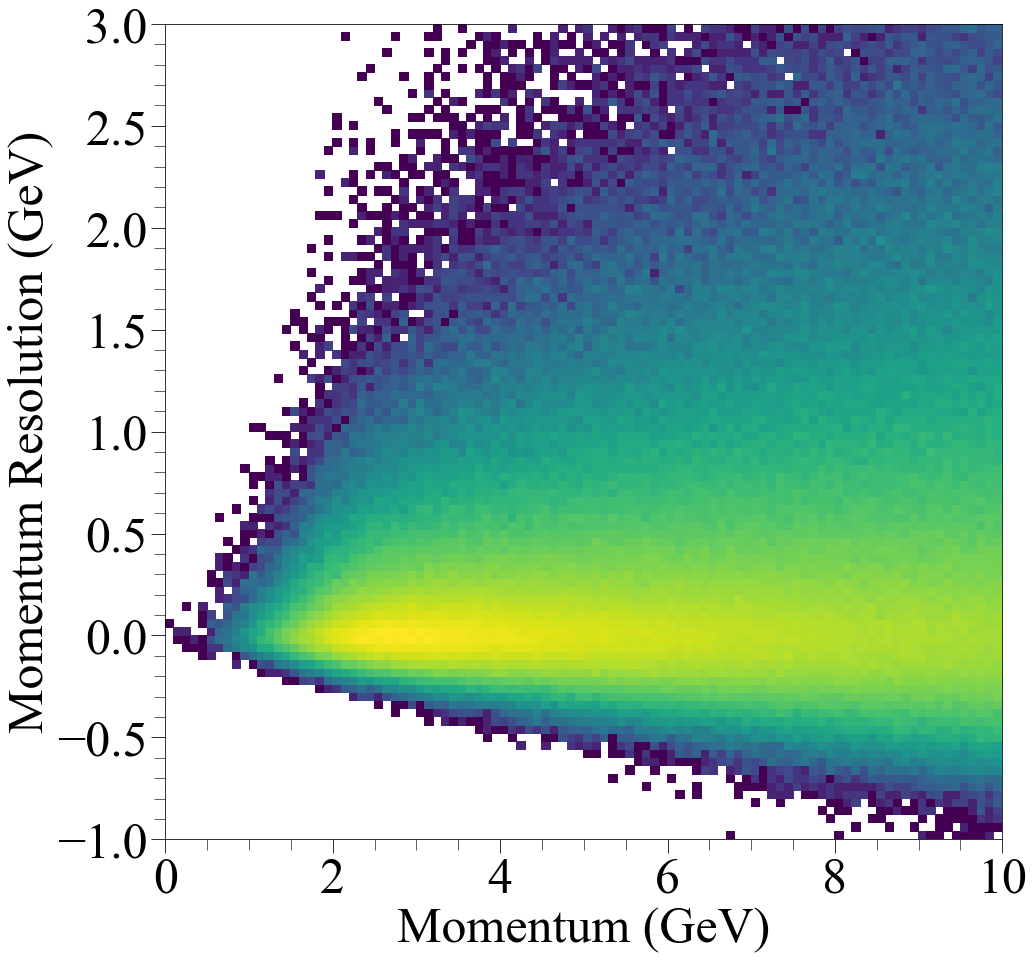}}
  \caption[]{ \label{fig:toy_2DRes} Some of the multidimensional correlations in the toy detector reconstruction. It is important that the machine learned simulation can reproduce these features.}
 \end{center}
\end{figure}

The resulting correlations between different resolutions and variables are shown in Fig. \ref{fig:toy_2DRes}. 

\section{Acceptance} \label{sec_accept}

\subsection{Detector acceptance}
High energy accelerator experiments produce a large amount of particles. However, only a fraction will be detected (accepted) and available for study. Detector acceptance is dependent on various factors. Detectors cannot produce responses to particles that do not hit them, but they generally cover only limited angular ranges. This is called geometric acceptance. Of course, it is not guaranteed that a particle will be registered as a hit even if it does physically interact with a detector subsystem. Limited resolution, imperfect trigger systems, material degradation, and inaccuracies in offline event reconstruction are amongst a number of effects that can prevent particle detection.  Any analysis has to take care to properly account for these effects. Failing to do so could create distortions in extracted parameters.

The standard approach to determining the acceptance is via Monte Carlo studies: events are generated and passed through a detailed detector simulation using frameworks such as GEANT4 (\cite{AGOSTINELLI2003250}). Assuming the simulation is sufficiently accurate, taking the ratio of accepted to generated events will give the desired answer. Unfortunately, this is computationally intensive. An alternative is to use Machine Learning by treating the problem as binary classification, which allows the derivation of per-event acceptance probabilities. In this paper, we detail a method using a fully-connected neural network with an optional Boosted Decision Tree model for corrections.

\subsection{Probability Classification for Density Ratios}
\label{sub:den_ratio}

When designing a detector simulation, one would like to construct the full probability distribution that describes the detector acceptance $p_a(\textbf{x})$, where \textbf{x} may consist of multiple independent variables. This is a difficult statistical inference task when no strict prior assumptions can be made about its form. Even with severe simplifications, the acceptance is likely to depend non-trivially on multiple variables.  Instead we opt to solve an easier problem of estimating the ratio of the probability densities for generated and accepted events, $\frac{p_a(\textbf{x})}{p_g(\textbf{x})}$. The knowledge of this ratio is sufficient for modelling acceptance. Once it has been calculated for each event, it can be used as the probability $w$ in a rejection sampling algorithm to simulate a pass through the detector. 

A convenient technique for density ratio estimation is to treat it as a binary classification problem. Given multivariate data \textbf{x} which is split into two 'classes', which we label i = 0 and i = 1, we aim to train a probabilistic classifier model \textbf{c} that can separate them successfully and output a probability $c(\textbf{x}) \approx p(i = 1|\textbf{x})$ that the data sample falls into class i = 1. For the purposes of this paper, i = 1 is used to denote events accepted by the detector and i = 0 to denote generated (truth) events. If the model is well calibrated, we can get the density ratio as

\begin{equation}\label{eqn_w}
   w =  \frac{p_a(\textbf{x})}{p_g(\textbf{x})} \approx \frac{c(\textbf{x})}{1 - c(\textbf{x})}
\end{equation}

Similar approaches have been used for different applications. \cite{Martschei_2012} used a proprietary neural network model to calculate weights used for reweighting Monte Carlo samples to more closely agree with data. \cite{rogozhnikov2016reweighting} used a Boosted Decision Tree model with a bespoke objective function to iteratively reweight histograms. 

\subsection{Models and experiments}
\label{sub:models}

We chose to test two different model architectures for the classifier. The first was a Gradient Boosted Decision Tree (BDT). This is an ensemble model based on the standard decision tree using the boosting paradigm. A normal decision tree attempts to split the training data by repeatedly splitting it into left and right child nodes. The splits are performed by selecting simple cuts on a random choice of variables such that some objective function is minimised. In the sklearn implementation (\cite{treeimpl}) this is the log loss,

\begin{equation*}
    L = \sum_i p_{mi}(1 - p_{mi}),
\end{equation*}

where $p_{mi}$ is the proportion of samples in node m belonging to class i. The splitting continues until either all terminal nodes contain 1 sample each or a specified maximum depth is reached.

Tree boosting is based on the intuition that it is easier to train several smaller models than a single large model. It constructs a series of regression decision trees that calculate the probability $p(c(\textbf{x}) = 1 | \textbf{x})$ as 

\begin{equation*}
    p(c(\textbf{x}) = 1 | \textbf{x}) = \sigma (F(\textbf{x})),
\end{equation*}
 where $\sigma$ is the sigmoid function and $F_i$ is the internal output of a combined model, i.e. the sum of $i$ weak learners. The boosted ensemble is trained stagewise such that for each new learner $t_k$, it minimises $L(F_{k-1}(\textbf{x}) + at_k(\textbf{x}))$ via gradient descent. $a$ is a parameter called the learning rate, which controls the size of steps used in the gradient descent procedure.  
 
The second model was a fully-connected neural network (\cite{bengio2017deep}). Neural networks are architectures composed of layers of nodes often called neurons that are connected by weighted edges. The inputs to a neuron are a weighted sum of all the edges connecting to it. This is then passed through an activation function, such as tanh or ReLU. The latter is especially common and is defined as $A(x) = max(0,x)$.  This serves as input for the next layer in the network:

\begin{equation*}
    o_k = A\left(\sum_i w_{ik}o_i\right),
\end{equation*}

where $o_i$ is the output of neuron i and A is the activation function.

In binary classification tasks, the results in the output layer are passed through a sigmoid function to obtain the probability that the sample belongs to class i = 1. Neural networks are trained using variations of Stochastic Gradient Descent algorithms -- a popular choice is ADAM (\cite{kingma2014adam}). The gradients are used to modify the network weights using the backpropagation algorithm, aiming to minimise a loss function. As in the case of the BDT we used the binary log loss. 

An important part of using Machine Learning models is hyperparameter optimisation. The performance of a classifier will depend in large part on whether the settings were set appropriately.  A wide range of settings for both the BDT and the neural network models were tested. Discussion of the minute experimental details is omitted for brevity, as the goal of the paper is simply to present an empirically successful model, but we describe some of the general tendencies here. 

Boosting algorithms are usually based on shallow Trees with a low maximum depth. However, this proved to decrease the model's ability to reproduce angular correlations. This makes sense intuitively, as more layers in the Trees allow for combinations of splits that are based on multiple variables. The number of weak learners did not seem to make a difference past a limit of about 100, only increasing the training time, but anything below 50 performed poorly. Similarly, the learning rate did not seem to have a large effect as long as it remained around the default value of 0.1.

The space of parameters for neural network architectures is even harder to search than that of 'classic' models such as decision trees. Taking into account the low dimensional nature of the data, we chose simple multilayer networks with ReLU activation functions and a sigmoid output. The networks seemed to require a certain size before they were able to produce 1D distributions without significant deviations. Empirically, we found that networks without at least approximately 500 nodes in some of the layers were inadequate. The networks were trained using the ADAM optimizer with batch sizes of 1024. We performed some tuning on the optimizer's learning rate; performance seemed to improve if it was reduced to values of $10^{-3} - 10^{-4}$. Optimal number of passes over the training data (epochs) was chosen by an early stopping criterion and depended on the size of the network. 

  A data preprocessing step that we found useful for modelling multidimensional correlations was the Gaussian transform. We applied a quantile transformation $G^{-1}(F(X))$ where $G^{-1}$ is the Gaussian quantile function and $F(x)$ is the cumulative distribution function for each variable. This transform forced each variable to follow a Gaussian distribution. 

The computational performance of both models is summarised in Table \ref{tab:time}. The CPU used for the measurement was an Intel 1145G7, while the GPU was an NVIDIA T4 accessed via a Google Colaboratory instance. Training time for the neural network were improved significantly by a GPU, but it performed worse during inference. This is likely explainable by the relatively small size of the network making it inefficient to run on a GPU, with data I/O dominating the computation time of the inference. The neural networks were also generally slower than the BDT for inference.

\begin{table}[hbt!]
\caption{Table showing the runtimes for training and inference of the BDT and neural network models. Results are presented as the mean and standard deviation from 50 iterations. The neural network used for these measurements consisted of 4 hidden layers with 512, 384, 256, and 16 neurons respectively, and was trained for 50 epochs. All times are reported for 1 million training and testing samples. }
\scalebox{0.893}{
\begin{tabular}{|l|l|l|l|l|}
\hline
                              & Training time (CPU) & Training time (GPU) & Inference time (CPU) & Inference time (GPU) \\ \hline
GBDT                          & 984 $\pm$ 10 s        &   -                  & 4.0 $\pm$ 0.04         &       -               \\ \hline
neural network                & 600 $\pm$ 23.7 s      & 175 $\pm$ 9.9 s                    & 12.01 $\pm$ 0.93 s     &             28 $\pm$ 8.64 s         \\ \hline
neural network  + Corrections & 1015 $\pm$ 18.4 s     &  576 $\pm$ 13.1 s                   & 13.47 $\pm$ 0.07 s     &   32.1 $\pm$ 8.5 s                   \\ \hline
\end{tabular}}

\label{tab:time}
\end{table}

\subsection{Toy Simulation}

\begin{figure}[hbt!]
    \centering
    \includegraphics[width=0.66\linewidth]{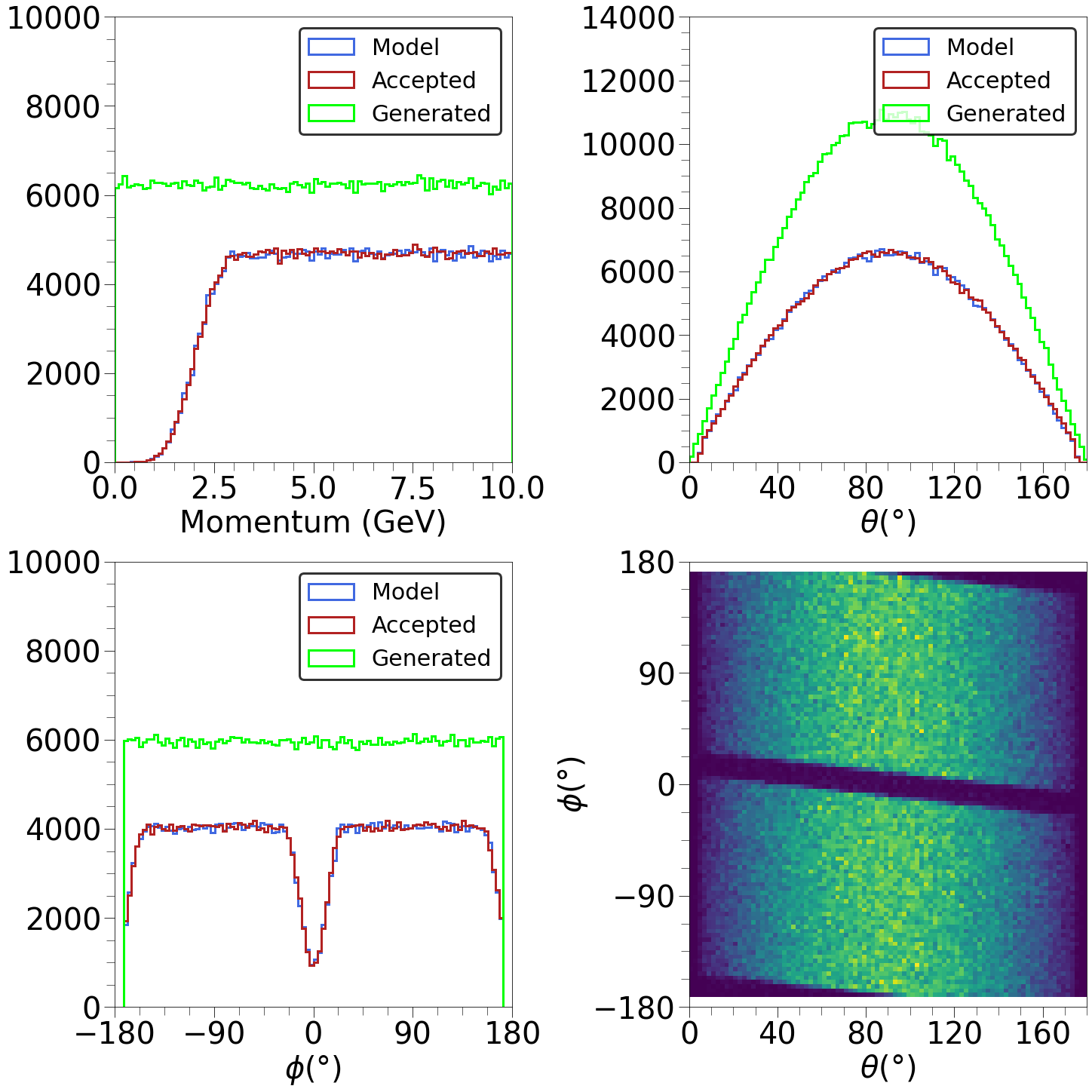}
    \caption{Results of the acceptance modelling with a BDT. The BDT used 100 weak learners with a maximum depth of 10 and a learning rate of 0.1. The 1D components are generally modelled well, with subtle features such as the polar angle acceptance holes in the 0\degree-5\degree and 175\degree-180\degree regions being captured. However, it does not capture the 2D correlations in the angular distributions perfectly as shown in the bottom right plot.}
    \label{fig:acceptance_gbdt}
\end{figure}

\begin{figure}[hbt!]
    \centering
    \includegraphics[width=0.33\linewidth]{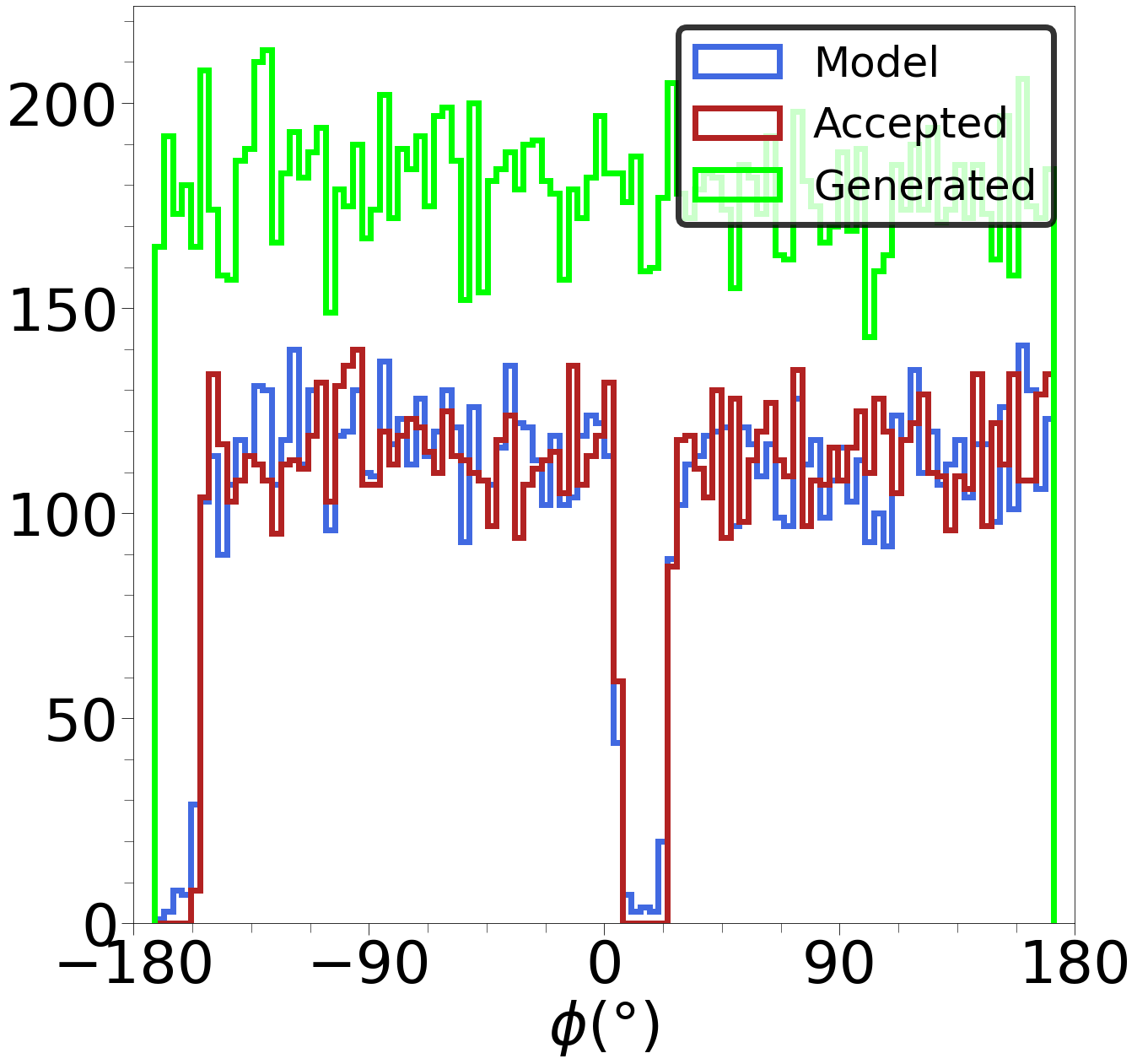}
    \includegraphics[width=0.33\linewidth]{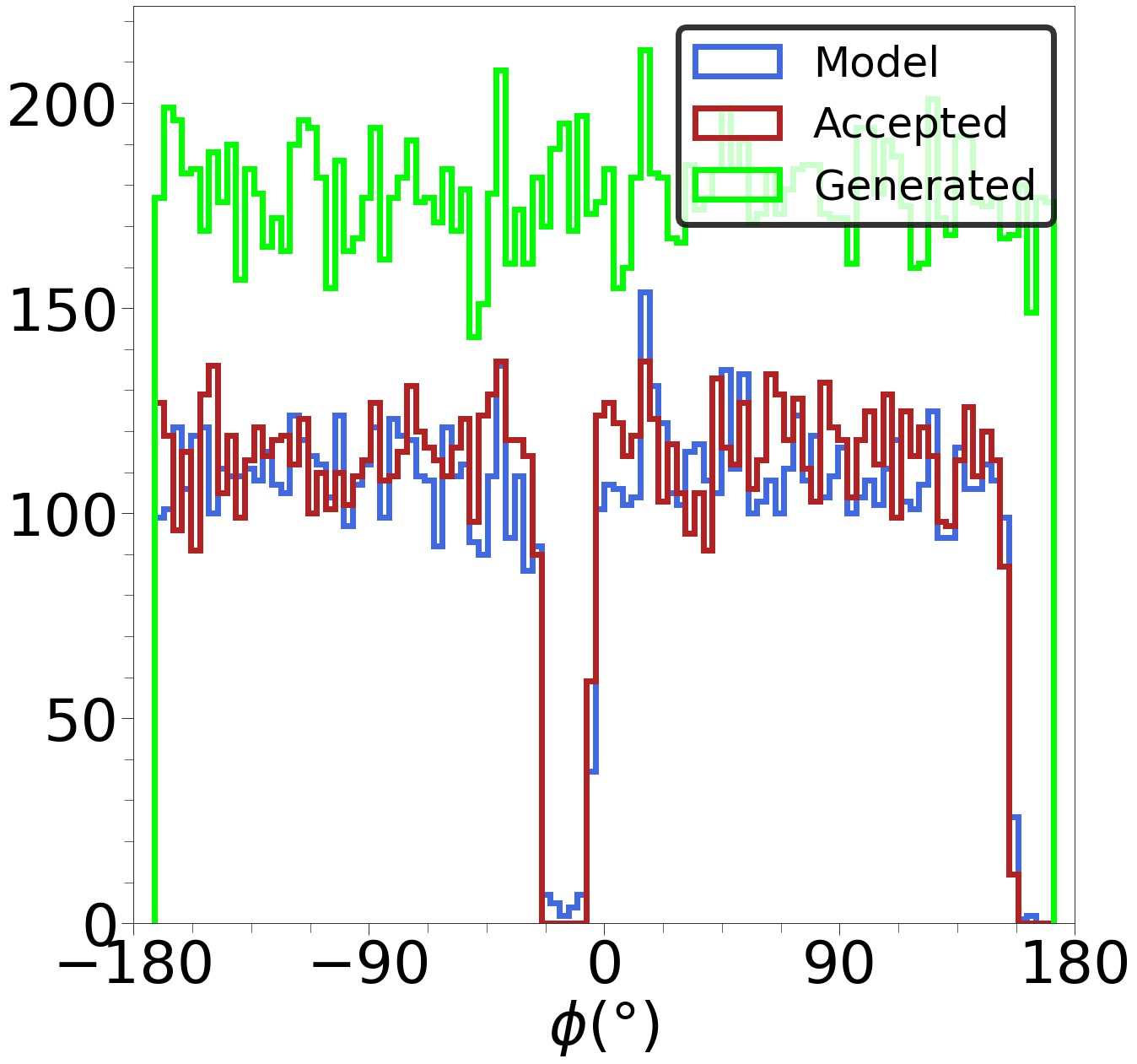}
    \caption{Slices of the azimuthal distribution in bins of the polar angle of the BDT model. First plot is binned in the range 0\degree-20$\degree$, while the second plot is binned in the range 160\degree-180$\degree$. The dips in the centres of the distributions are not well modelled.}
    \label{fig:acceptance_gbdt_slices}
\end{figure}

The classifiers were trained on the generated and accepted toy data described in Section \ref{sec_toy}. Prior to training all data was normalized to lie in the range [0-1].  We used 1 million samples for fitting the networks, with 80\% of this data set being used as training data and the remaining 20\% as a validation set for tracking loss. Another set of 1 million samples was used for testing final performance and producing the figures shown in this paper. The models were used to calculate the weights $w$ described in Eqn. \ref{eqn_w}.  When evaluating performance, we placed importance on good agreement of both the individual 3-vector components, especially the $5\degree$ gaps at the edges of the polar angle, and accurate reproduction of correlation between the polar and azimuthal angles. The latter can be a tricky problem for probabilistic classifiers, as many models tend to avoid predicting a probability of 0, (\cite{prob}). This would lead to a noticeable smearing between the two non-zero regions of the lower right 2D histogram in  Fig. \ref{fig:toy_1DDist}, with the diagonal empty band being contaminated with wrongly accepted events. Indeed, this proved to be an issue for the BDT. Regardless of hyperparameters used, we were unable to have it faithfully reproduce the correlation in the zero probability regions. A representative example can be seen in  Fig. \ref{fig:acceptance_gbdt}. The projection in bins of the polar angle as shown in  Fig. \ref{fig:acceptance_gbdt_slices} make the failure clearer. However, it did model the 1D components well, with the small gaps at the limits of the polar angle being well represented.

\begin{figure}[hbt!]
    \centering
    \includegraphics[width=0.66\linewidth]{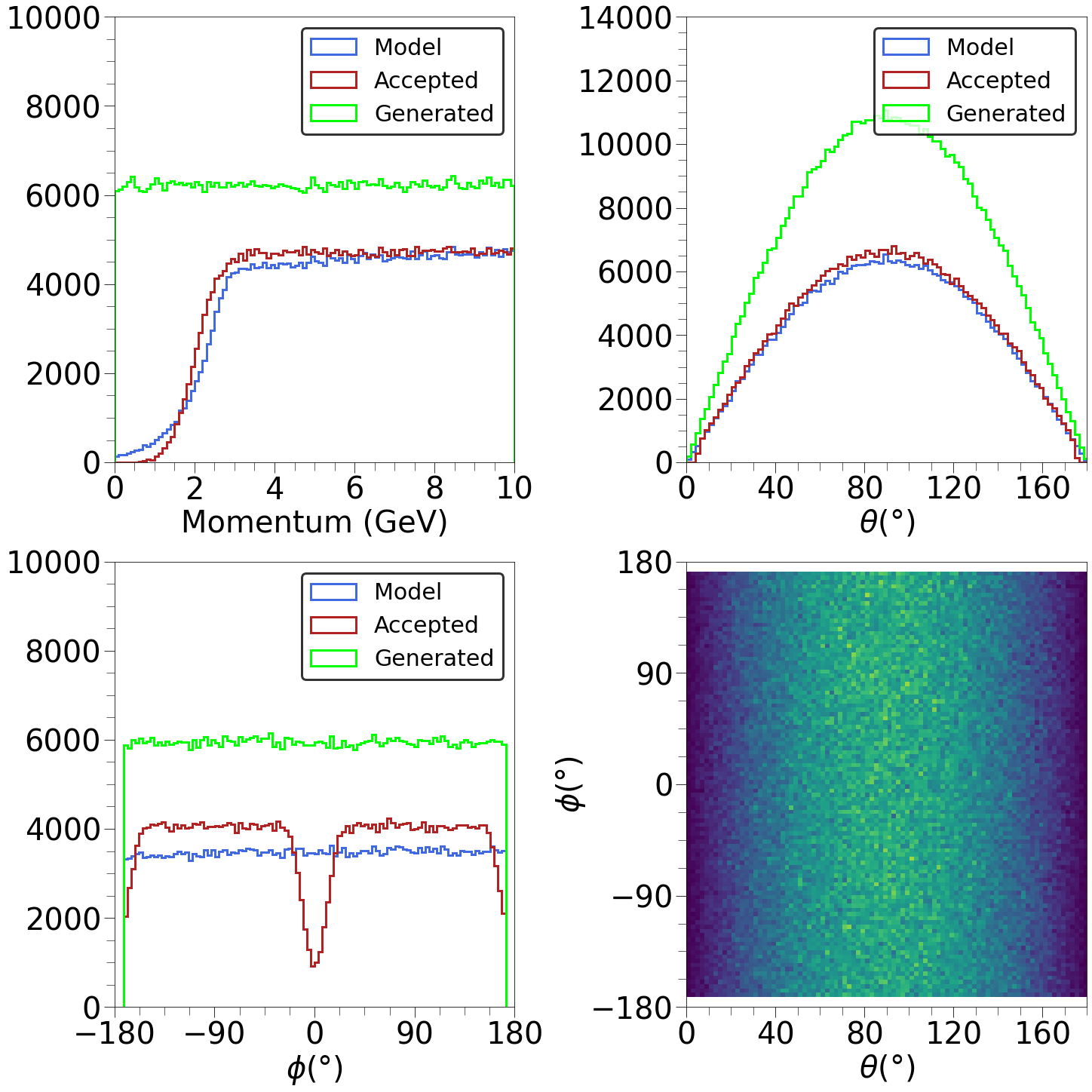}
    \caption{Results of the acceptance modelling with a neural network without sufficient capacity: one hidden layer of size 16 with a ReLU activation function. While the network seems to do acceptably well modelling the polar angle distribution and, to a lesser extent, the momentum distribution, it completely fails to learn the azimuthal angle distribution. This also naturally results in failing to model the angular correlation.}
    \label{fig:acceptance_dnn_bad}
\end{figure}

\begin{figure}[hbt!]
    \centering
    \includegraphics[width=0.66\linewidth]{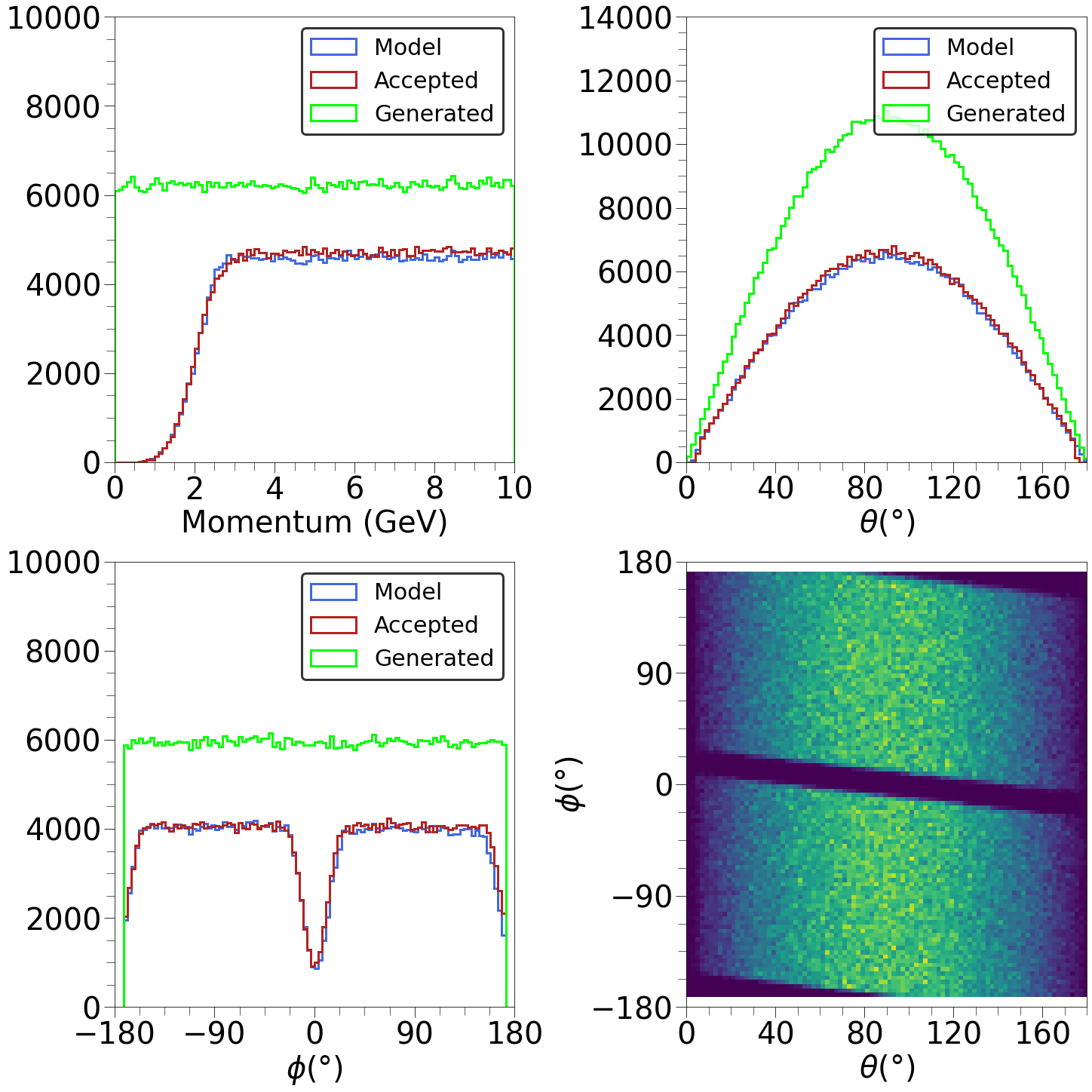}
    \caption{Results of the acceptance modelling with a neural network with sufficient capacity: hidden layer sizes of 512, 256, 128, 16 with ReLU activation functions. The results are generally good, but the gap at the edges of the polar angle acceptance between 175\degree and 180\degree, as well as the dip in azimuthal angle acceptance around 0\degree are not properly captured. }
    \label{fig:acceptance_dnn_good_nongauss}
\end{figure}

Neural network performance heavily depended on the architecture used. An illustrative example of results from a 'poor' network without enough capacity, using just a single 16 node intermediate layer, is shown in  Fig. \ref{fig:acceptance_dnn_bad}, while a 'better' network architecture's results are shown in  Fig. \ref{fig:acceptance_dnn_good_nongauss}. This larger network contains 4 hidden layers with 512, 256, 128, and 16 nodes respectively. Larger capacity seems to effect clear improvements in the results. In general with such a neural network we still observed a failure to precisely model the gaps at the edges of the polar angle distribution. 

\begin{figure}[hbt!]
    \centering
    \includegraphics[width=0.66\linewidth]{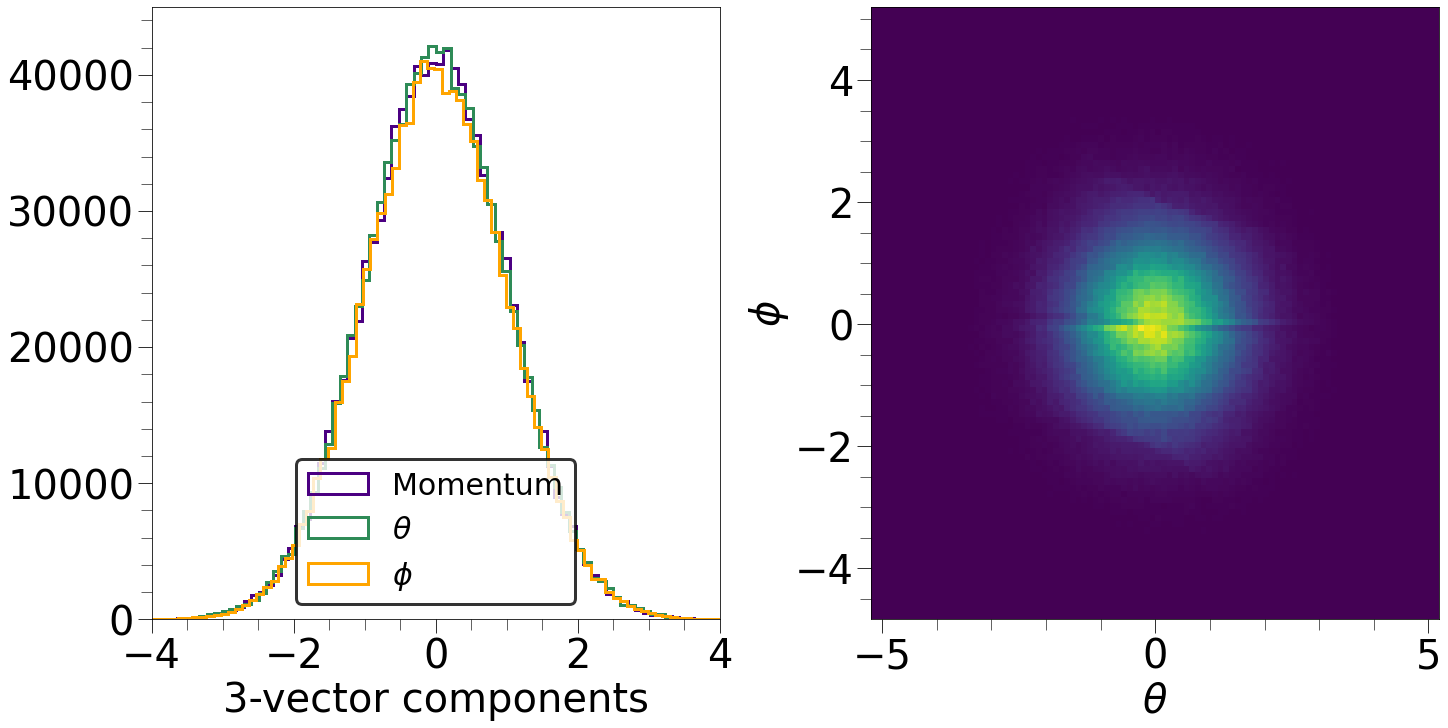}
    \caption{The first plot shows 3-vector components after applying the quantile transformation. All three components adopt a distribution very close to a Gaussian. The second plot shows that after the transformation, the 2D angular distribution no longer has a discontinuous topology. }
    \label{fig:gauss_transform}
\end{figure}

The problem of accurately modelling the small gaps was fixed by using the Gaussian transform introduced in Section \ref{sub:models}. The effects of applying it to the data can be seen in  Fig. \ref{fig:gauss_transform}. All of the 3-momentum component distributions became close to perfect Gaussians after the transformation. One can also see that the distribution of the polar and azimuthal angles no longer has two distinct regions with a zero probability band in the middle, but is instead a single 2D Gaussian-like with curved tails. Such a distribution is likely easier to model for the network and we found it performed better on our toy detector after the transformation.

\begin{figure}[hbt!]
    \centering
    \includegraphics[width=0.66\linewidth]{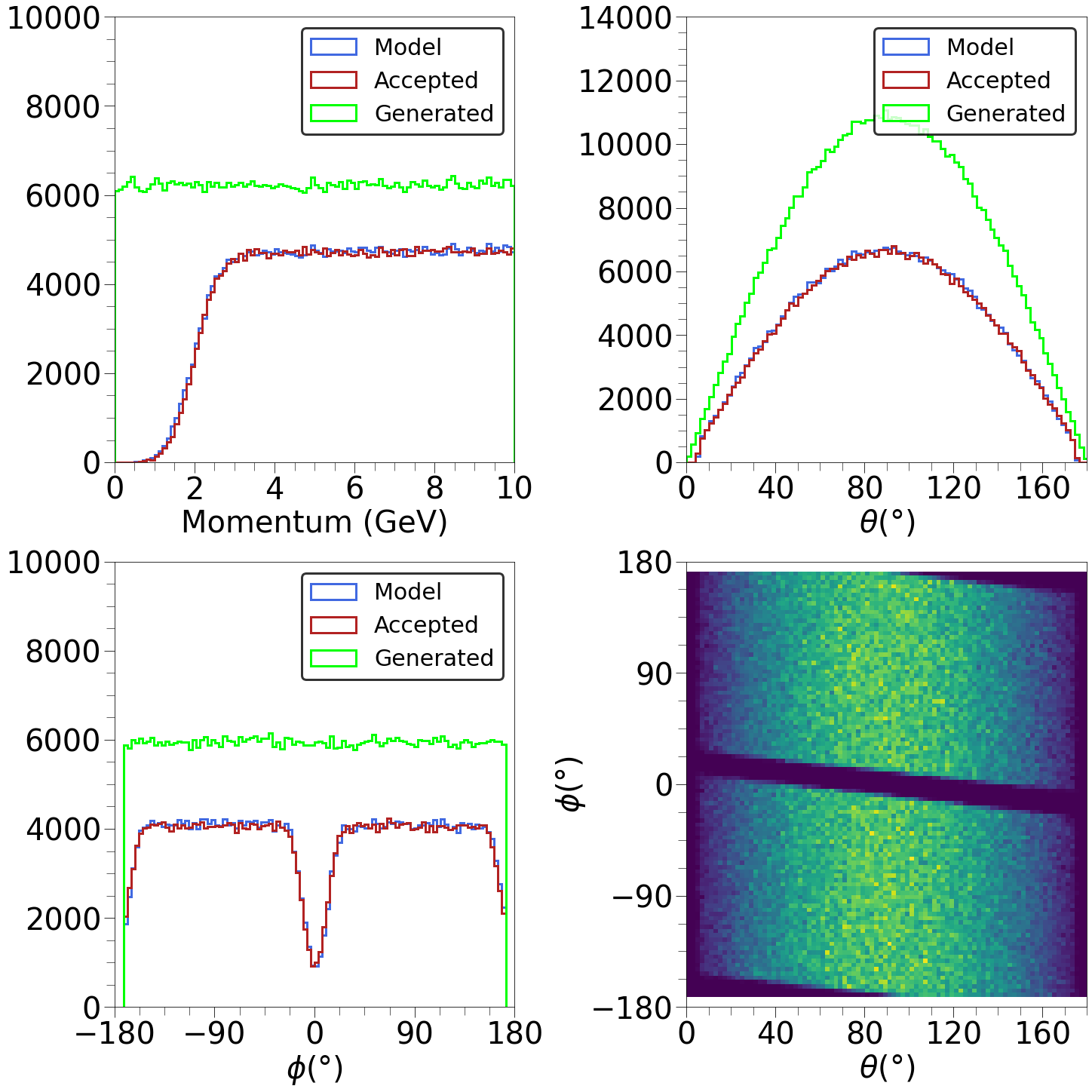}
    \caption{Results of applying a neural network with a Gaussian transform for acceptance modelling. The first three plots show the 1D components of the particle 3-vectors, while the final plot shows the 2D histogram of the two angular components. We see generally good agreement between the accepted and simulated distributions, though discrepancies in $\phi$ remain.}
    \label{fig:acceptance_dnn_good}
\end{figure}

\begin{figure}[hbt!]
    \centering
    \includegraphics[width=\linewidth]{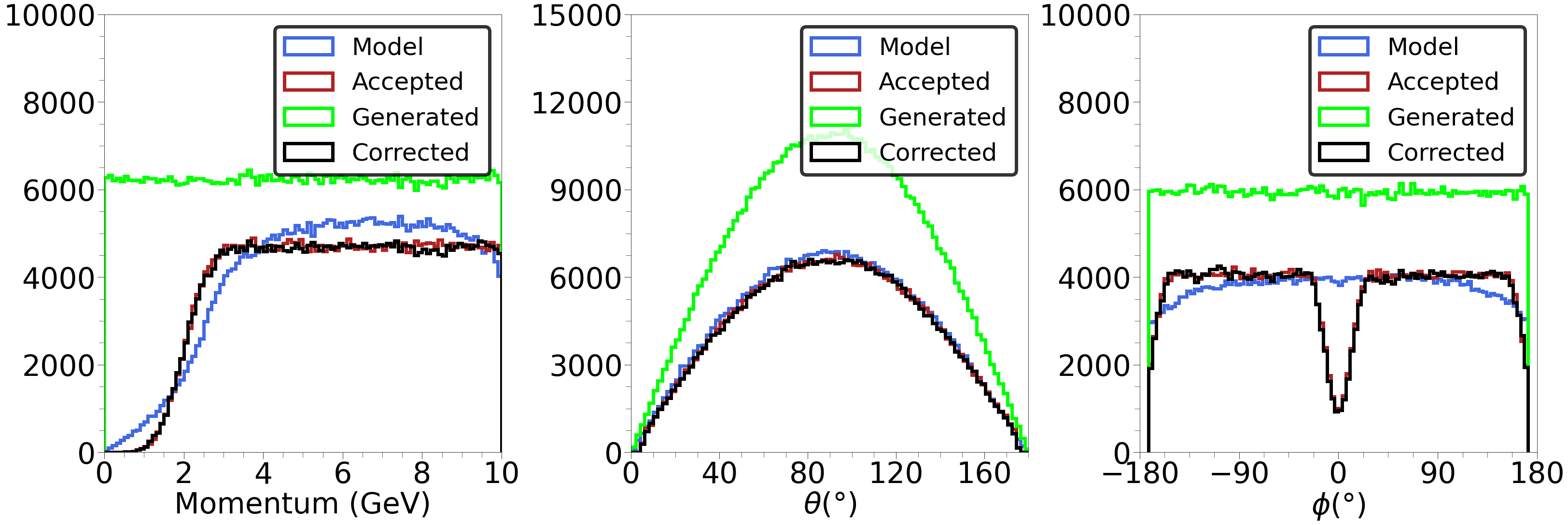}
    \caption{Results of applying a neural network with a Gaussian transform for acceptance modelling with a BDT correction. The BDT used 100 weak learners with a maximum depth of 10 and a learning rate of 0.1. The network used is the low capacity model with one hidden 16 neuron layer. The three plots show the 1D components of the particle 3-vectors. We see that the network by itself fails to model the distributions successfully, but the BDT corrections improve agreement significantly.}
    \label{fig:acceptance_dnn_corrected}
\end{figure}

\begin{figure}[hbt!]
    \centering
    \includegraphics[width=\linewidth]{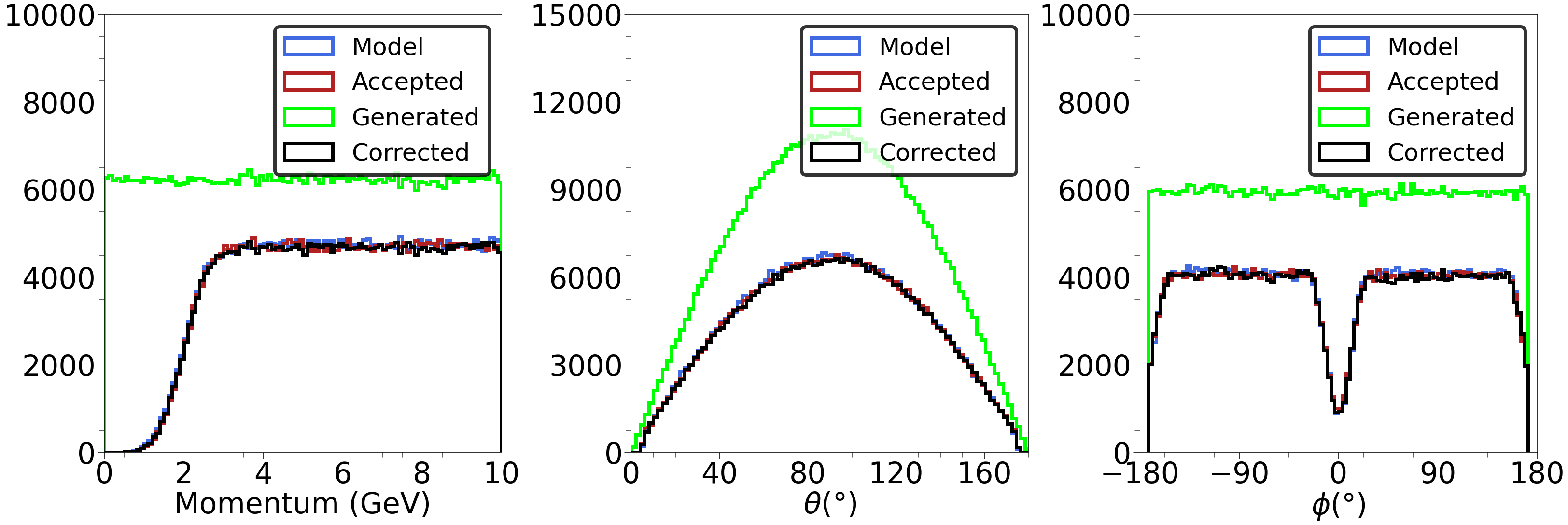}
    \caption{Results of applying a neural network with a Gaussian transform for acceptance modelling with a BDT correction. The BDT used 100 weak learners with a maximum depth of 10 and a learning rate of 0.1. The network used is the higher capacity model with 4 hidden layers of 512, 256, 128, and 16 neurons respectively. The improvement in the 3-vector component distributions is smaller than in the case of the low capacity network.}
    \label{fig:acceptance_dnn_good_corrected}
\end{figure}

With the Gaussian transform applied before using the 'good' 4 hidden layer network, the angular correlations were successfully reproduced, as shown in Fig. \ref{fig:acceptance_dnn_good}. In particular, the acceptance holes in the 0\degree-5\degree and 175\degree-180\degree regions of the polar angle are now correctly modelled, a feature which the networks without the transform could not reproduce. Despite this, it still exhibited some minor deviations in the 1D 3-vector components, especially the azimuthal angle. To improve the results, we applied a reweighting step. We trained a GBDT model on the output of the neural network and the generated data. The GBDT could then calculate a new set of weights $w_{corr}$ that were treated as corrections to the network weight so that $w_{new} = w*w_{corr}$. To show the effectiveness of this reweighting procedure, we used it to correct the output of the previously shown 'bad' network, although now with a Gaussian transform being applied to the data first. The results are shown in  Fig. \ref{fig:acceptance_dnn_corrected}. The 1D components are improved significantly. When the reweighting is applied to a 'good' network, the corrections are naturally smaller, but still improve the results. These can be seen in  Fig. \ref{fig:acceptance_dnn_good_corrected}.

We note for the training procedure of the models described above that both the BDT and neural network contain stochastic elements in the optimisation. For the BDT, this comes in the form of the split selection being randomised. In the case of the neural network the weight initialisation is random. Another source of randomness is the random splitting of the originally generated data into the sets of 1 million samples used for training and testing.  Often when training the optimisation may not find the best results given an initial random number seed, so one should re-run training several times and select the model with the best agreement.

\section{Reconstruction} \label{sec_recon}

\subsection{Detector Resolution}

Detector resolution refers to how accurately a particle's true momentum components are determined given detector effects and reconstruction algorithms. This can be measured with simulated data by taking the difference between the true and reconstructed values of the three components. As in the Toy detector of Section \ref{sec_toy}, the resolution on a given variable can depend on itself and other variables, and despite often being modelled as normal distributions centered on 0, the resolution distributions can present tails or asymmetries. Simulations need to accurately reproduce these dependencies and shapes of the resolution so as to accurately reproduce the detector response to reconstructing a particle. 

In typical simulations of hadron physics experiments, the resolution introduced by a detector is reproduced by simulating the various interactions of a particle with the detector, but this can be very expensive computationally. With machine learning methods there is the potential to use generative models to produce synthetic data, trained to look like full simulation data. These models must be conditional on the true value of one or more variables to encode the dependency of the resolution on those. These generative models would be trained on a feature vector containing the true values of the variables and additional random variables to allow sampling from the distributions, with the target being either the reconstructed values or the resolution itself. Generative Adversarial Networks (GANs, \cite{goodfellow2014generative}), and the extension to conditional GANs (\cite{mirza2014conditional}), are the state of the art in terms of machine learning based generative models, but these are typically complicated to train and due to large architectures also have slower training and prediction rates than compared to other
machine learning approaches. Instead, we decided to focus on simpler, faster algorithms that can adequately sample from the resolution distribution given and conditional on the true momentum components.

\subsection{k Nearest Neighbours}

The k Nearest Neighbours algorithm (kNN) when using one neighbour, k=1, is essentially a fast nearest value list search, returning the closest value to the input variables in the training data. For this application, using the Toy simulation data from Section \ref{sec_toy}, the kNN algorithm was trained with the true values of particle momentum in p, $\theta$, and $\phi$ scaled between 0 and 1 as input and the difference between true and reconstructed values as output. This was done using the KNeighborsRegressor taken from the scikit-learn library in python (\cite{scikit-learn}), setting the number of nearest neighbours, k, to 1.

\begin{figure}[hbt!]
 \begin{center}
   \raisebox{0.5mm}{\includegraphics[trim = 2 2 2 2, clip, width=0.33\linewidth]{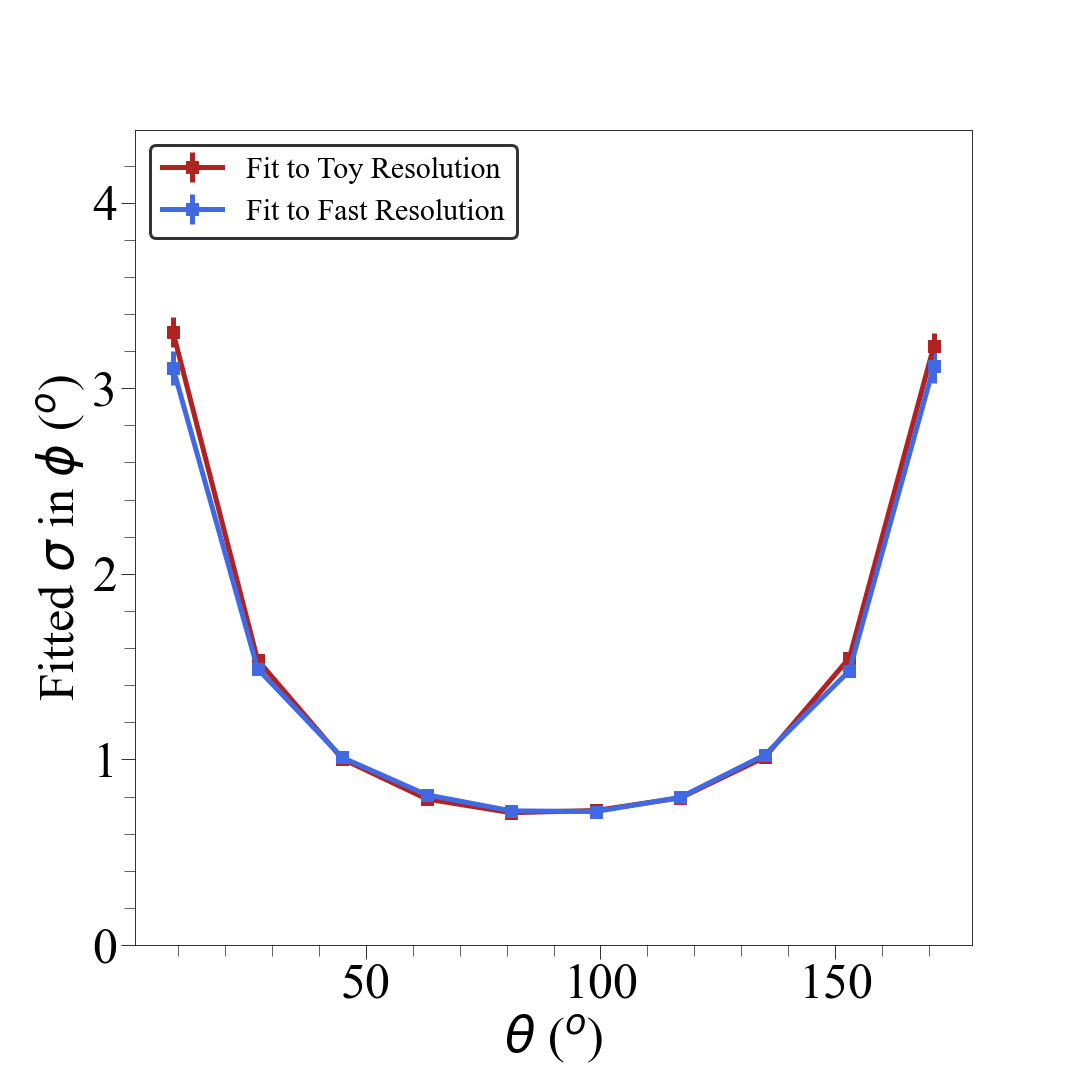}
   \includegraphics[trim = 2 2 2 2, clip, width=0.33\linewidth]{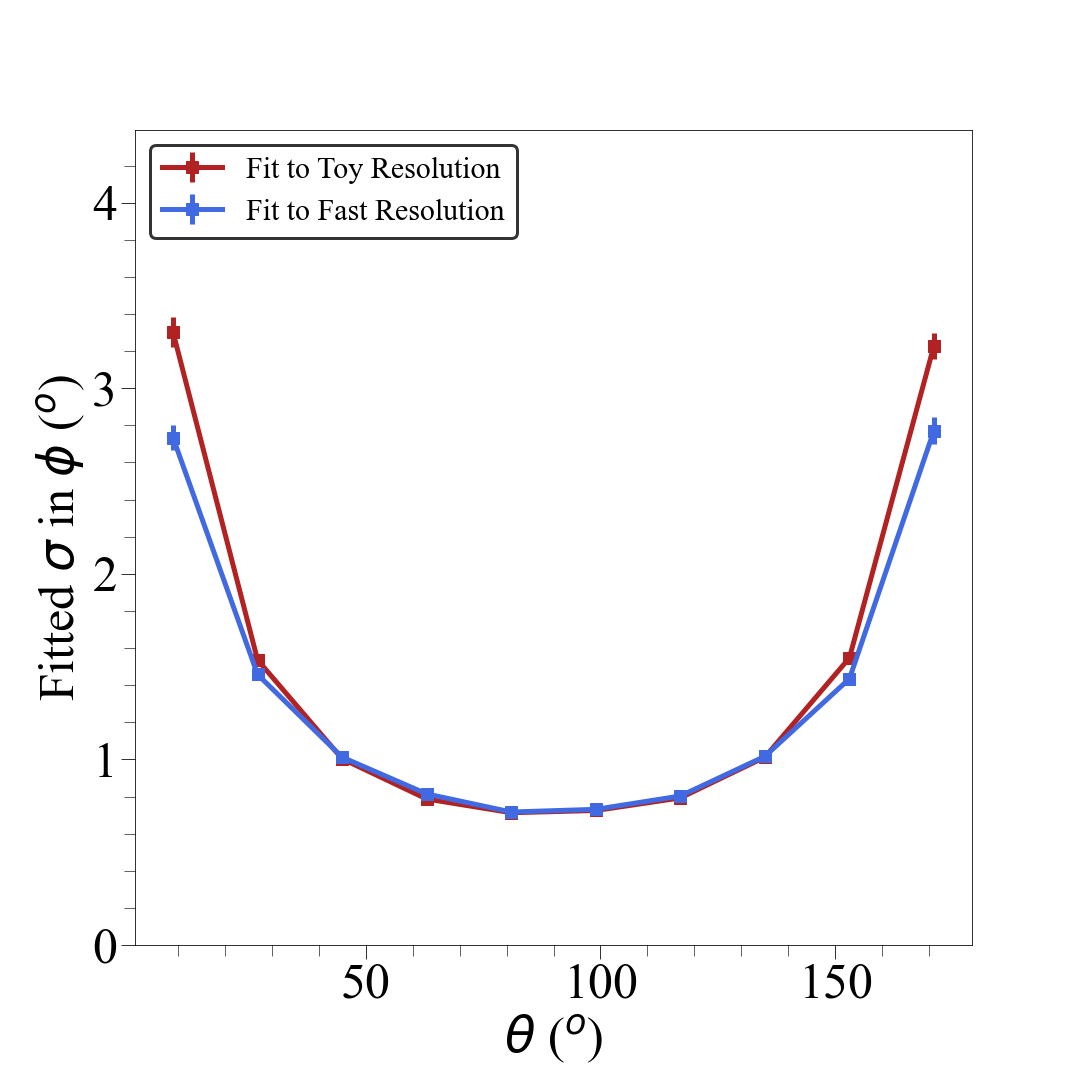}
   \includegraphics[trim = 2 2 2 2, clip, width=0.33\linewidth]{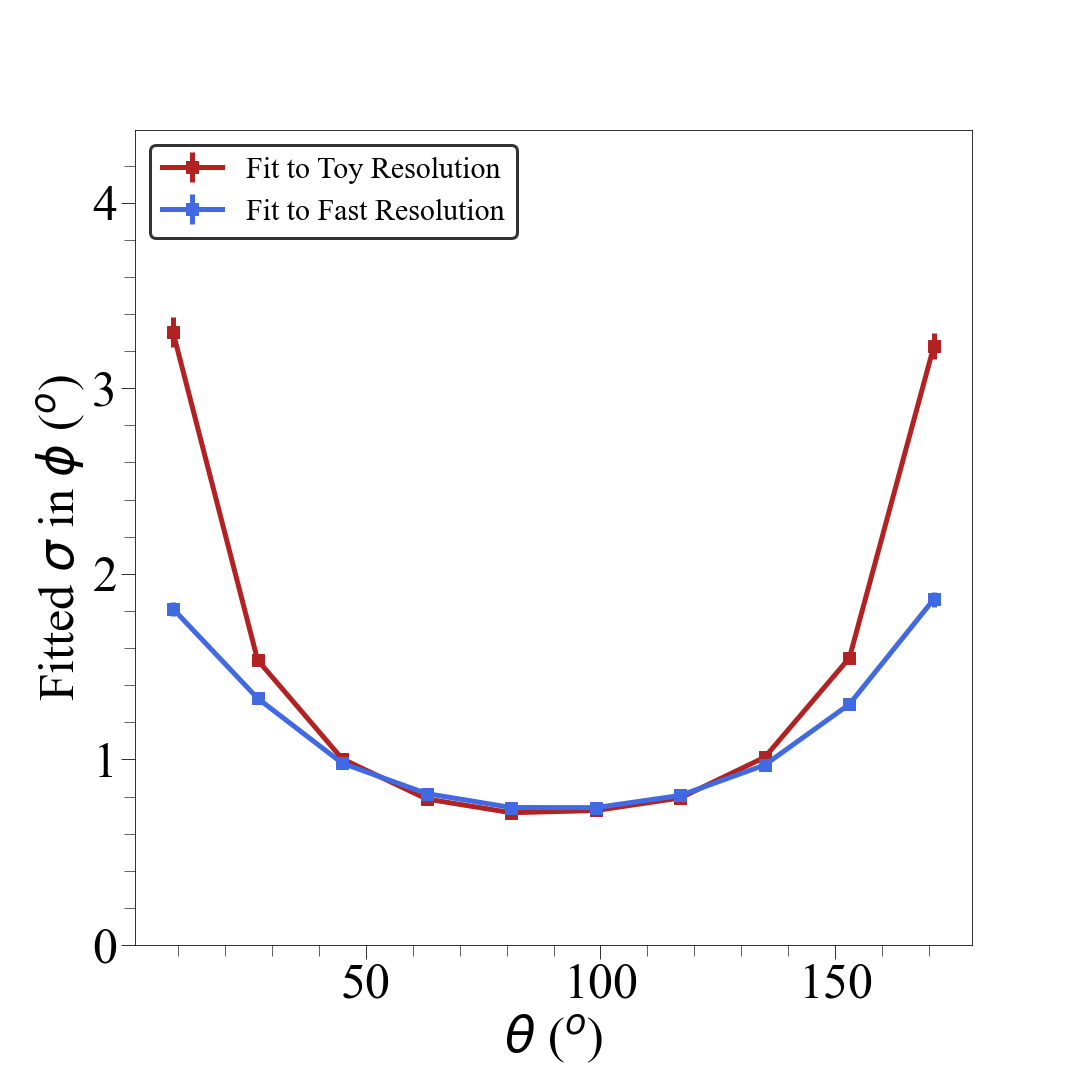}}
  \caption[]{ \label{fig:knn_resPhi} The fitted standard deviation of the resolution of $\phi$ in bins of $\theta$ for the Toy (red) and Fast (blue) simulation resolutions. The kNN was trained without any random inputs (left), one random input (centre) ranging from 0 to 1 and 4 random inputs (right) ranging from 0 to 1.}
 \end{center}
\end{figure}

Figure \ref{fig:knn_resPhi} shows the extracted width of a fitted normal distribution to the resolution of $\phi$ in bins of $\theta$. The red curves were made by fitting the Toy simulation resolution. The blue curves were made by fitting the Fast simulation resolution produced by the kNN. As shown, the fast changing resolution on $\phi$ as a function of $\theta$, re-scaled to its original size, is then almost perfectly reproduced in the leftmost plot of Fig \ref{fig:knn_resPhi}. Furthermore, the prediction rate of the kNN is of order $10^4 \mbox{s}^{-1}$ and therefore can be much faster than full detector simulations.

\begin{figure}[hbt!]
 \begin{center}
   \raisebox{0.5mm}{\includegraphics[trim = 2 2 2 2, clip, width=0.33\linewidth]{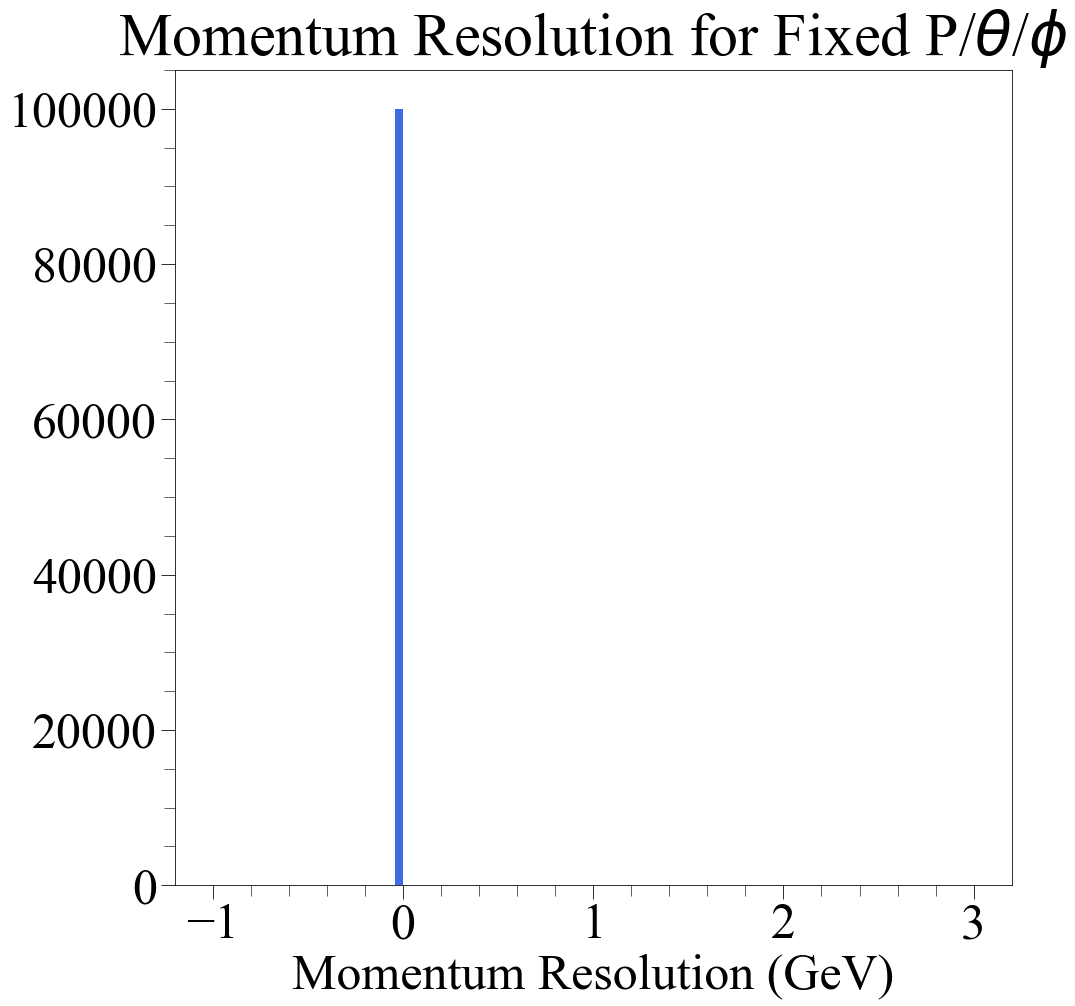}
   \includegraphics[trim = 2 2 2 2, clip, width=0.33\linewidth]{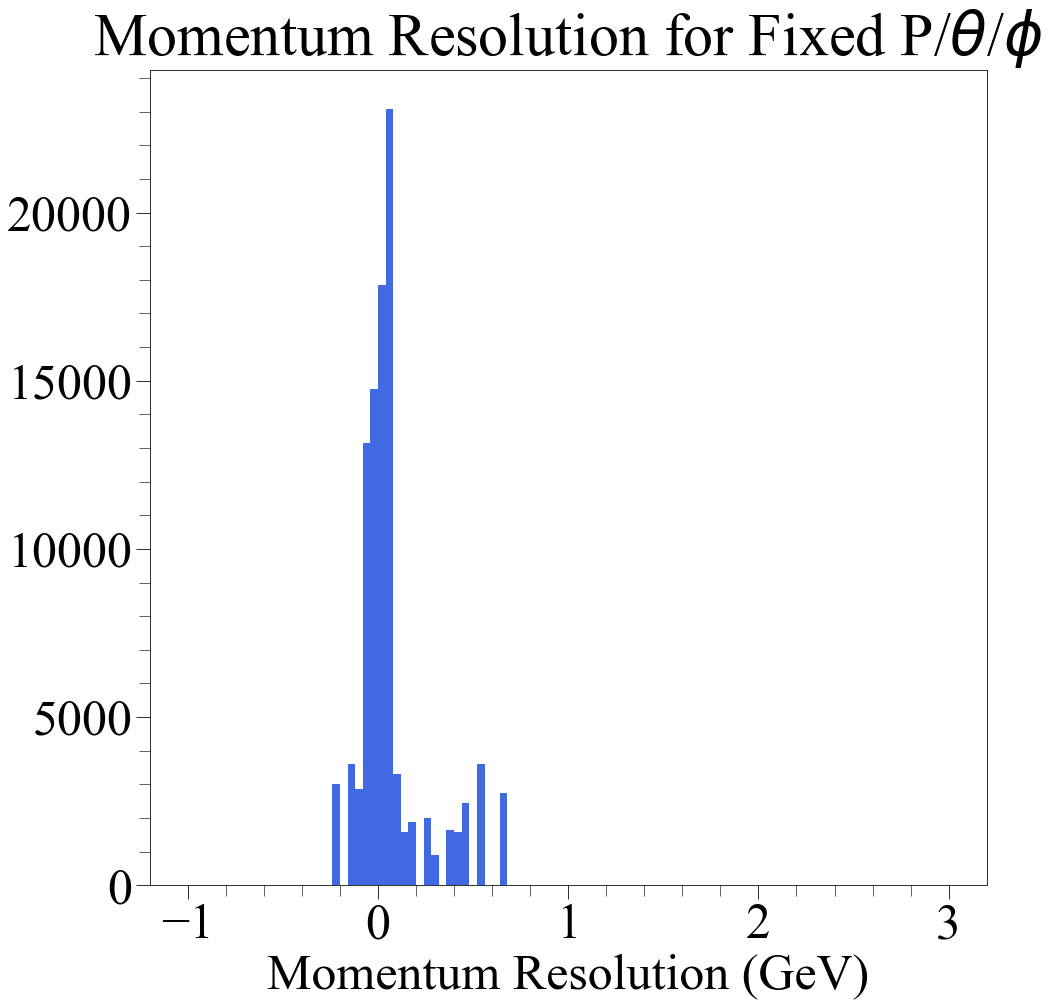}
   \includegraphics[trim = 2 2 2 2, clip, width=0.33\linewidth]{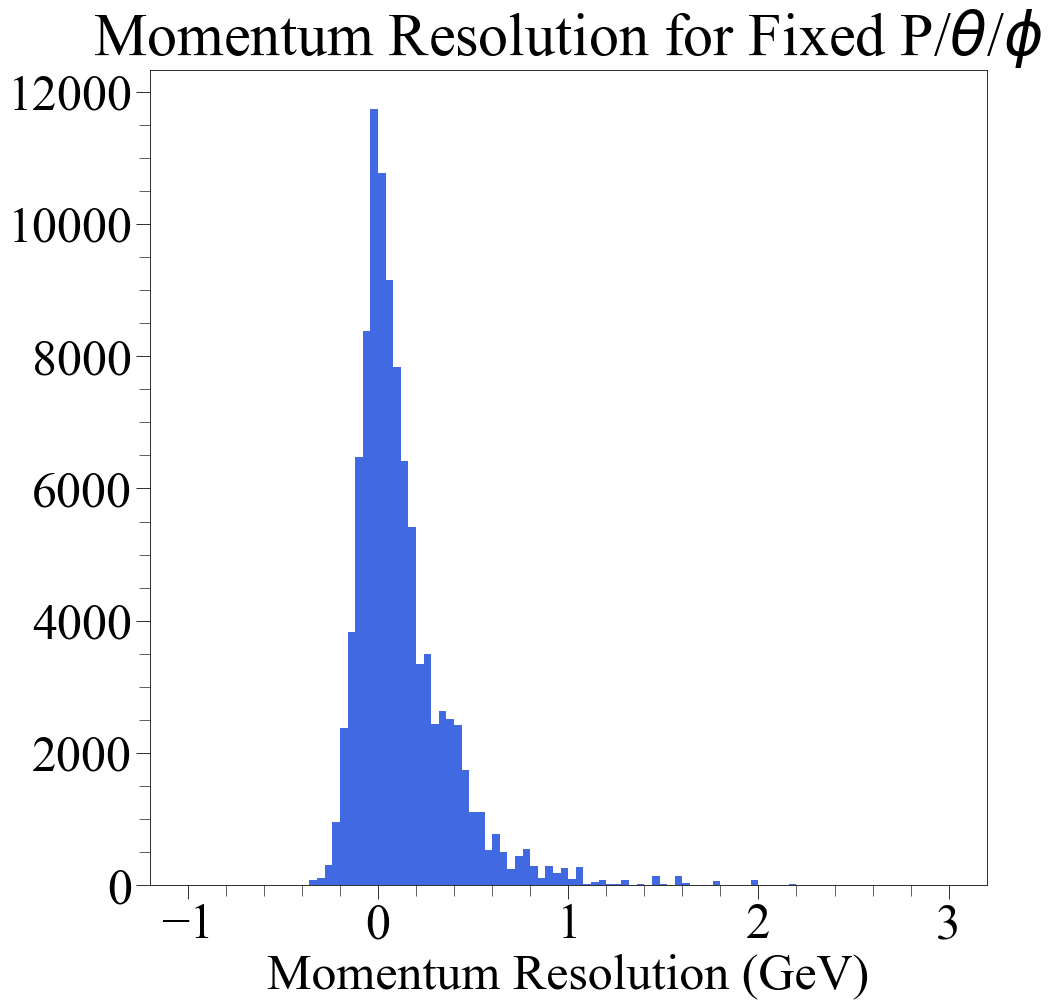}}
  \caption[]{ \label{fig:knn_spikesP} The resolution in momentum when the kNN was applied $10^5$ times to the same values of momentum, $\theta$, $\phi$. The kNN was trained without any random inputs (left), one random input (centre) ranging from 0 to 1 and 4 random inputs (right) ranging from 0 to 1.}
 \end{center}
\end{figure}

However, as the kNN is then effectively working as a nearest value list search, it can only return one value for the difference given p, $\theta$, and $\phi$, whereas for a full simulation you would generate a smooth resolution distribution if you used the same p, $\theta$, and $\phi$ repeatedly. Figure \ref{fig:knn_spikesP} left plot shows the resolution distribution predicted by the kNN for $10^5$ events with the same value inputs. As we expect only a single value for the difference is returned.

Adding an extra random input variable added some randomness to the predictions given the same values of p, $\theta$ and $\phi$, as seen in Figures \ref{fig:knn_spikesP} middle and right plots. Adding an extra random input gives another variable to be close to and there now is more than one nearest neighbour for fixed p, $\theta$, $\phi$ leading to multiple output values dependent on the random variable.
However, adding up to four random inputs whilst leading to a smooth distribution for repeated predictions significantly decreases the accuracy of the predictions, as shown in the corresponding plots of Figure \ref{fig:knn_resPhi}.

\begin{figure}[hbt!]
 \begin{center}
   \raisebox{0.5mm}{\includegraphics[trim = 2 2 2 2, clip, width=0.33\linewidth]{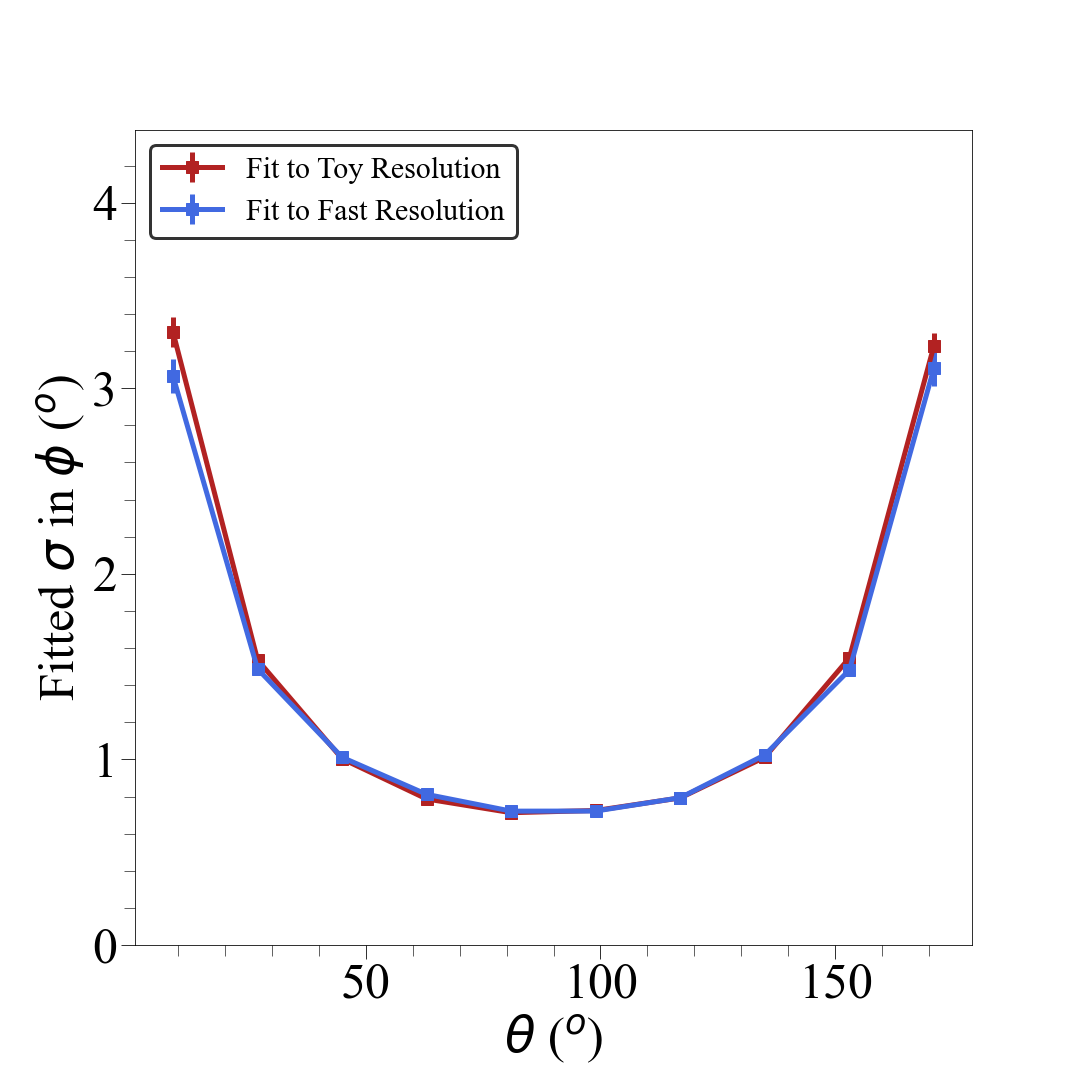}
   \includegraphics[trim = 2 2 2 2, clip, width=0.33\linewidth]{figs/reconstructionFigs/Fitsigma_Phi_Theta_1Reg_1rIN_0rOUT_1rINLim_kNN.png}
   \includegraphics[trim = 2 2 2 2, clip, width=0.33\linewidth]{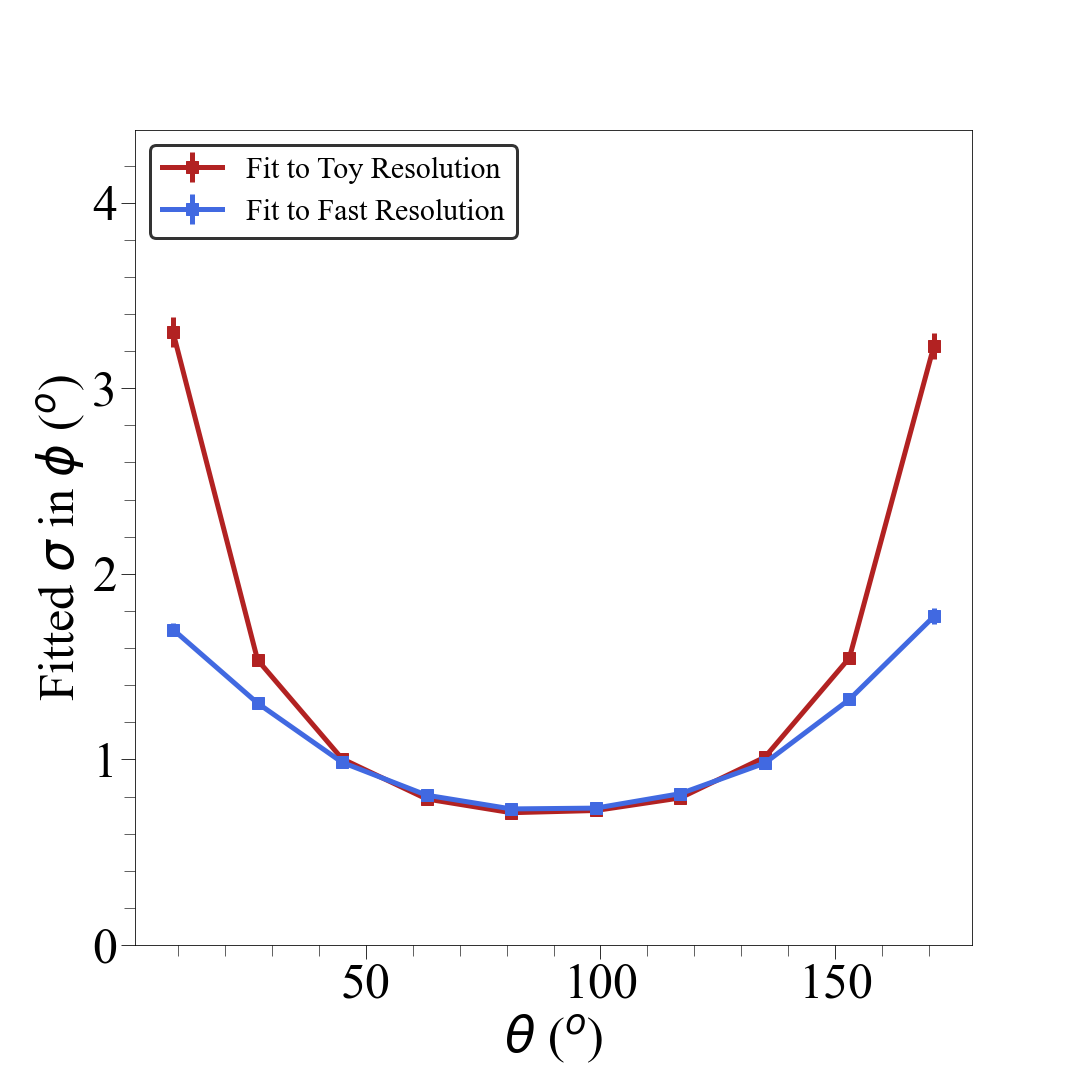}}
  \caption[]{ \label{fig:knn_resPhi_varRange} The fitted standard deviation of the resolution of $\phi$ in bins of $\theta$ for the Toy (red) and Fast (blue) simulation resolutions. Here the kNN was trained with one random input ranging from 0 to 0.01 (left), 0 to 1 (centre) and 0 to 100 (right).}
 \end{center}
\end{figure}

\begin{figure}[hbt!]
 \begin{center}
   \raisebox{0.5mm}{\includegraphics[trim = 2 2 2 2, clip, width=0.33\linewidth]{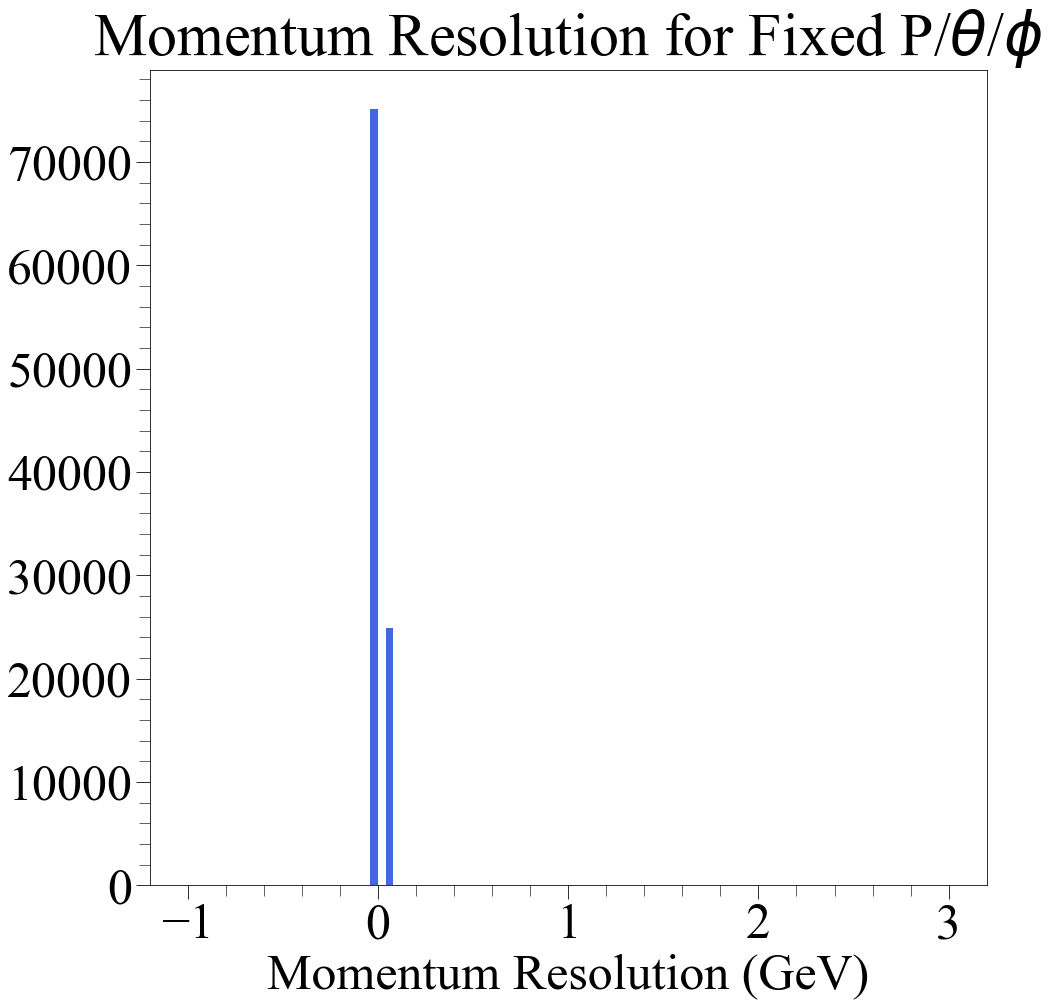}
   \includegraphics[trim = 2 2 2 2, clip, width=0.33\linewidth]{figs/reconstructionFigs/P_spikes_1Reg_1rIN_0rOUT_1rINLim_kNN.png}
   \includegraphics[trim = 2 2 2 2, clip, width=0.33\linewidth]{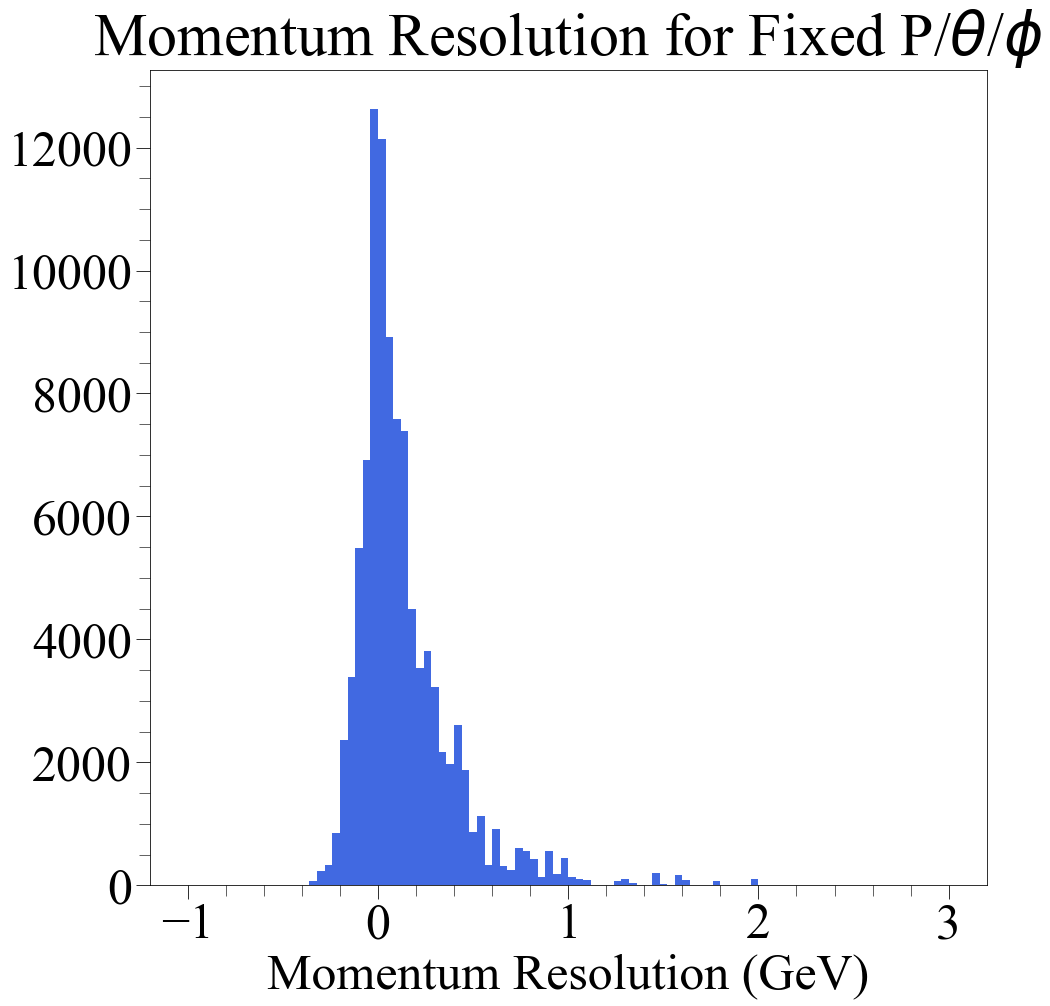}}
  \caption[]{ \label{fig:knn_spikesP_varRange} The resolution in momentum when the kNN was applied $10^5$ times to the same values of momentum, $\theta$, $\phi$. The kNN here was trained with one random input ranging from 0 to 0.01 (left), 0 to 1 (centre) and 0 to 100 (right).}
 \end{center}
\end{figure}

An alternative to this is to change the scale of the random input; as p, $\theta$ and $\phi$ are scaled between 0 and 1, a larger range for the random input gives it more weight in the evaluation of nearest neighbours with the Euclidean distance between data points. The effect, as seen in Figures \ref{fig:knn_resPhi_varRange} and \ref{fig:knn_spikesP_varRange}, is comparable to adding more random inputs: a larger range increases the smoothness of repeated predictions but decreases the accuracy of the predictions.

\subsection{Decision Trees}

\begin{figure}[hbt!]
 \begin{center}
   \raisebox{0.5mm}{\includegraphics[trim = 2 2 2 2, clip, width=0.33\linewidth]{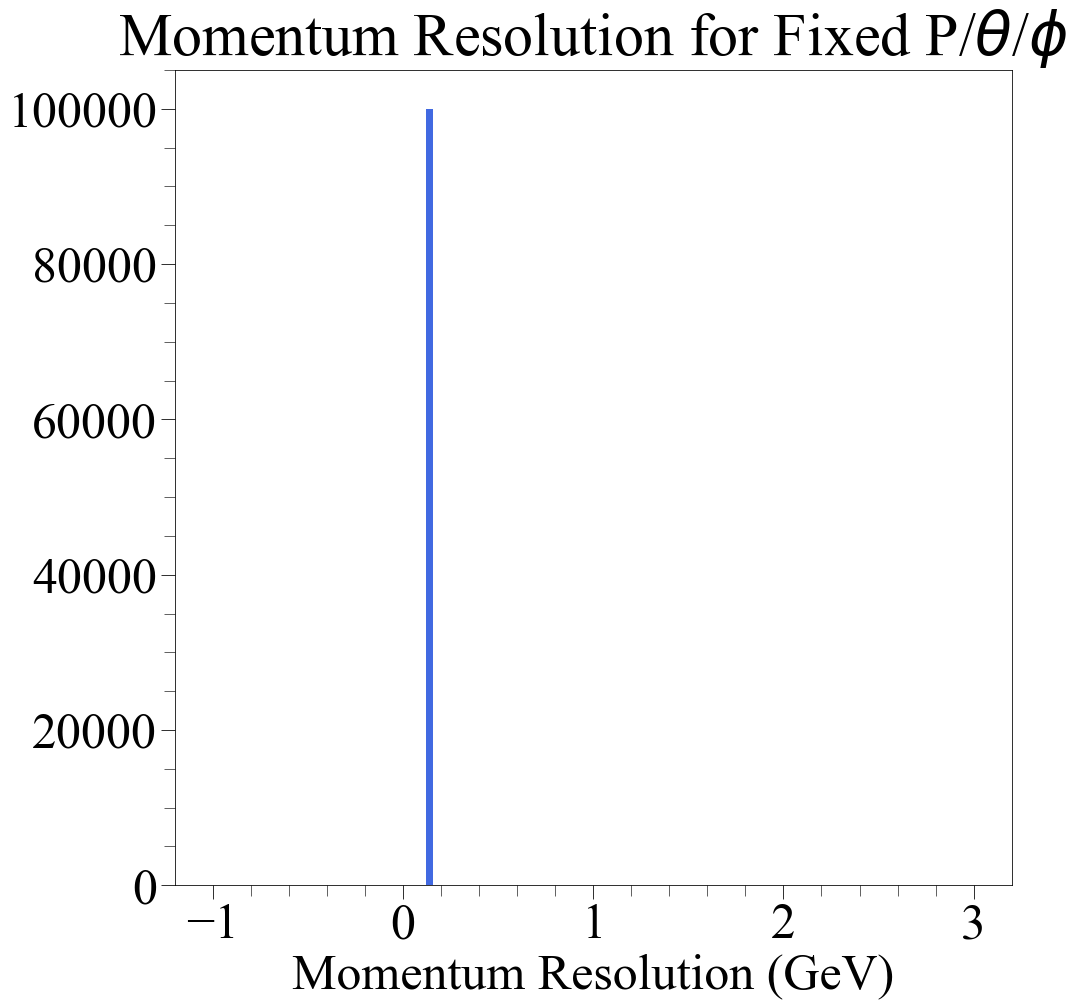}
   \includegraphics[trim = 2 2 2 2, clip, width=0.33\linewidth]{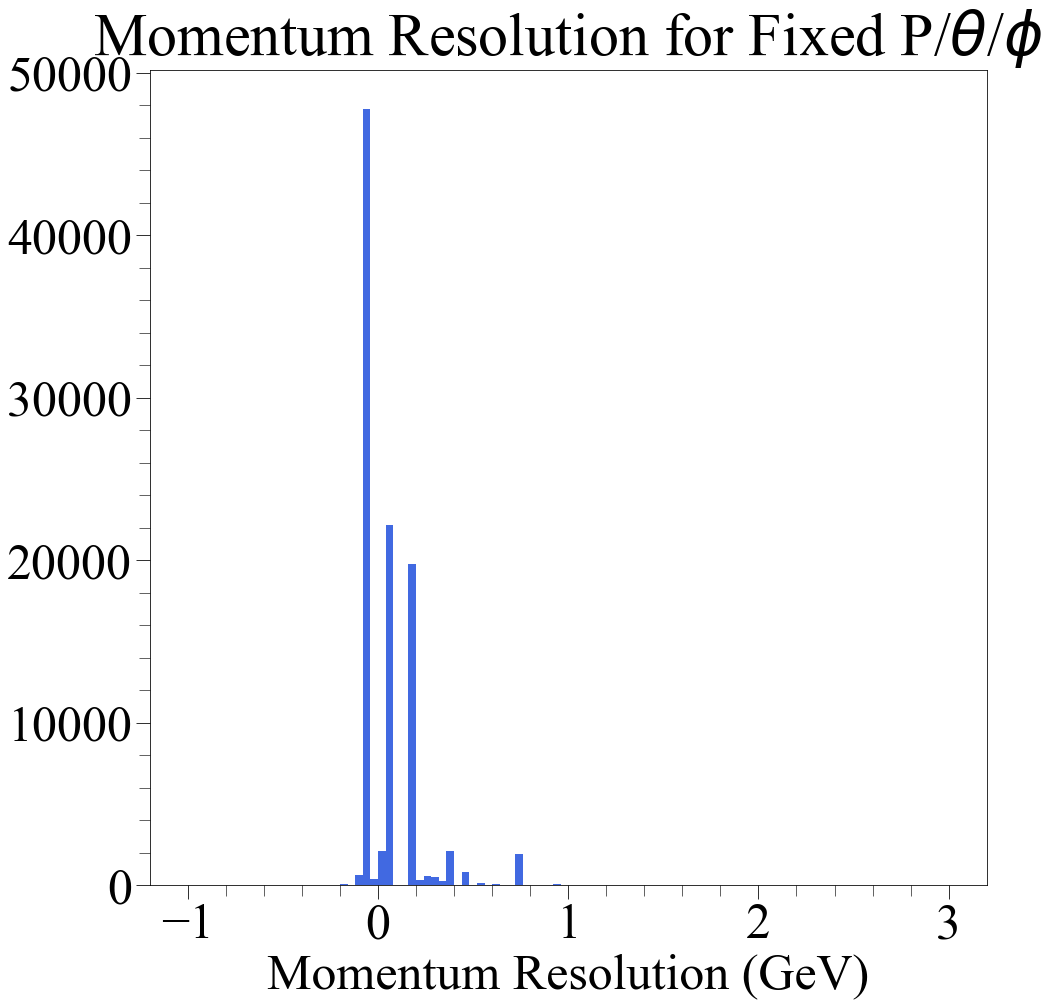}
   \includegraphics[trim = 2 2 2 2, clip, width=0.33\linewidth]{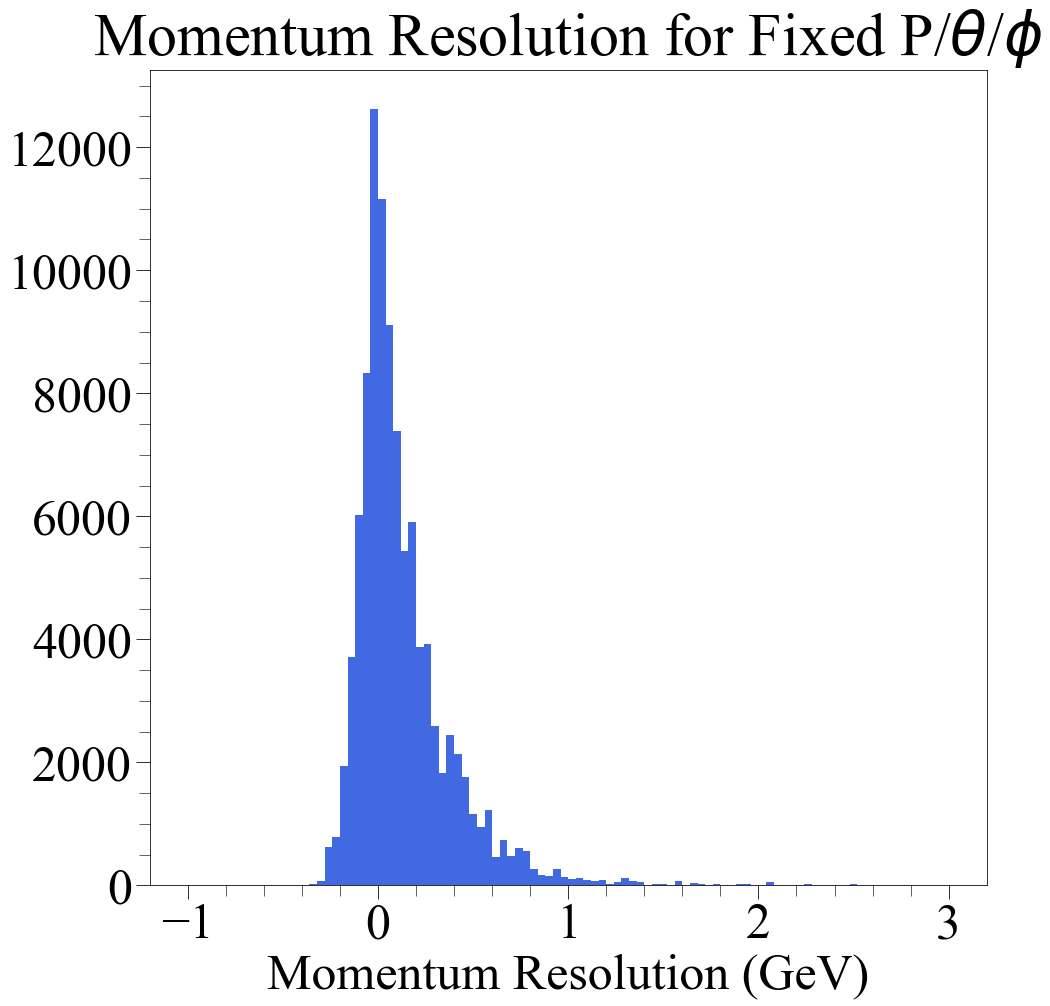}}
  \caption[]{ \label{fig:dt_spikesP} The resolution in momentum when the DT was applied $10^5$ times to the same values of momentum, $\theta$, $\phi$ with DTs trained without any random inputs (left), one random input (centre) and 10 random inputs (right) ranging from 0 to 1.}
 \end{center}
\end{figure}

\begin{figure}[hbt!]
 \begin{center}
   \raisebox{0.5mm}{\includegraphics[trim = 2 2 2 2, clip, width=0.33\linewidth]{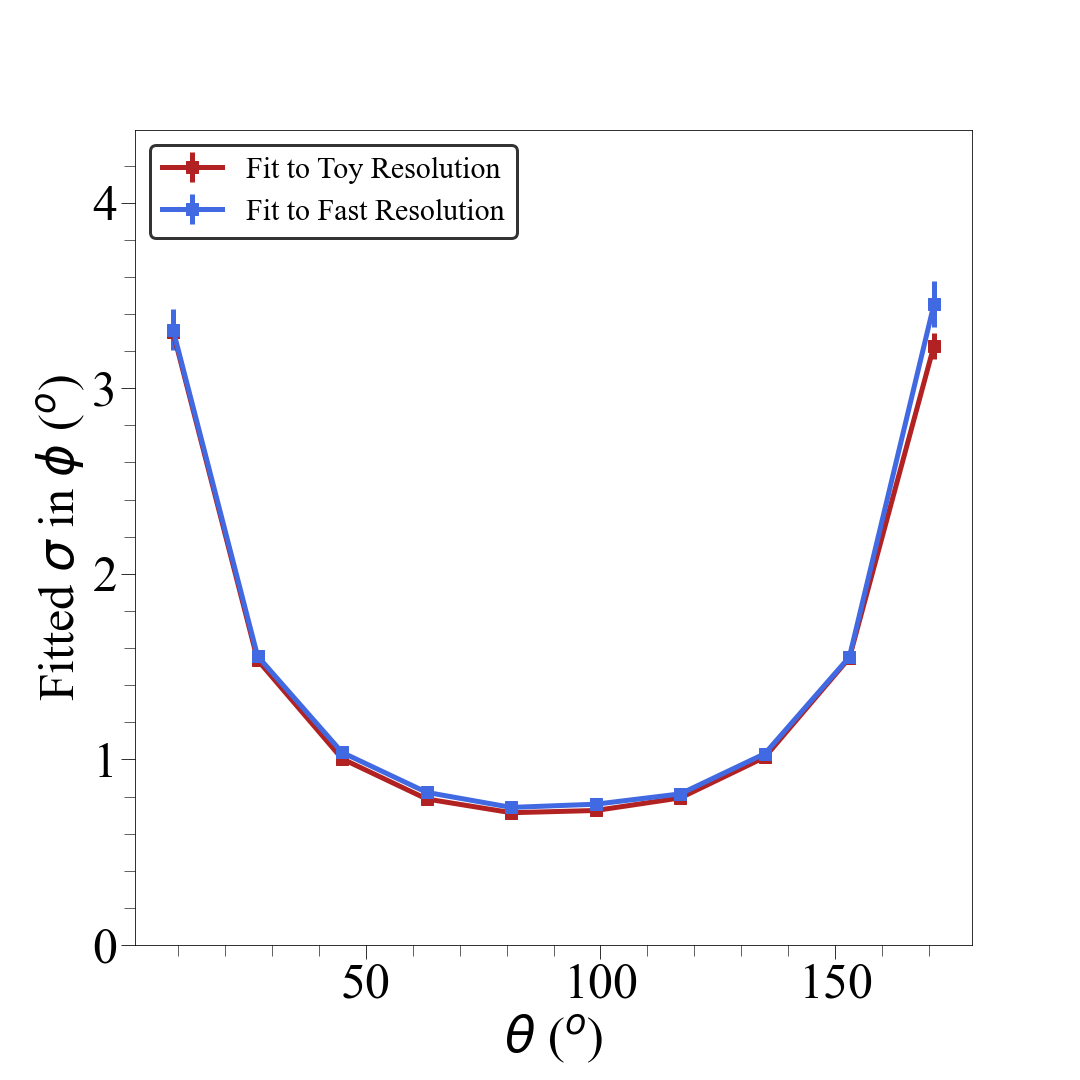}
   \includegraphics[trim = 2 2 2 2, clip, width=0.33\linewidth]{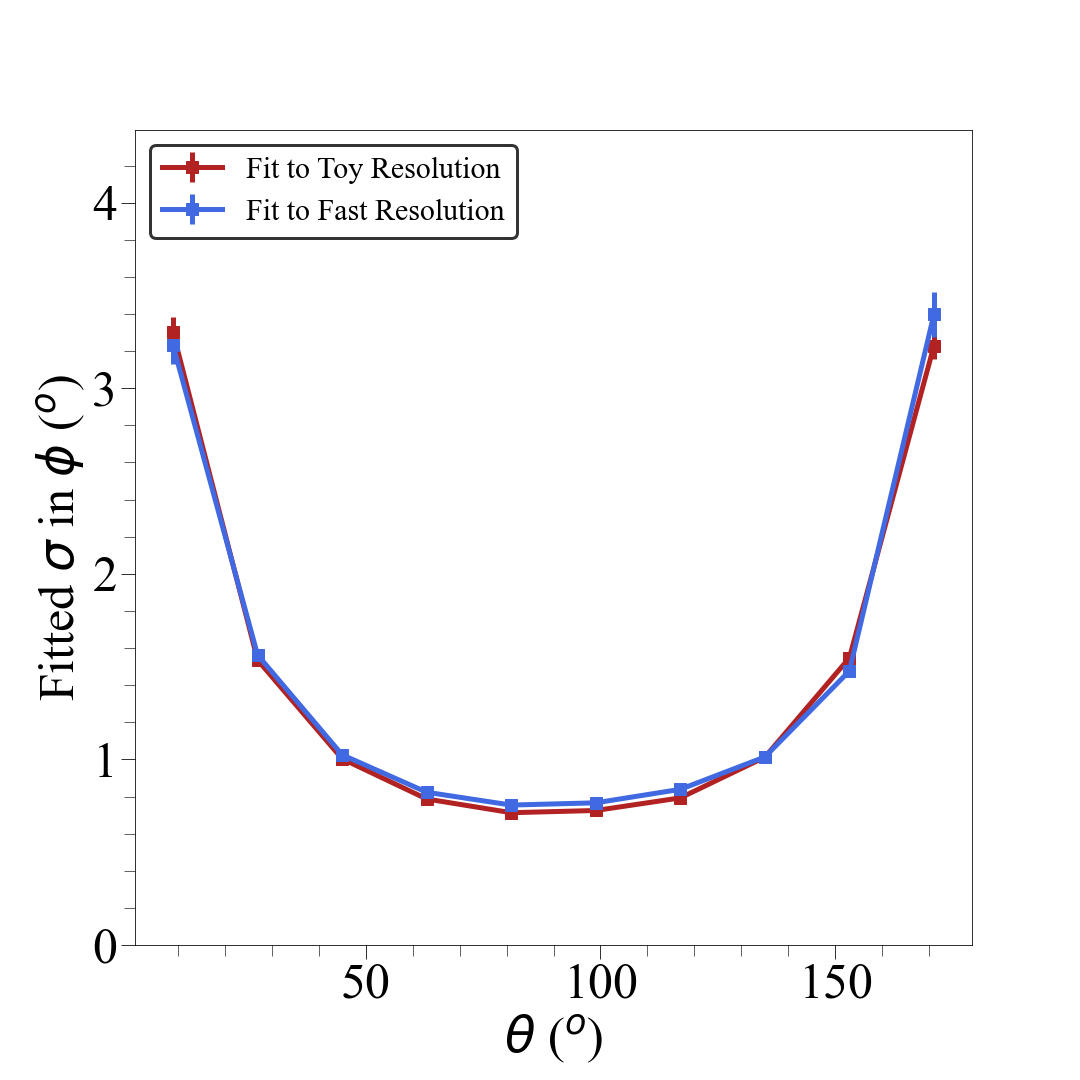}
   \includegraphics[trim = 2 2 2 2, clip, width=0.33\linewidth]{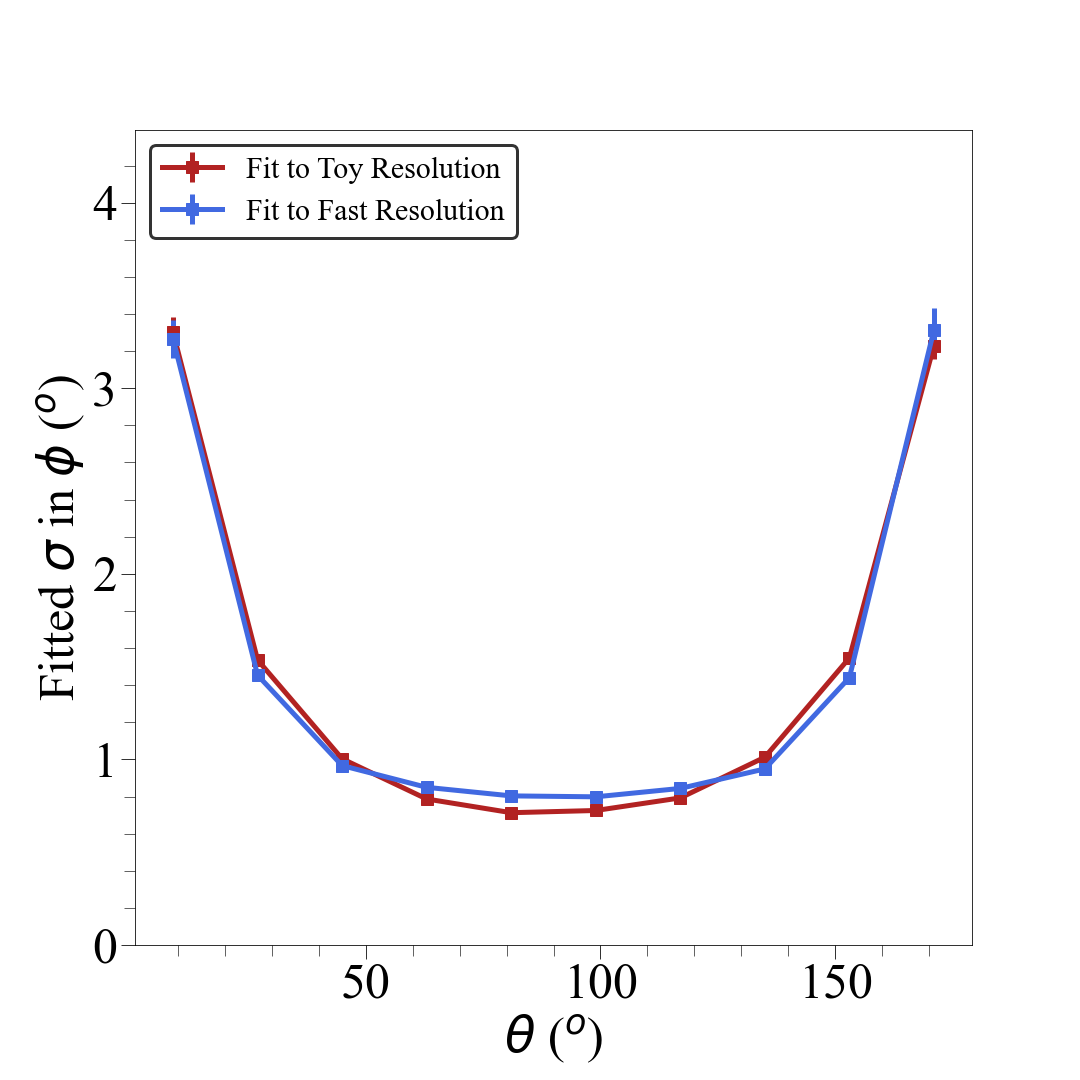}}
  \caption[]{ \label{fig:dt_resPhi} The fitted standard deviation of the resolution of $\phi$ in bins of $\theta$ for the Toy (red) and Fast (blue) simulation resolutions with DTs trained without any random inputs (left), one random input (centre) and 10 random inputs (right) ranging from 0 to 1.}
 \end{center}
\end{figure}

As an alternative to the kNN we used regression with a decision tree (DT) to work as a nearest value list search by over-fitting it to the training data, essentially creating as many leaves in the decision tree as there are training events. This was done with the scikit-learn DecisionTreeRegressor (\cite{scikit-learn}), with the default architecture. Similarly to the kNN, when training only using p, $\theta$, and $\phi$ as inputs the DT will only predict one value of the resolution difference for fixed values of inputs. Once again, by adding one or ten random input variables we add some randomness to these predictions, as shown for these three cases in Figure \ref{fig:dt_spikesP}. As before, adding more random inputs decreases the accuracy of the predictions, as shown in Figure \ref{fig:dt_resPhi} on the right. However, changing the range on the random input does not have an impact on the DT as they are essentially a series of yes or no conditions on the inputs. These conditions do not depend on the scale of the inputs as they can be scaled the same way.

\begin{figure}[hbt!]
 \begin{center}
   \raisebox{0.5mm}{\includegraphics[trim = 2 2 2 2, clip, width=0.33\linewidth]{figs/reconstructionFigs/P_spikes_1Reg_1rIN_0rOUT_1rINLim.png}
   \includegraphics[trim = 2 2 2 2, clip, width=0.33\linewidth]{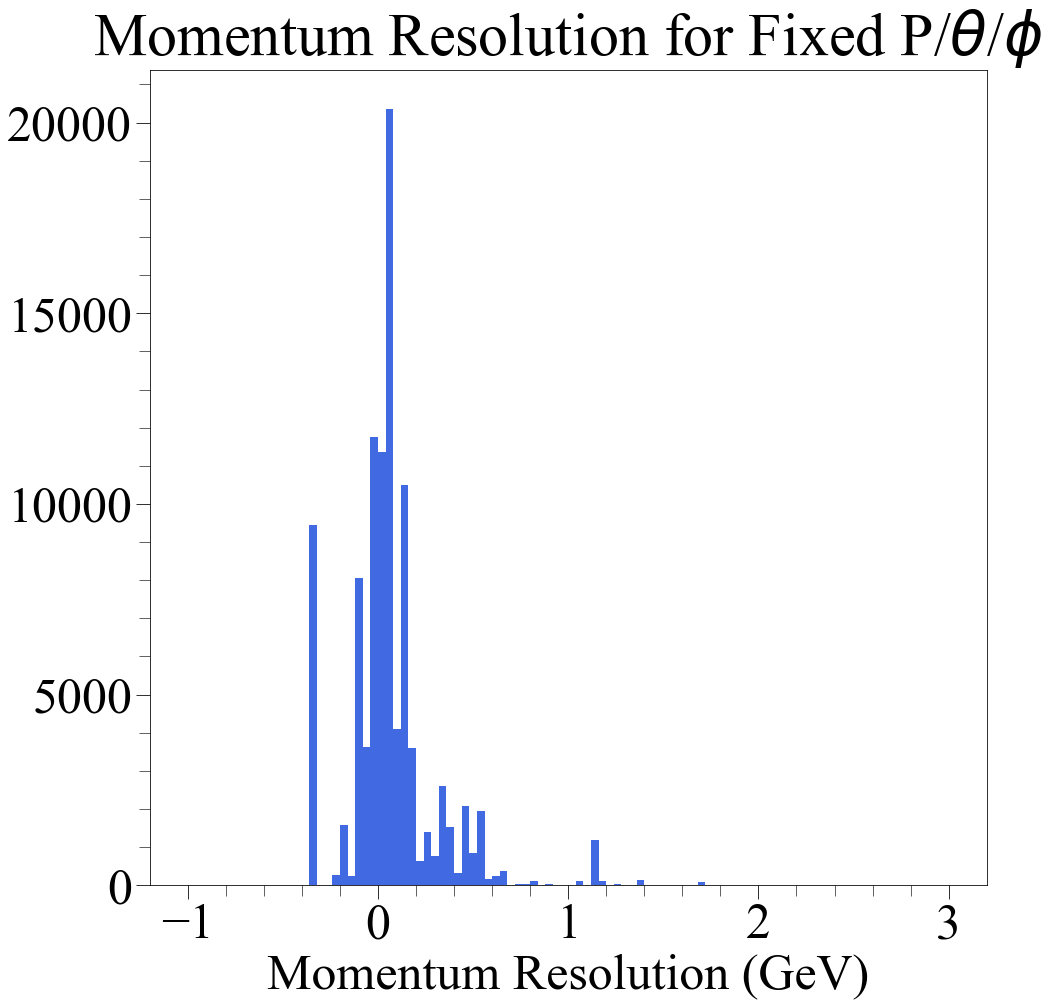}
   \includegraphics[trim = 2 2 2 2, clip, width=0.33\linewidth]{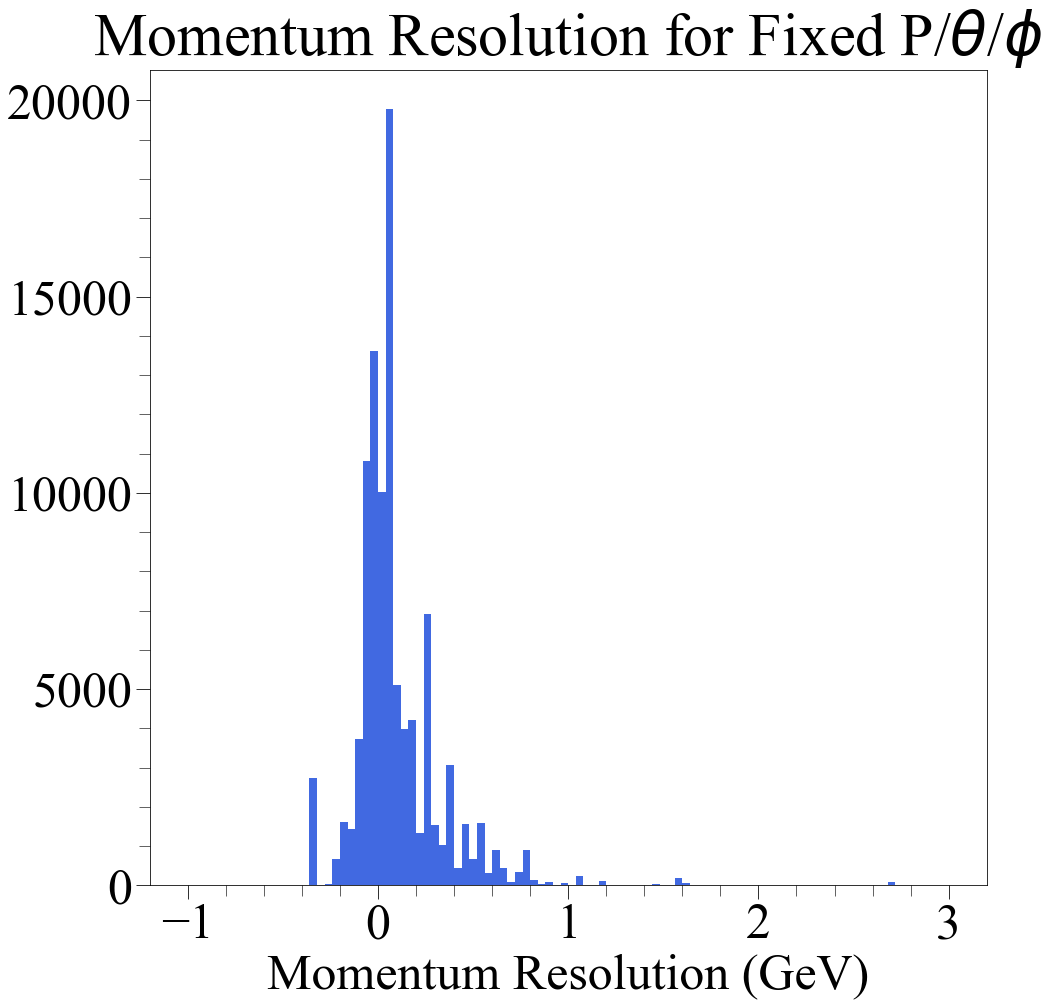}}
  \caption[]{ \label{fig:dtNReg_spikesP} The momentum resolution when 1 DT (left), 5 DTs (centre) and 50 DTs (right) each with one random input were applied $10^5$ times to the same values of momentum, $\theta$, $\phi$.}
 \end{center}
\end{figure}

\begin{figure}[hbt!]
 \begin{center}
   \raisebox{0.5mm}{\includegraphics[trim = 2 2 2 2, clip, width=0.33\linewidth]{figs/reconstructionFigs/Fitsigma_Phi_Theta_1Reg_1rIN_0rOUT_1rINLim.png}
   \includegraphics[trim = 2 2 2 2, clip, width=0.33\linewidth]{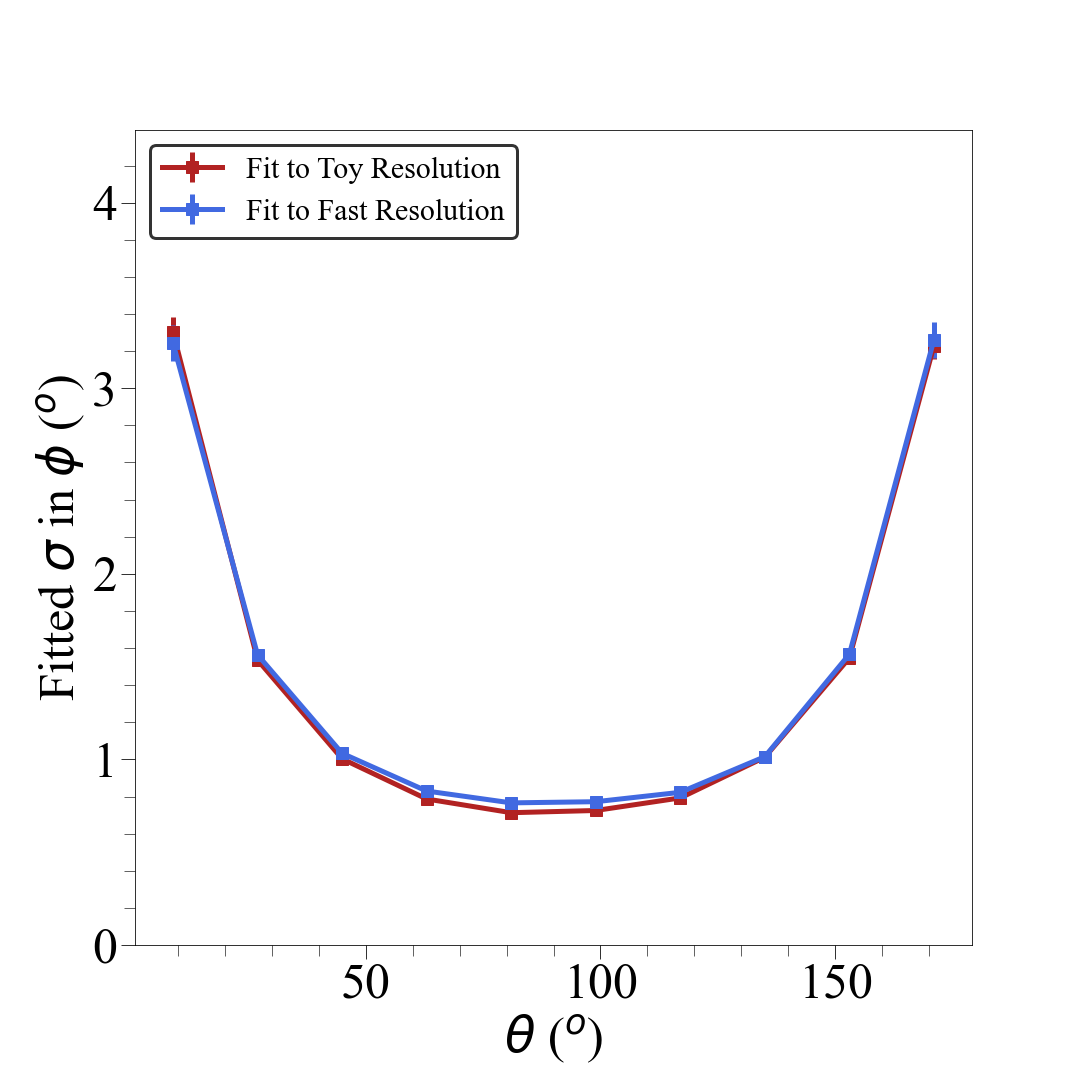}
   \includegraphics[trim = 2 2 2 2, clip, width=0.33\linewidth]{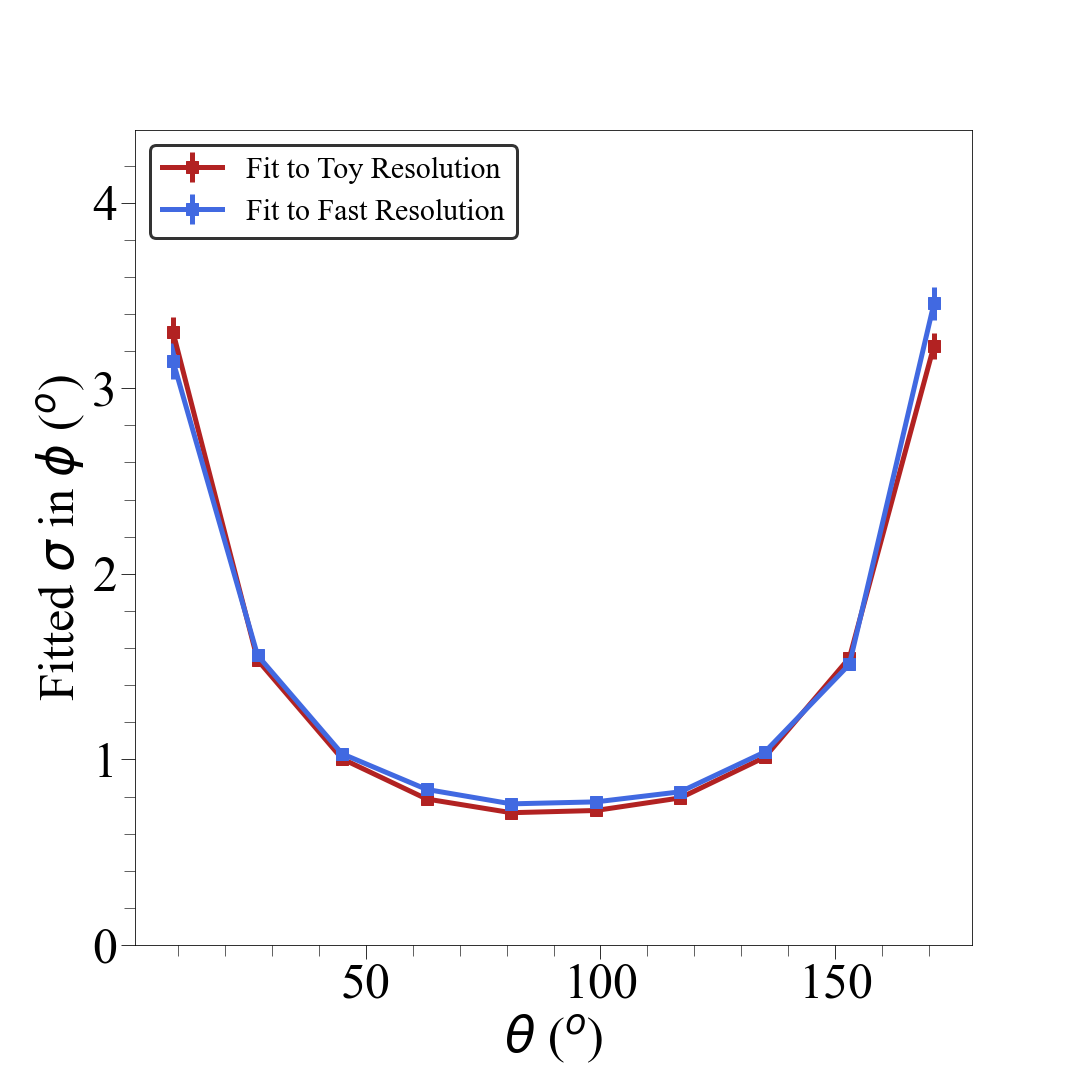}}
  \caption[]{ \label{fig:dtNReg_resPhi} The fitted standard deviation of the resolution of $\phi$ in bins of $\theta$ for the Toy (red) and Fast (blue) simulation resolutions with 1 DT (left), 5 DTs (centre) and 50 DTs (right) each with one random input.}
 \end{center}
\end{figure}

In order to improve the smoothness of the distribution, without compromising its reproduction of dependencies, we instead trained N DTs with the same original input values of p, $\theta$, $\phi$ and output differences, but with different sets of random inputs. During prediction a DT was picked at random for each event. As each DT was trained with different random inputs they predicted different outputs helping smooth the distribution when predicting on fixed values of p, $\theta$ and $\phi$, as shown in Figure \ref{fig:dtNReg_spikesP} for 1, 5 and 50 DTs. However, as they were trained on the same data set they mapped the resolution and its dependencies equally well, as shown in Figure \ref{fig:dtNReg_resPhi} for 1, 5 and 50 DTs. As only one DT is used for each prediction, picking from N DTs does not impact the prediction rate which is of order $10^5\mbox{s}^{-1}$, an order of magnitude faster than the kNN and again much faster than full detector simulations.

\clearpage

\subsection{Quantifying Performance}

We performed a number of quantitative checks on how the smoothness and quality of the predictions vary as a function of random inputs and number of kNNs or DTs. We also tried adding random outputs to see if adding additional outputs affected the overall quality of prediction, and varying the upper limit of the range of the random input. Finally we varied the scaling of outputs.

The $S$ value measured the smoothness of the histogrammed resolution prediction on fixed values of momentum components and was defined as 

\[ S = \sum_n \frac{|b_{n} - b_{n+1}|}{2*N} \]

for $b_{n}$ the number of counts in the $n^{th}$ bin and $N$ the number of times the prediction was repeated. A value of $S$=1 means that all the predictions are in the same bin. 

To measure the quality of the prediction we used a least squares error ($LSE$) value defined as
\[ LSE=\sum \frac{(\sigma_{obs}- \sigma_{exp})^2}{\sigma_{exp}} \]

where $\sigma_{obs}$ was the fitted resolution width of the fast simulation, and $\sigma_{exp}$ is that for the Toy detector simulation; the sum was made over several bins in p, $\theta$ and $\phi$ and summed over all three outputs. A value of $LSE=0$ would denote a perfectly accurate fast simulation

These tests were then performed by changing the number of DTs, number of random inputs, number of random outputs, upper limit of the range of the random input and scaling of non random outputs on $10^5$ training and testing events. The plots of $LSE$ and the $S$ metric can be found in Appendix \ref{appendix_a}. To summarise, these plots show that the smoothness of predictions improves when adding up to 50 DTs before reaching a plateau, whilst this does not affect the quality or predictions. Adding additional kNNs however did not affect either the $LSE$ or $S$ metrics. For both DTs and kNNs, adding additional random inputs improved the smoothness of predictions whilst decreasing the accuracy of predictions, leading to a balancing game between the two metrics. Varying the range of the random input for the kNN lead to a similar balancing game as adding more random inputs. For both DTs and kNNs, adding additional random outputs did not affect either metrics. This is important when considering the potential to additionally simulate lower level variables, such as the response of calorimeters or time-of-flight detectors. Finally, varying the scale of the non-random outputs affected the $LSE$ metric without impacting the smoothness of predictions. It was observed that when one output has a larger range than the other two the $LSE$ increased. Therefore, it is advised to scale the outputs to the same range prior to training, then perform the inverse transform to the output predictions.

\subsection{Comparing Fast to Toy simulations}

\begin{figure}[hbt!]
 \begin{center}
   \raisebox{0.5mm}{\includegraphics[trim = 2 2 2 2, clip, width=0.33\linewidth]{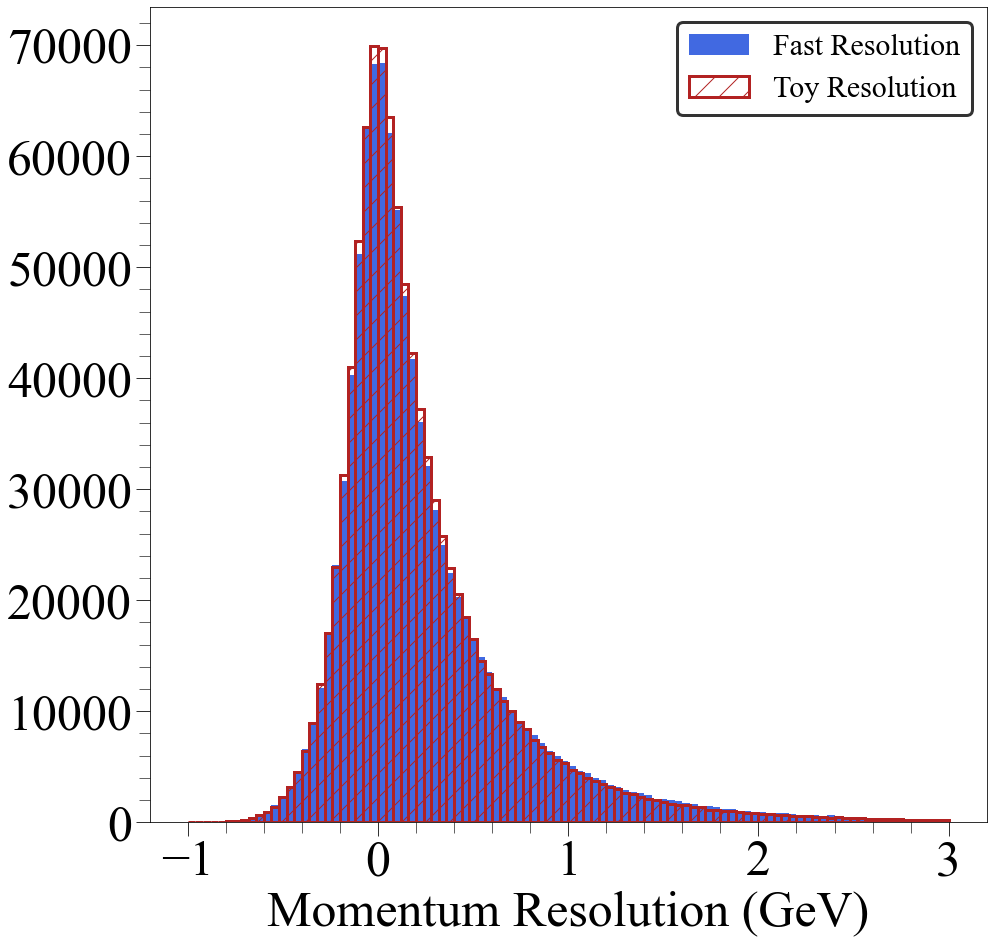}
   \includegraphics[trim = 2 2 2 2, clip, width=0.33\linewidth]{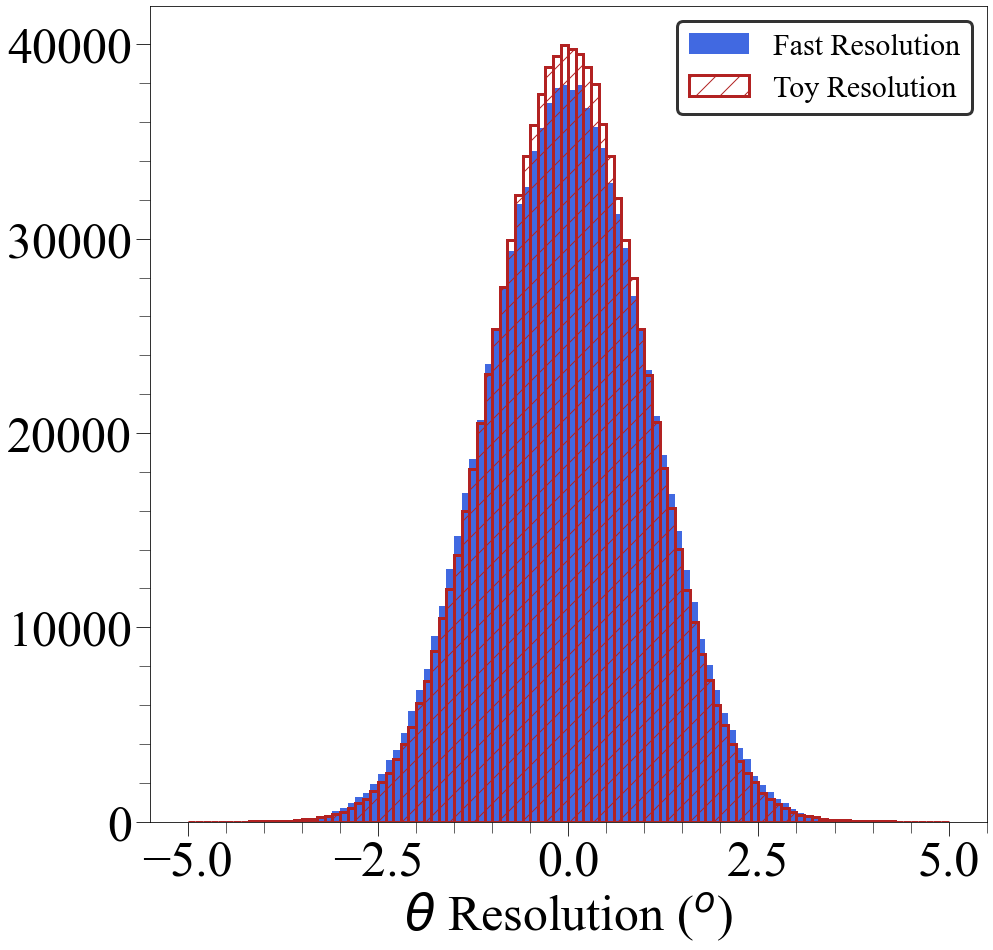}
   \includegraphics[trim = 2 2 2 2, clip, width=0.33\linewidth]{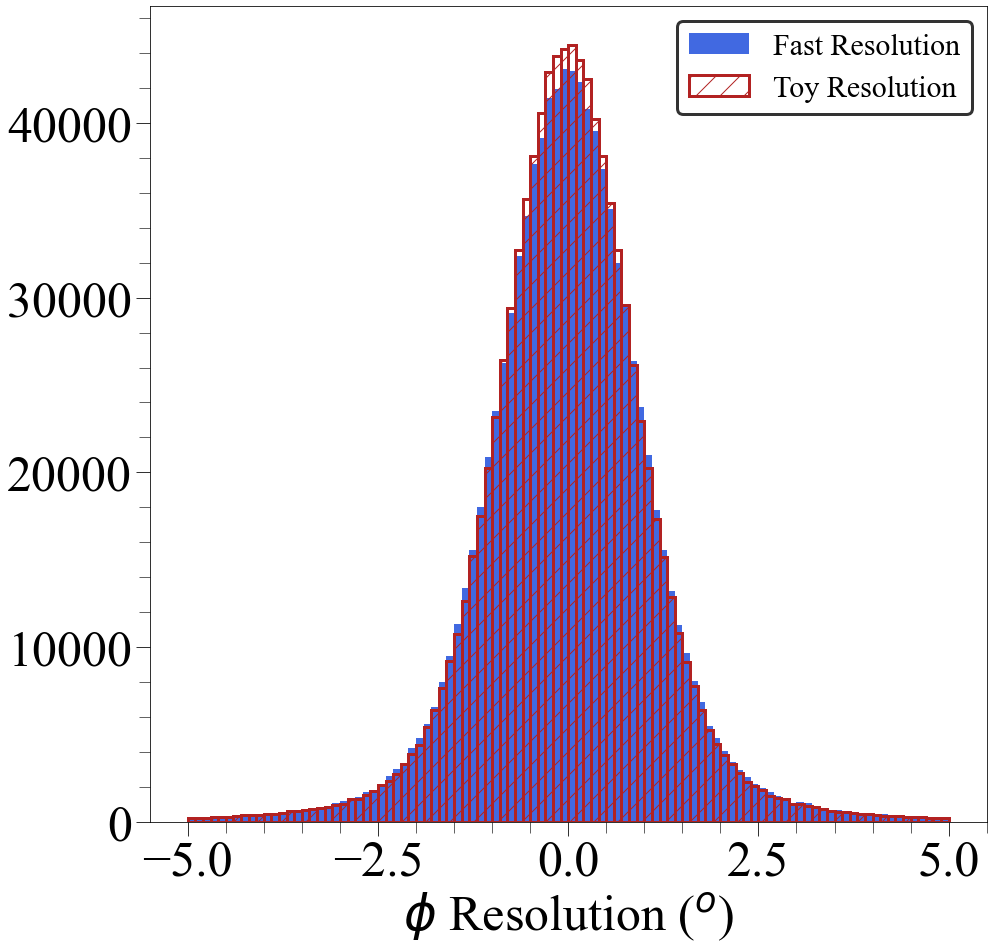}}
  \caption[]{ \label{fig:dt_resAll} The Fast (blue) and Toy (red) momentum, $\theta$, $\phi$ resolutions.}
 \end{center}
\end{figure}

\begin{figure}[hbt!]
 \begin{center}
   \raisebox{0.5mm}{\includegraphics[trim = 2 2 2 2, clip, width=0.33\linewidth]{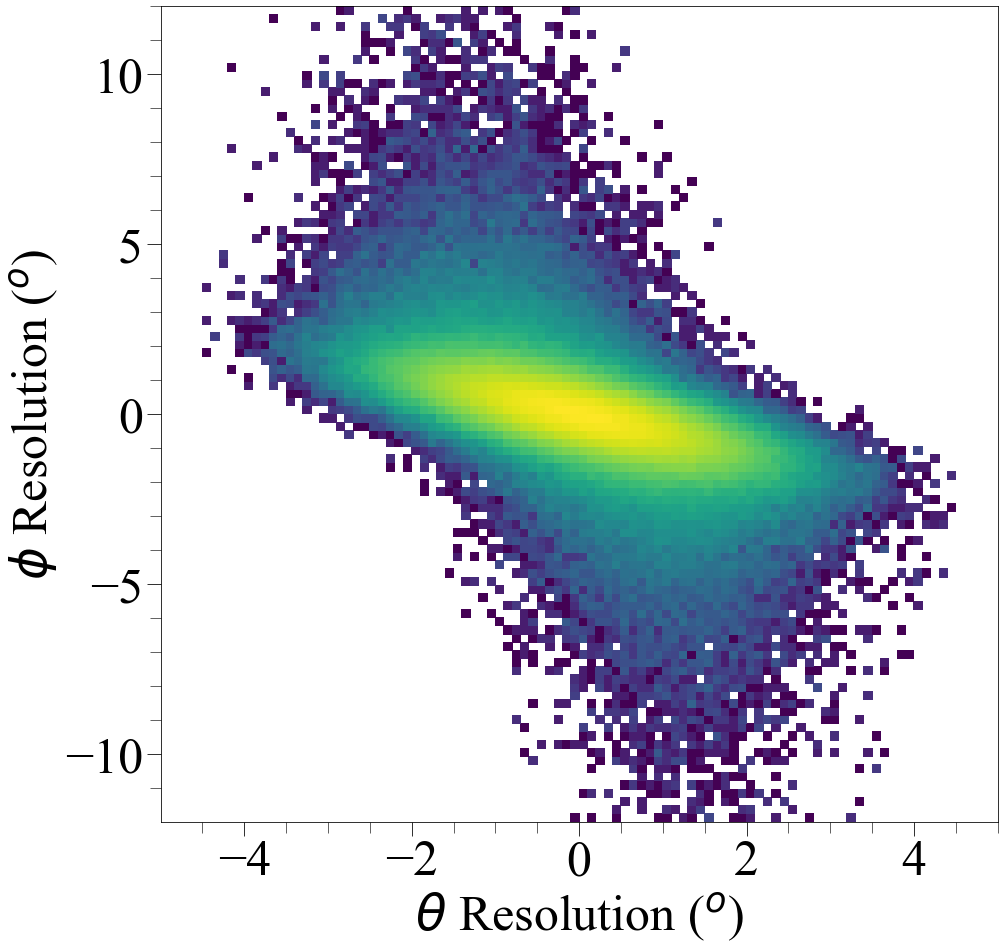}
   \includegraphics[trim = 2 2 2 2, clip, width=0.33\linewidth]{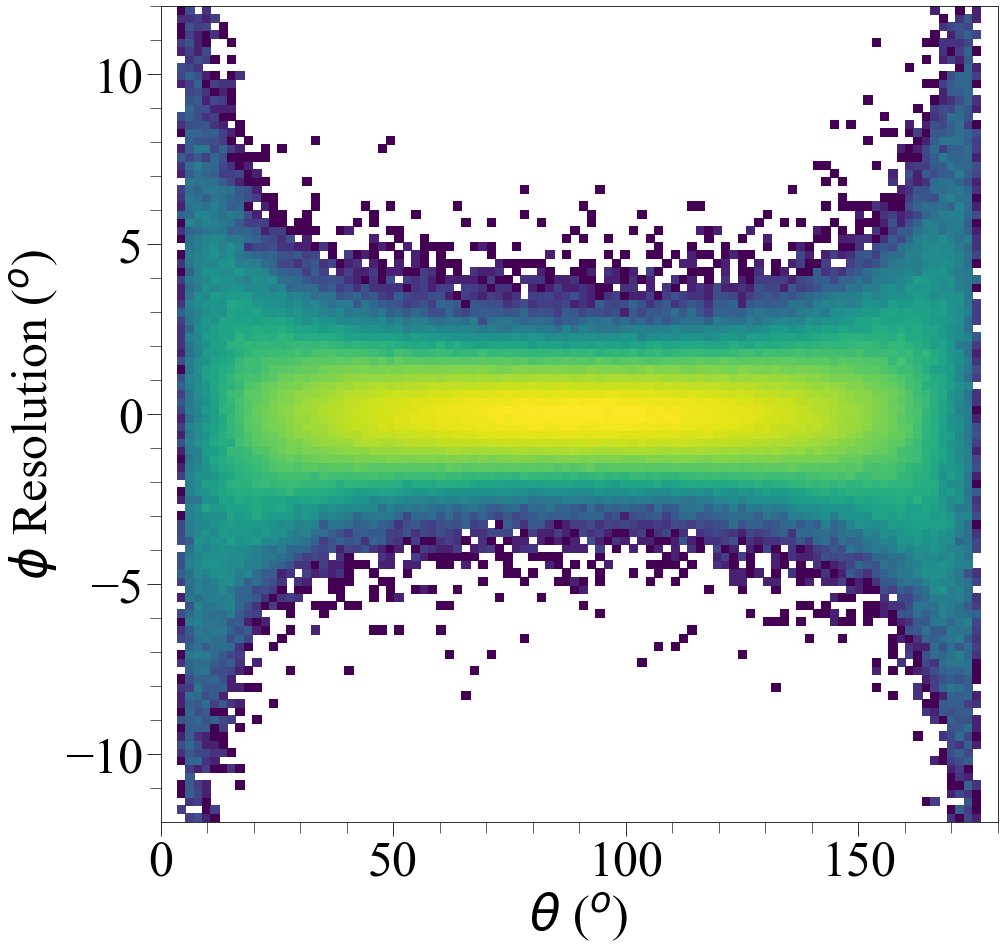}
   \includegraphics[trim = 2 2 2 2, clip, width=0.33\linewidth]{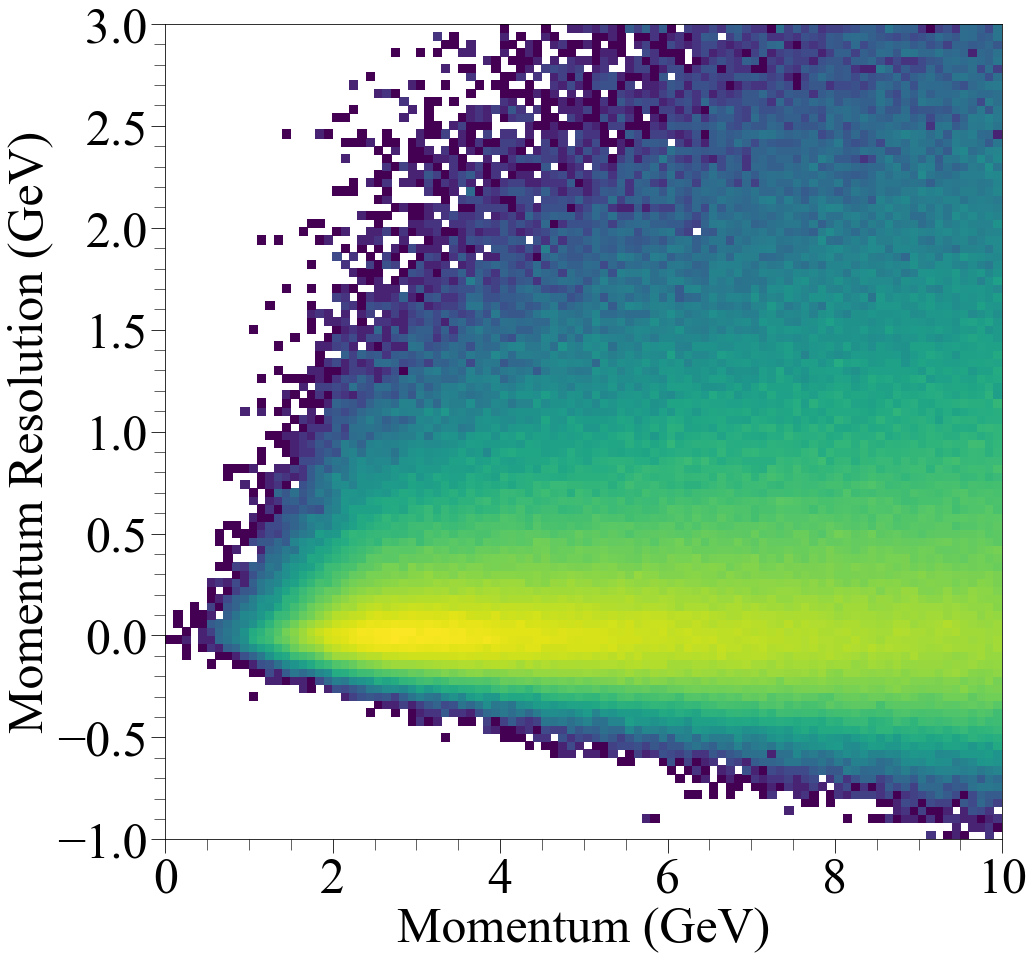}}
  \caption[]{ \label{fig:dt_resAll_2D} From left to right, the resolution of $\theta$ as a function of the resolution of $\phi$, the resolution of $\phi$ as a function of $\theta$ and the resolution of momentum as a function of momentum.}
 \end{center}
\end{figure}

After the tests above we chose 50 DTs, 5 random inputs ranging from 0 to 1, 0 random outputs and non random outputs all scaled between 0 and 100 to perform further studies. We increased the number of training events to $10^6$ which would be reasonable to attain with a full simulation. The algorithm shown here achieved an $S$ metric value of 0.09, and a $LSE$ of 0.2. In Figure \ref{fig:dt_resAll} we compare the momentum component differences for both Fast and Toy simulations integrated over all input events. The Fast simulation provided an excellent approximation of the Toy simulation, including tails in the momentum distribution. In Figure \ref{fig:dt_resAll_2D} we show the 2D distribution of the Fast simulation which replicate Figure \ref{fig:toy_2DRes} demonstrating that the multidimensional correlations were retained.

\begin{figure}[hbt!]
 \begin{center}
   \raisebox{0.5mm}{\includegraphics[trim = 2 2 2 2, clip, width=0.33\linewidth]{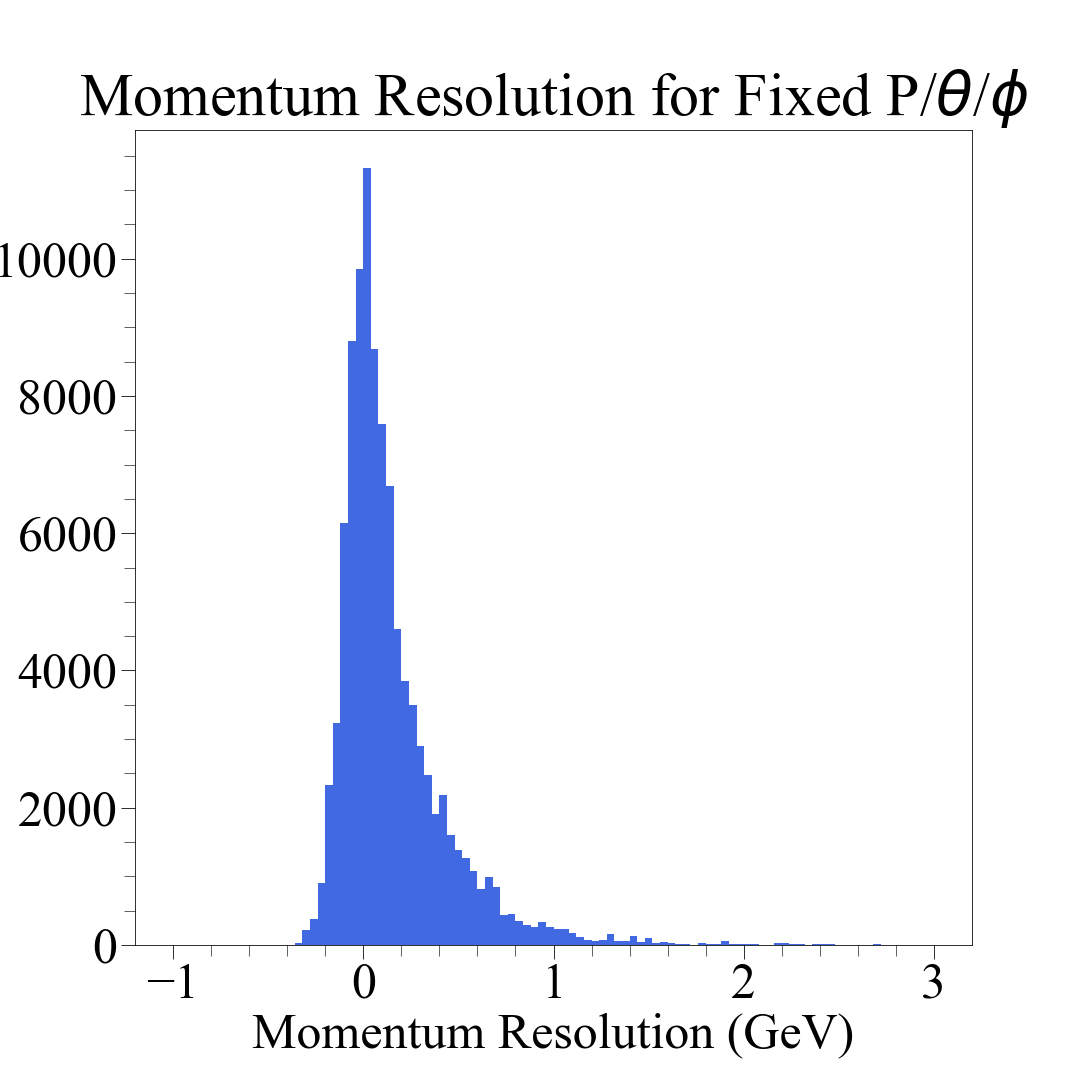}
   \includegraphics[trim = 2 2 2 2, clip, width=0.33\linewidth]{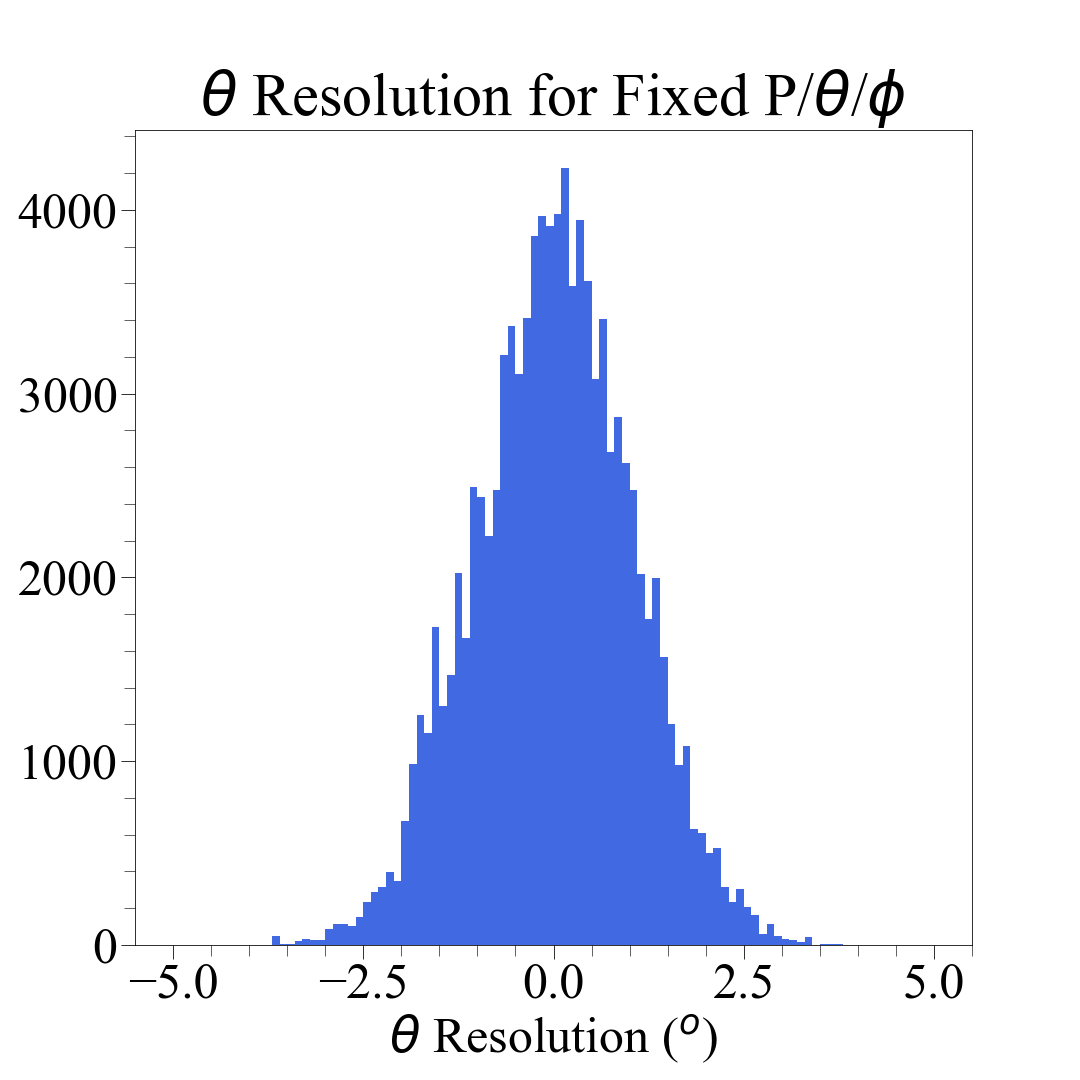}
   \includegraphics[trim = 2 2 2 2, clip, width=0.33\linewidth]{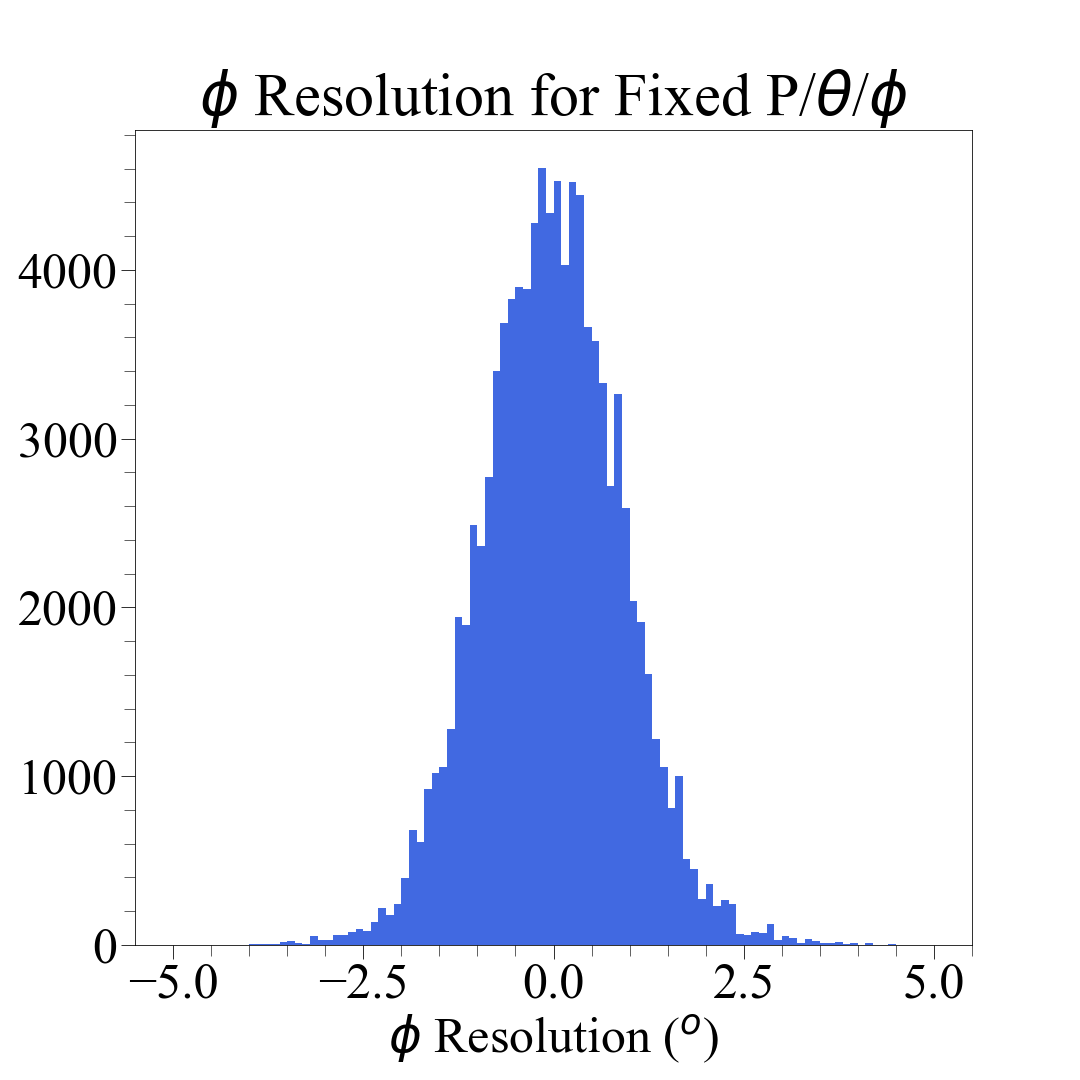}}\\
   \raisebox{0.5mm}{\includegraphics[trim = 2 2 2 2, clip, width=0.33\linewidth]{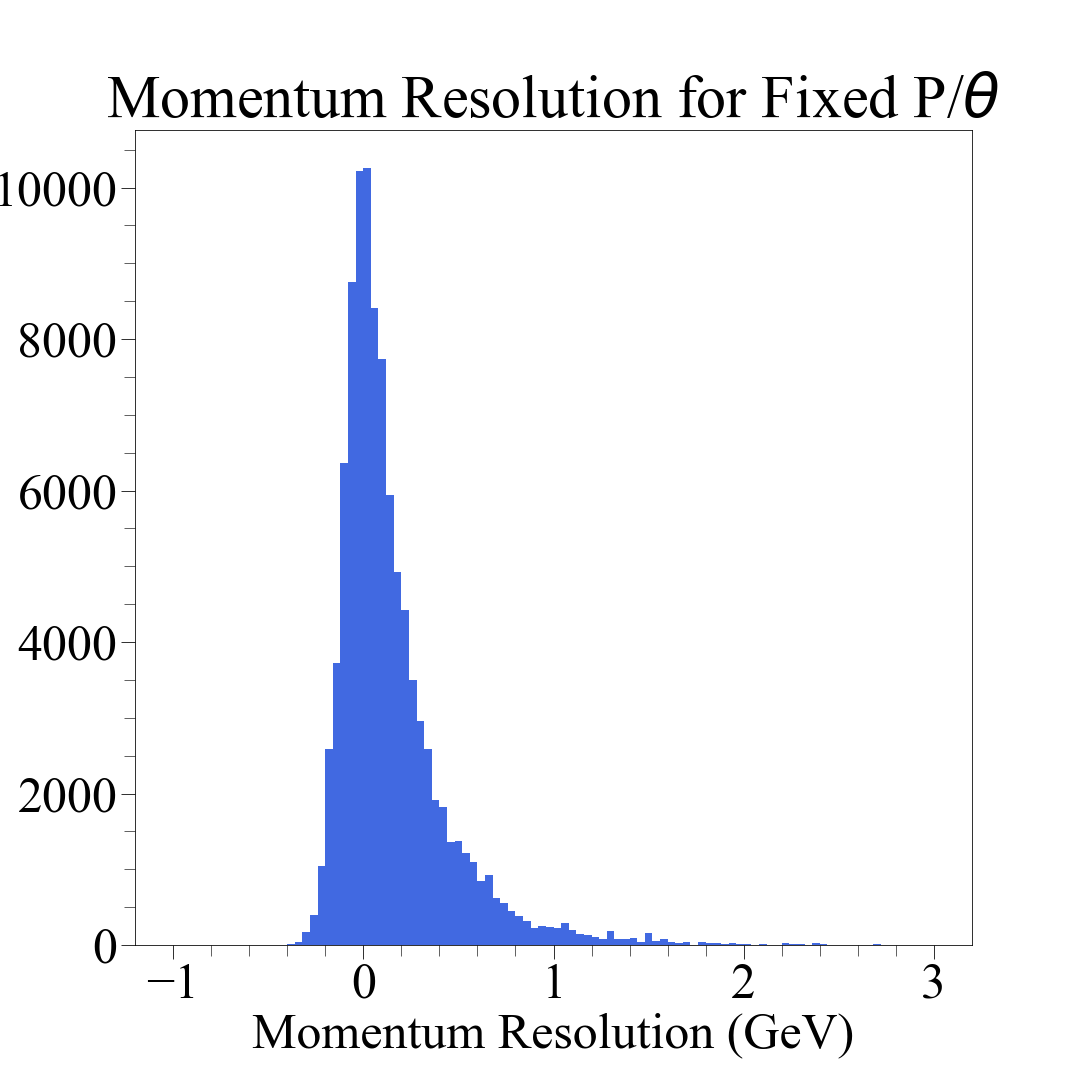}
   \includegraphics[trim = 2 2 2 2, clip, width=0.33\linewidth]{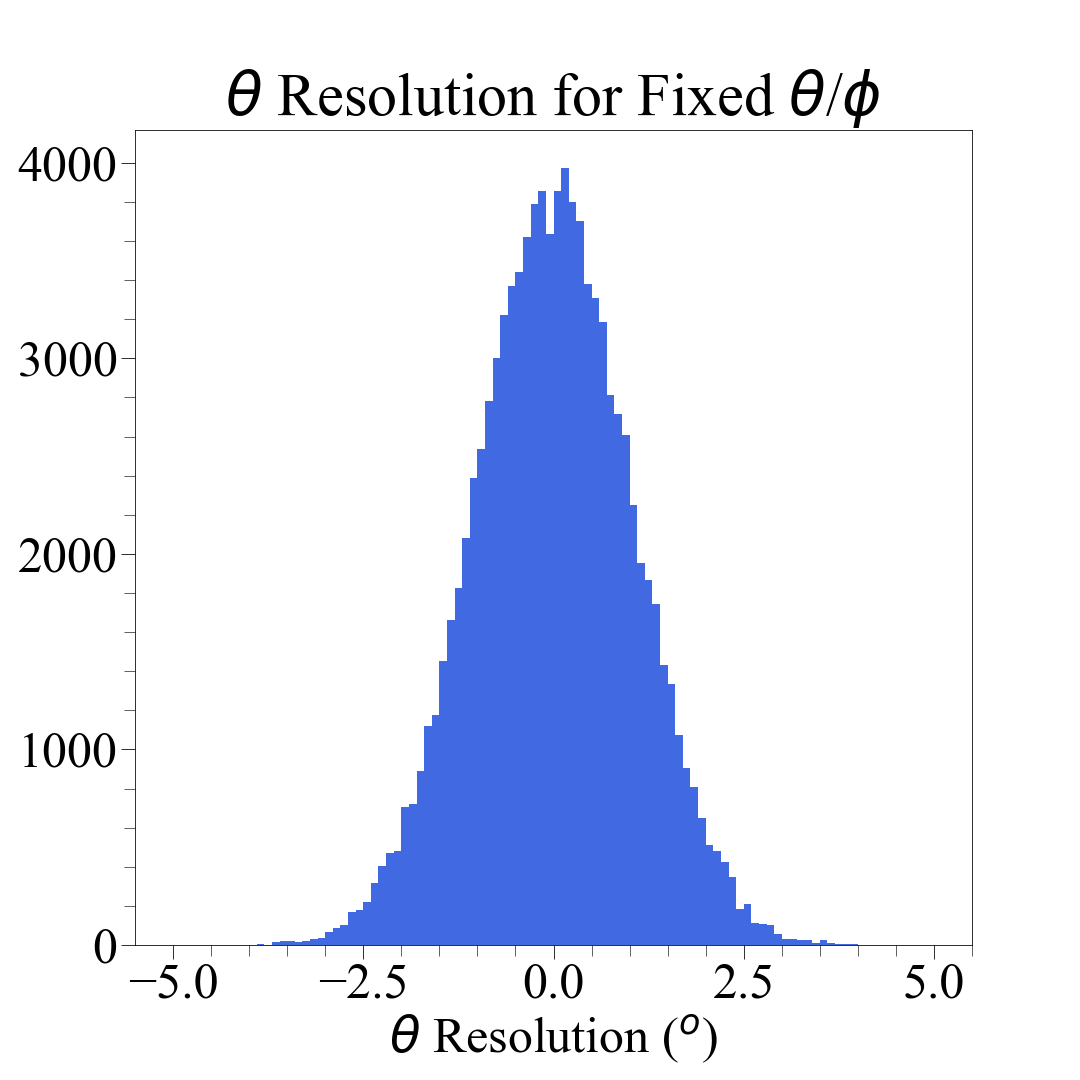}
   \includegraphics[trim = 2 2 2 2, clip, width=0.33\linewidth]{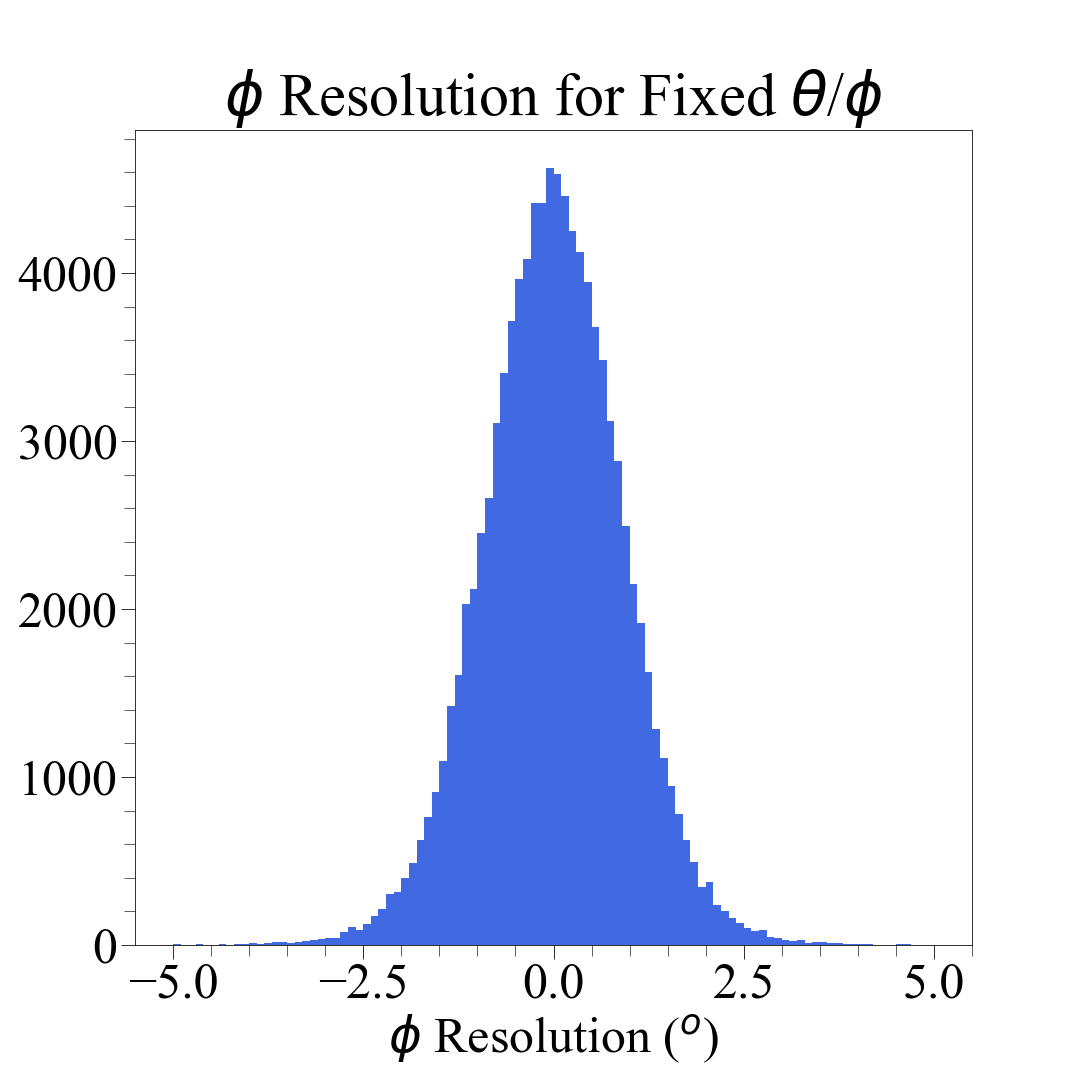}}\\
   \raisebox{0.5mm}{\includegraphics[trim = 2 2 2 2, clip, width=0.33\linewidth]{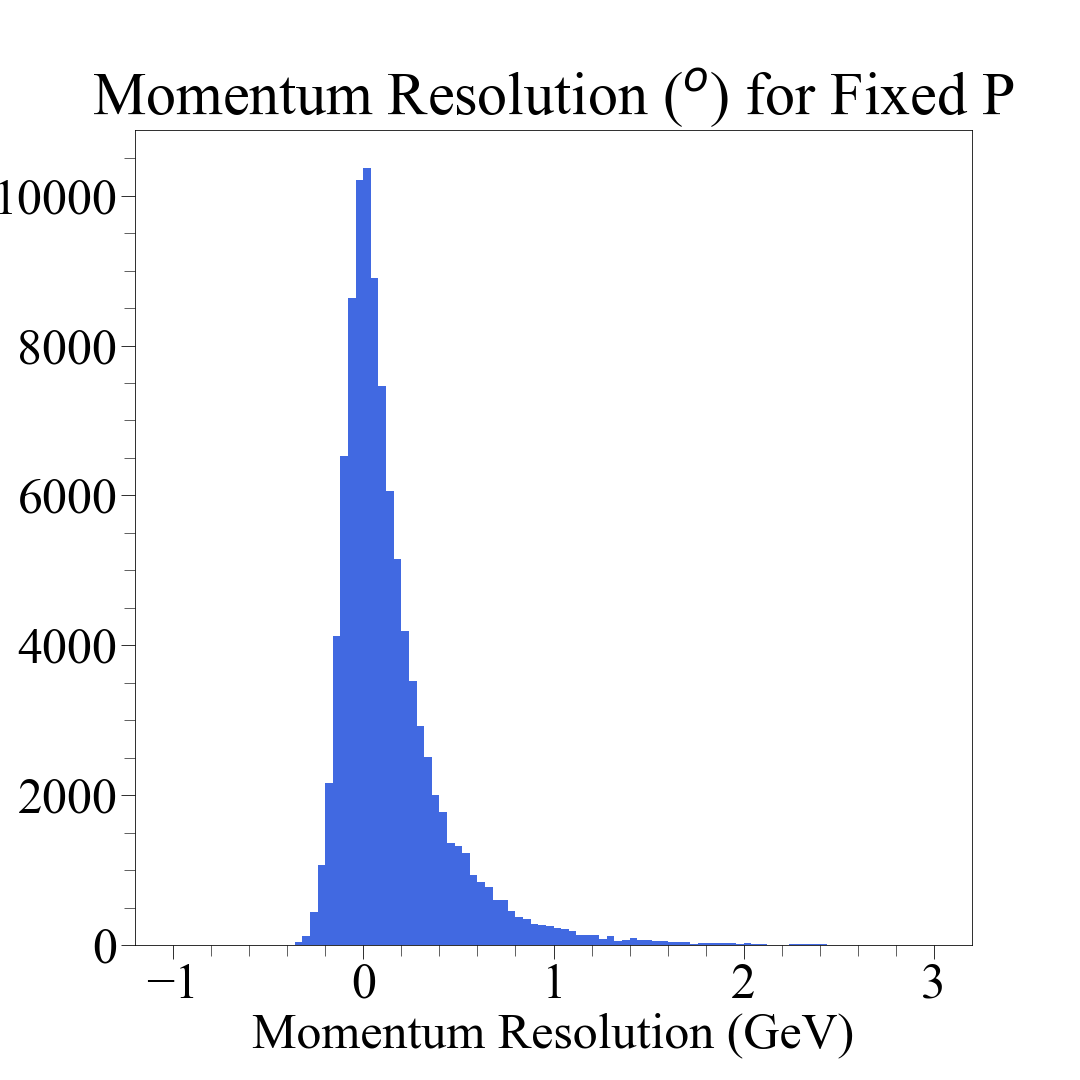}
   \includegraphics[trim = 2 2 2 2, clip, width=0.33\linewidth]{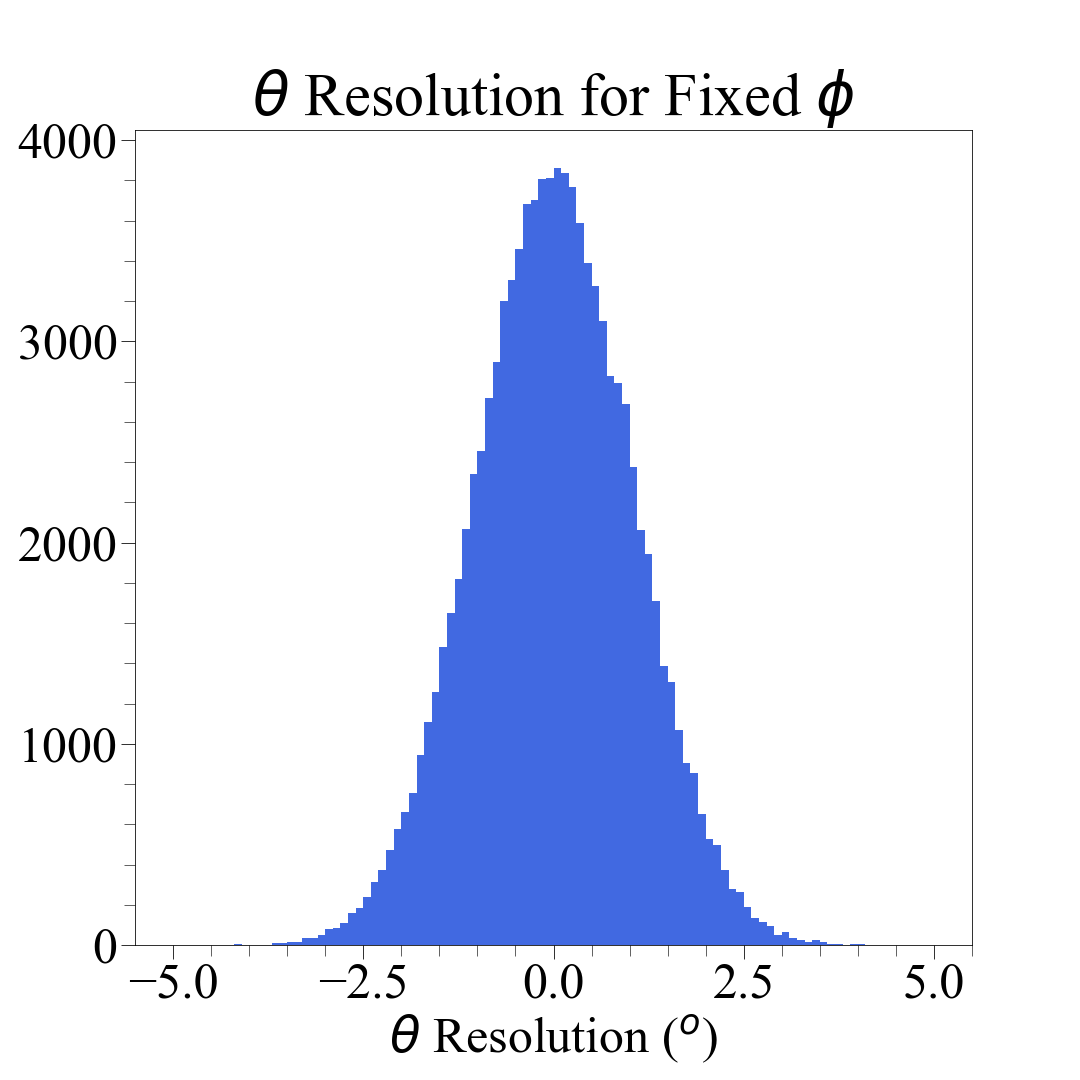}
   \includegraphics[trim = 2 2 2 2, clip, width=0.33\linewidth]{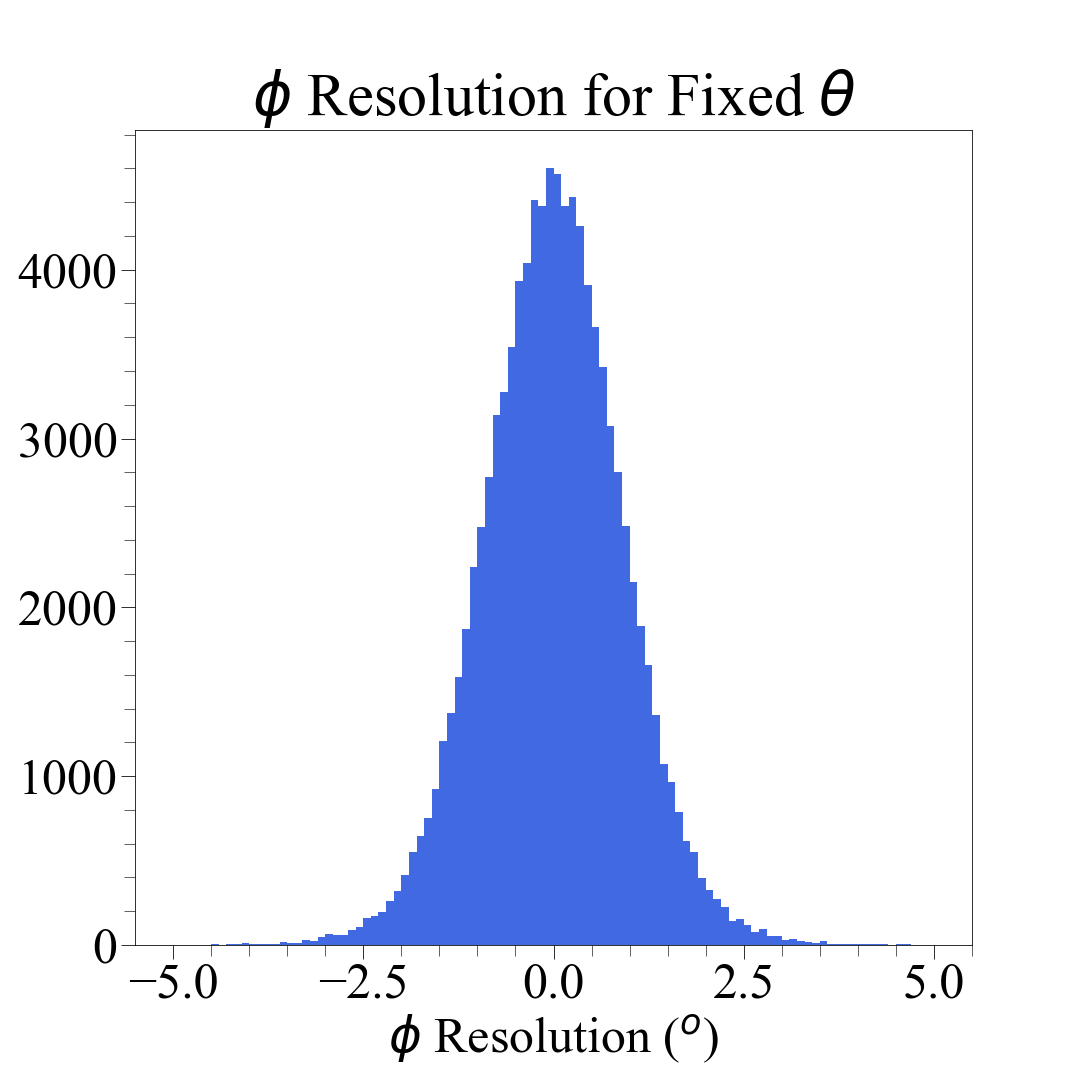}}
  \caption[]{ \label{fig:dt_spikesAll} The resolution in momentum, $\theta$, $\phi$ when the DT was applied $10^5$ times to the same values of momentum, $\theta$, $\phi$ on the first row, two fixed inputs on the second row and 1 fixed input on the third. When fixing one or two inputs we fix those for which the resolution of the variable plotted has a dependency, such as the resolution of momentum on momentum and that of $\theta$ and $\phi$ on $\theta$ and $\phi$.}
 \end{center}
\end{figure}

Using the same trained multiple DT predictions we plotted the Fast simulation differences for each component in Figure \ref{fig:dt_spikesAll} while fixing all three (first row), two (second row) and one (third row) of the input components. These plots demonstrate that even for fixed inputs we are able to attain a range of predictions mimicking the resolution of the Toy simulation for a given value of the input variables. As in full simulations the events will be generated with a range of momentum components, the residual spikiness at fixed points in phase space will be naturally smoothed out. In addition, typically for each event multiple particles will be tracked independently in the Fast simulation, further smoothing the higher level variables. This is demonstrated in our reaction simulation in Section \ref{sec_reaction}.

\begin{figure}[hbt!]
 \begin{center}
   \raisebox{0.5mm}{\includegraphics[trim = 2 2 2 2, clip, width=0.33\linewidth]{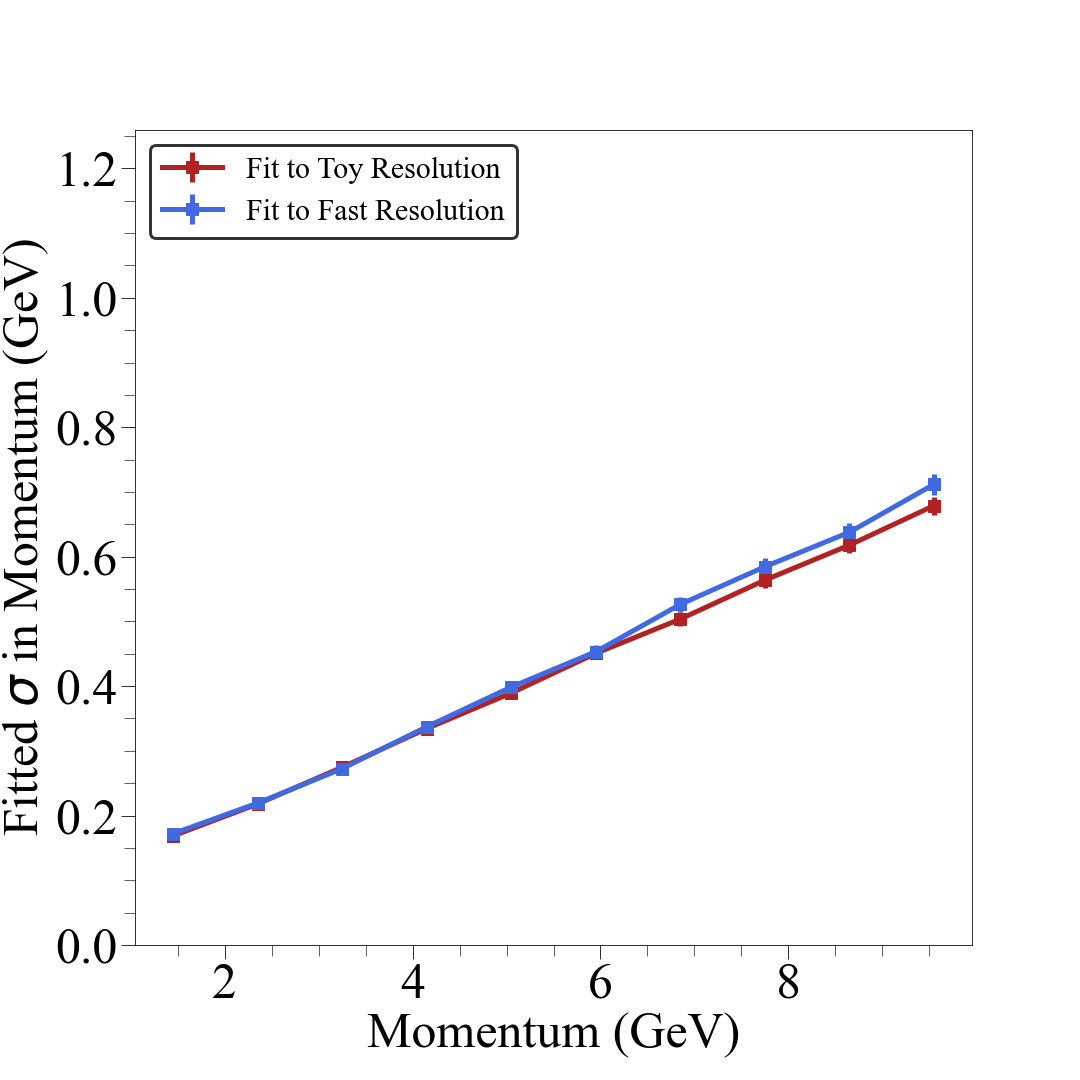}
   \includegraphics[trim = 2 2 2 2, clip, width=0.33\linewidth]{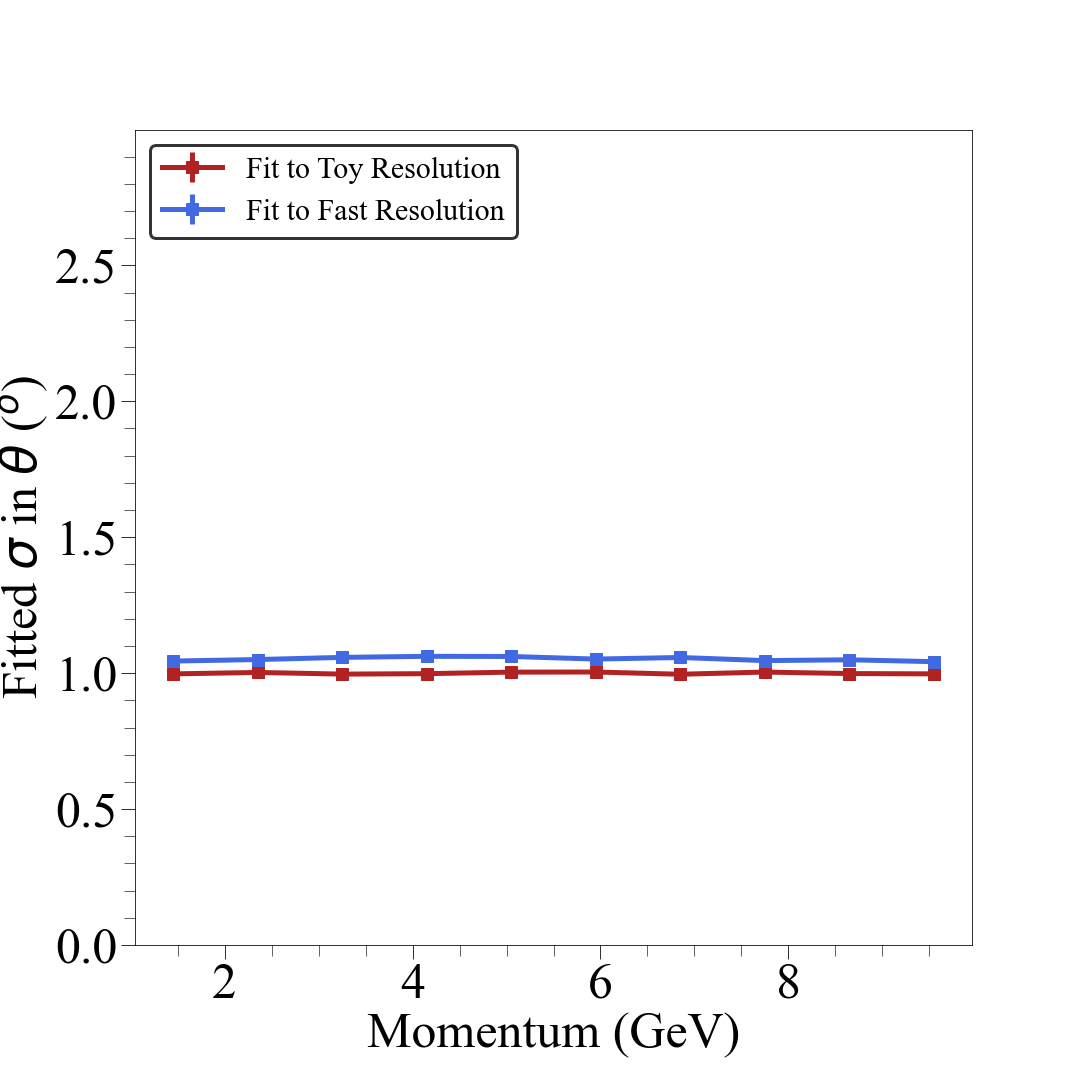}
   \includegraphics[trim = 2 2 2 2, clip, width=0.33\linewidth]{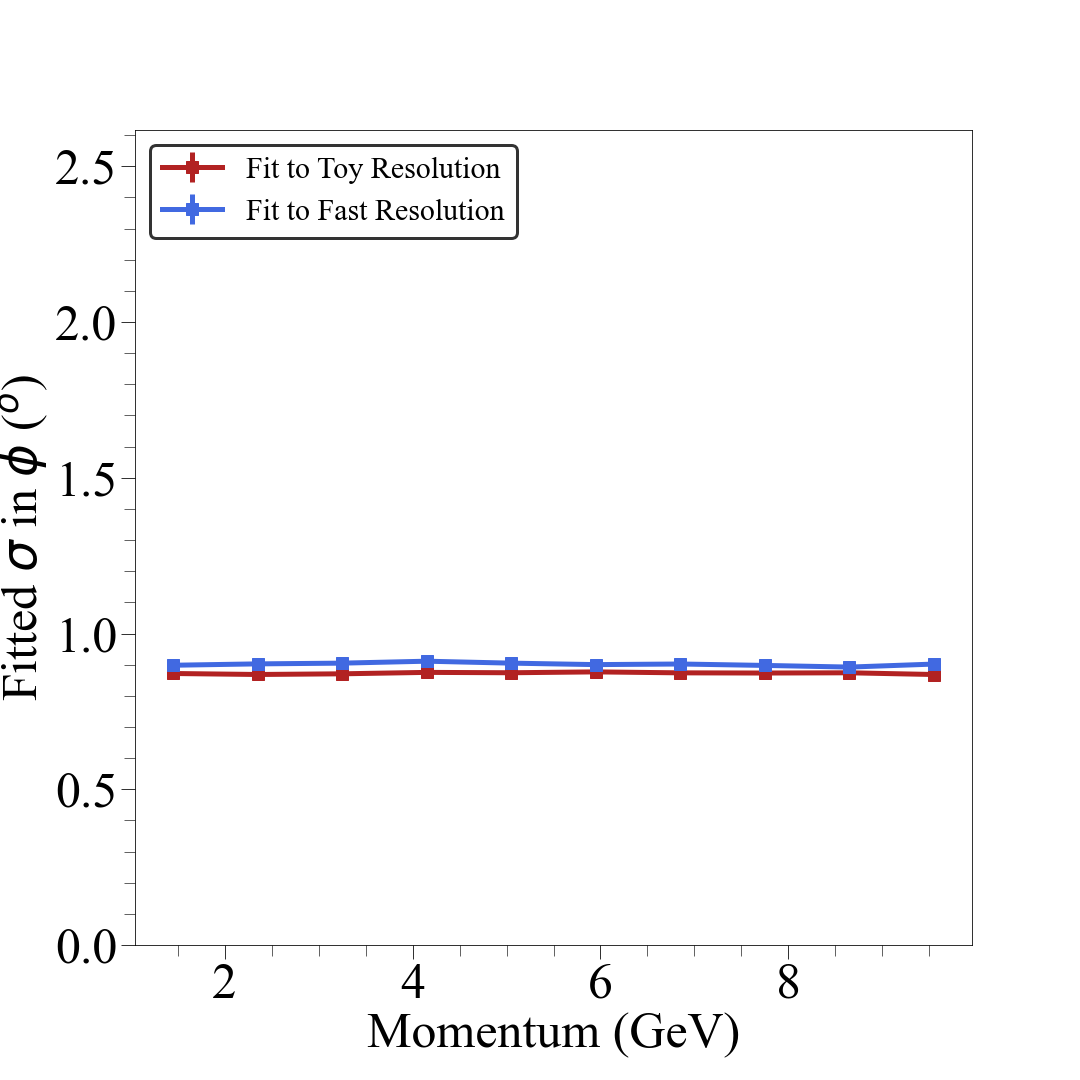}}\\
   \raisebox{0.5mm}{\includegraphics[trim = 2 2 2 2, clip, width=0.33\linewidth]{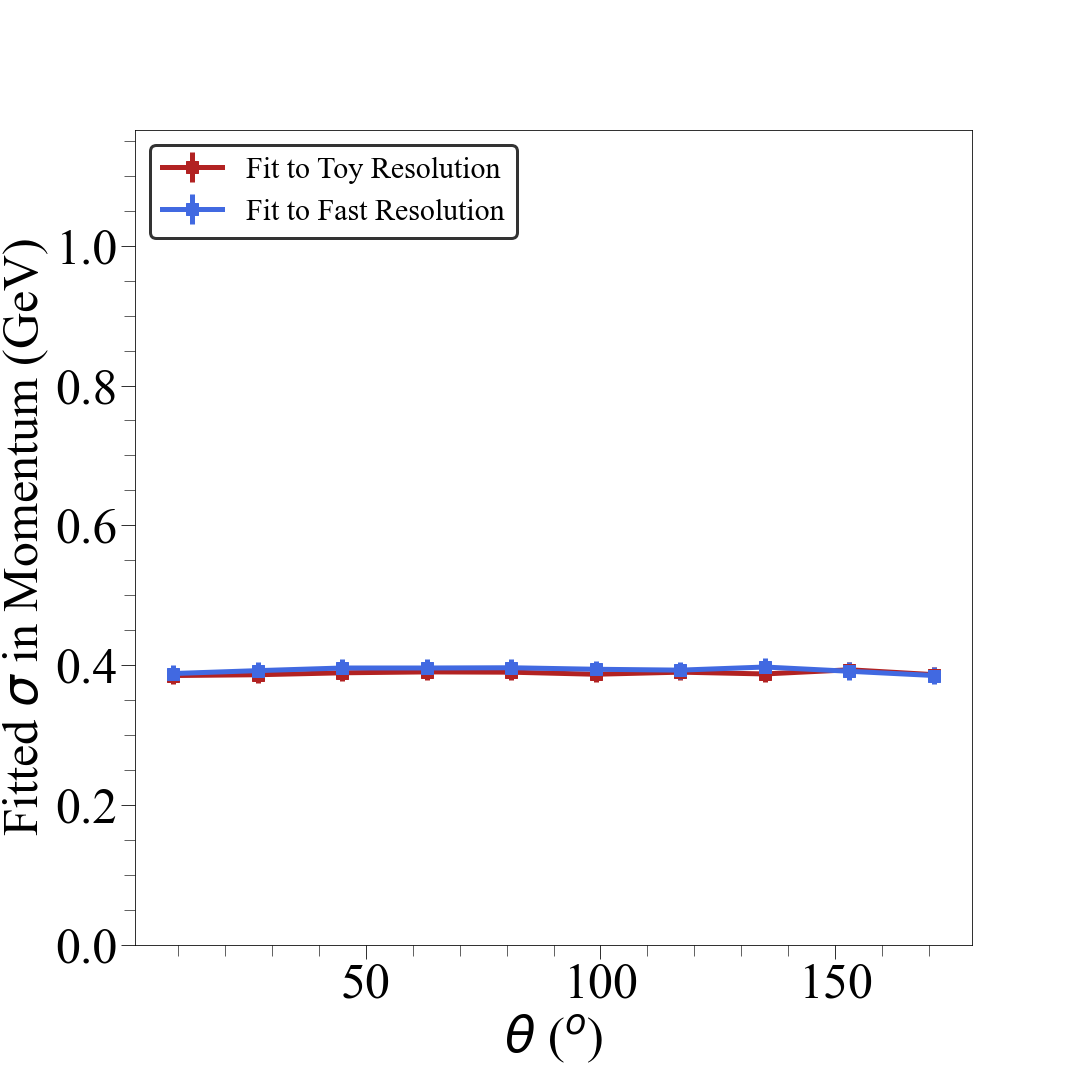}
   \includegraphics[trim = 2 2 2 2, clip, width=0.33\linewidth]{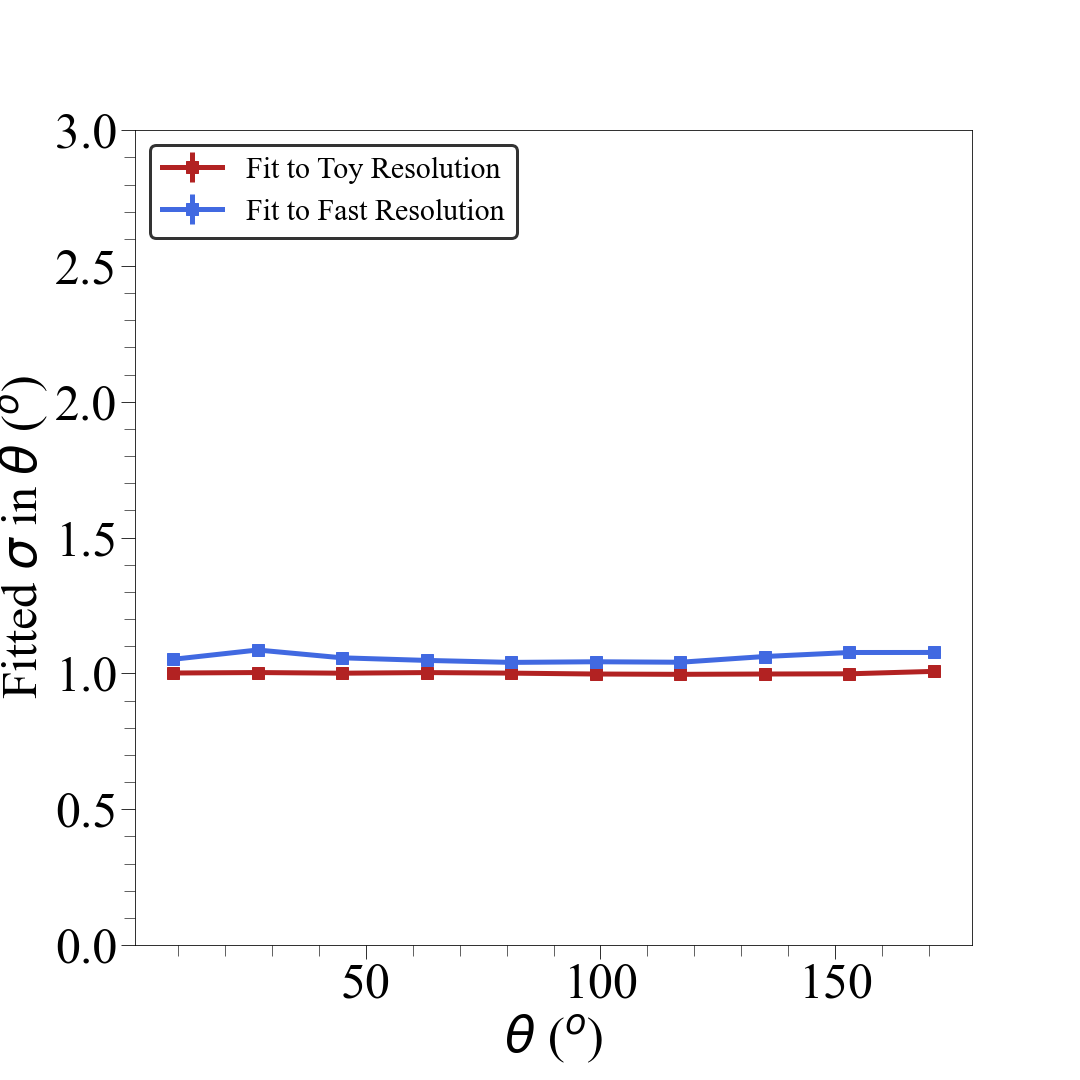}
   \includegraphics[trim = 2 2 2 2, clip, width=0.33\linewidth]{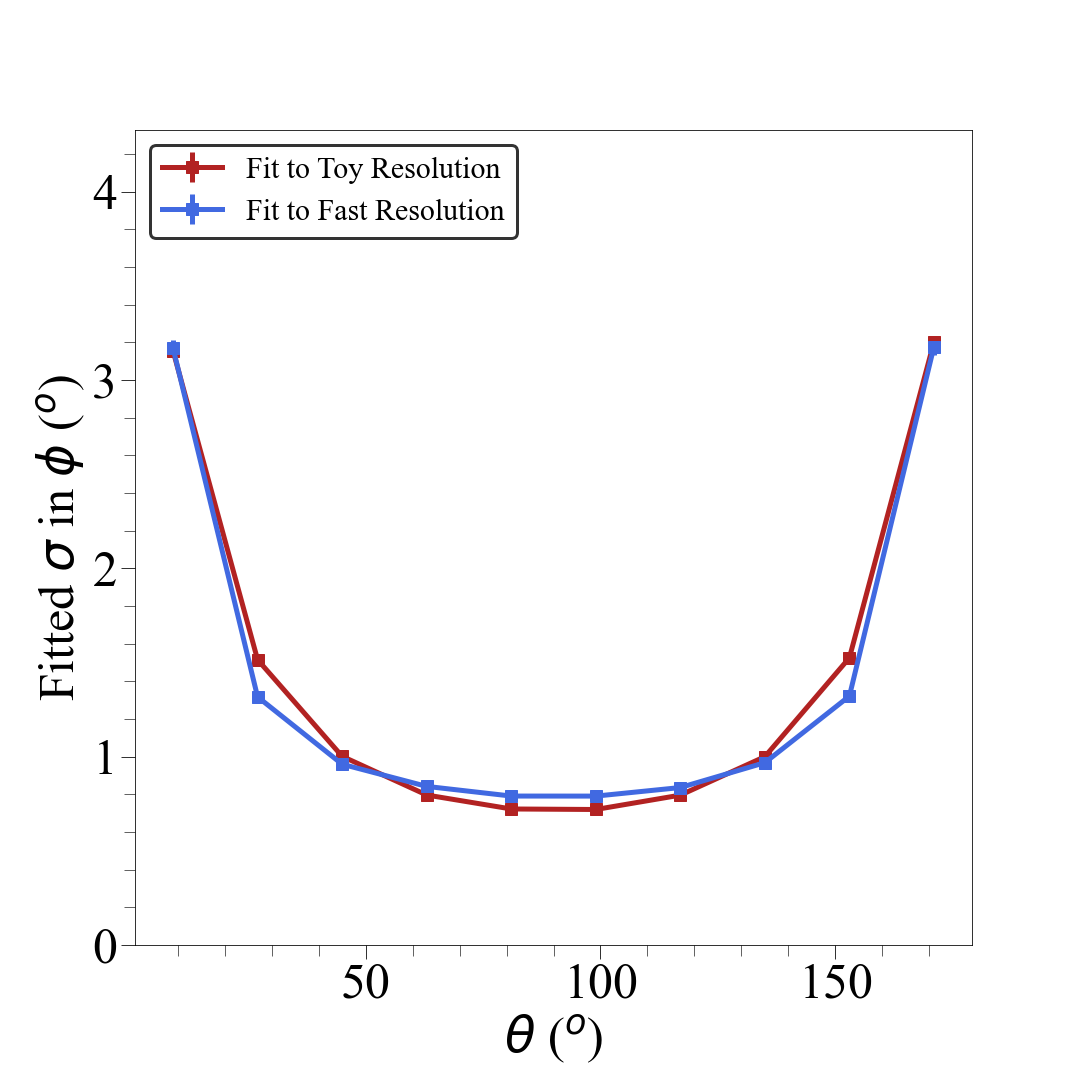}}\\
   \raisebox{0.5mm}{\includegraphics[trim = 2 2 2 2, clip, width=0.33\linewidth]{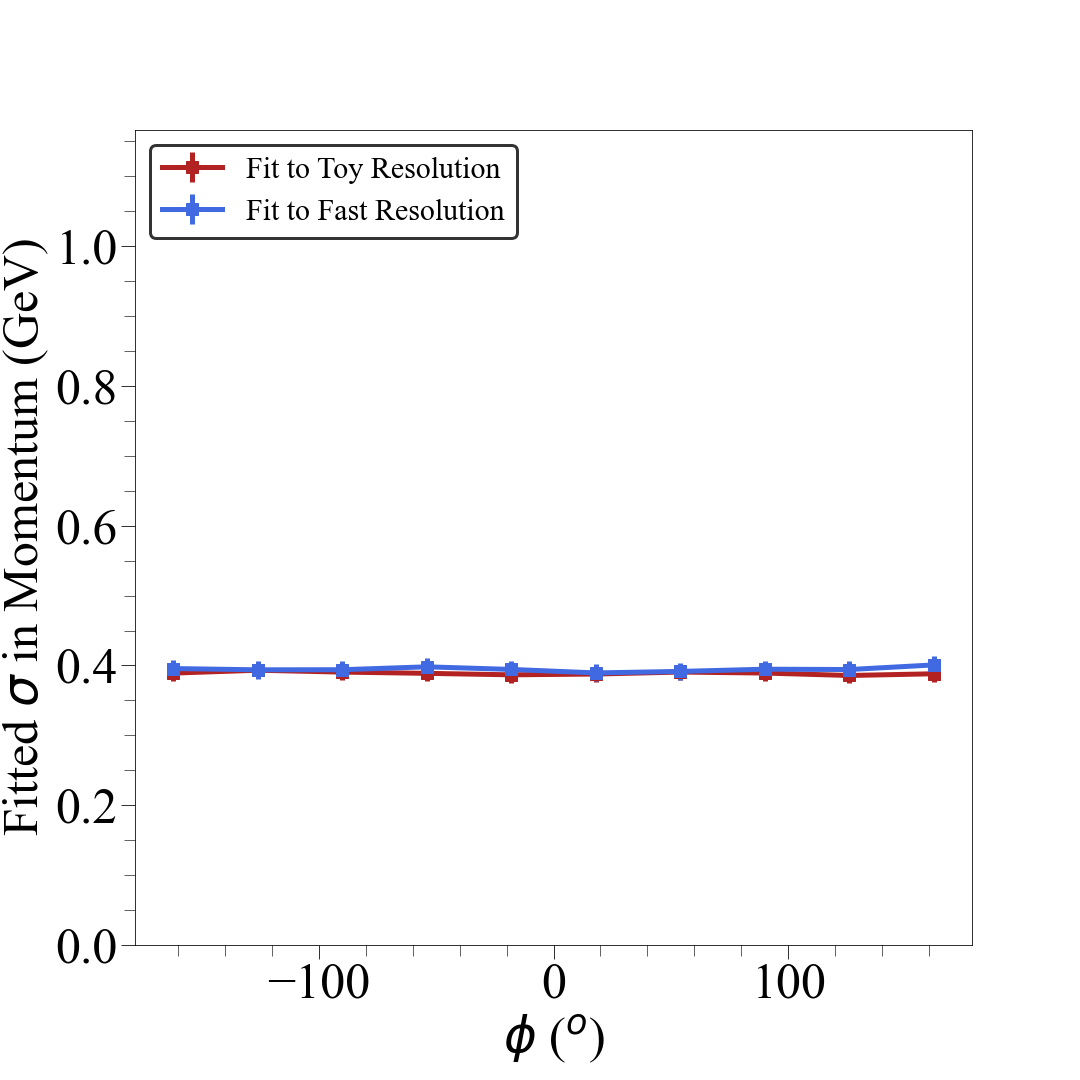}
   \includegraphics[trim = 2 2 2 2, clip, width=0.33\linewidth]{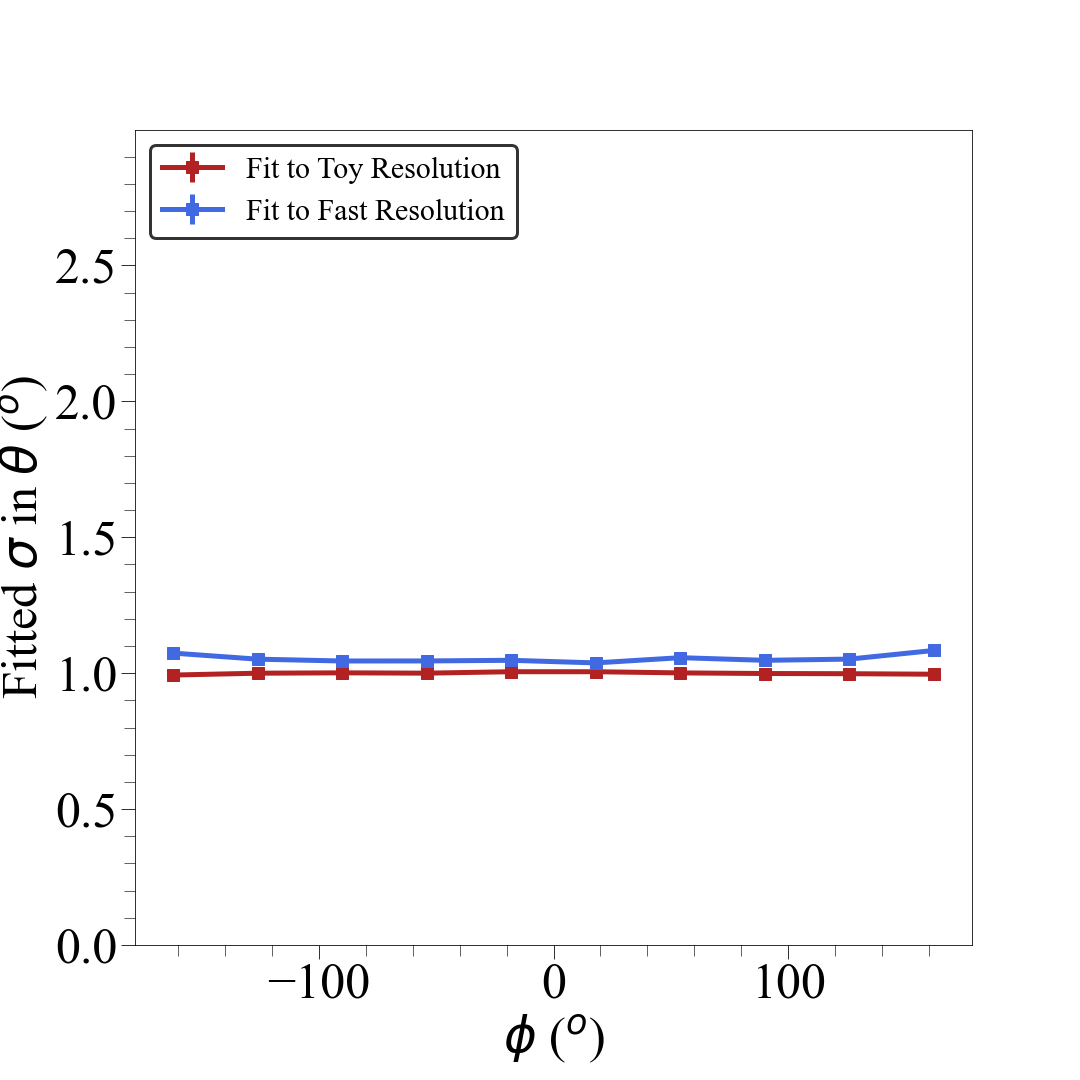}
   \includegraphics[trim = 2 2 2 2, clip, width=0.33\linewidth]{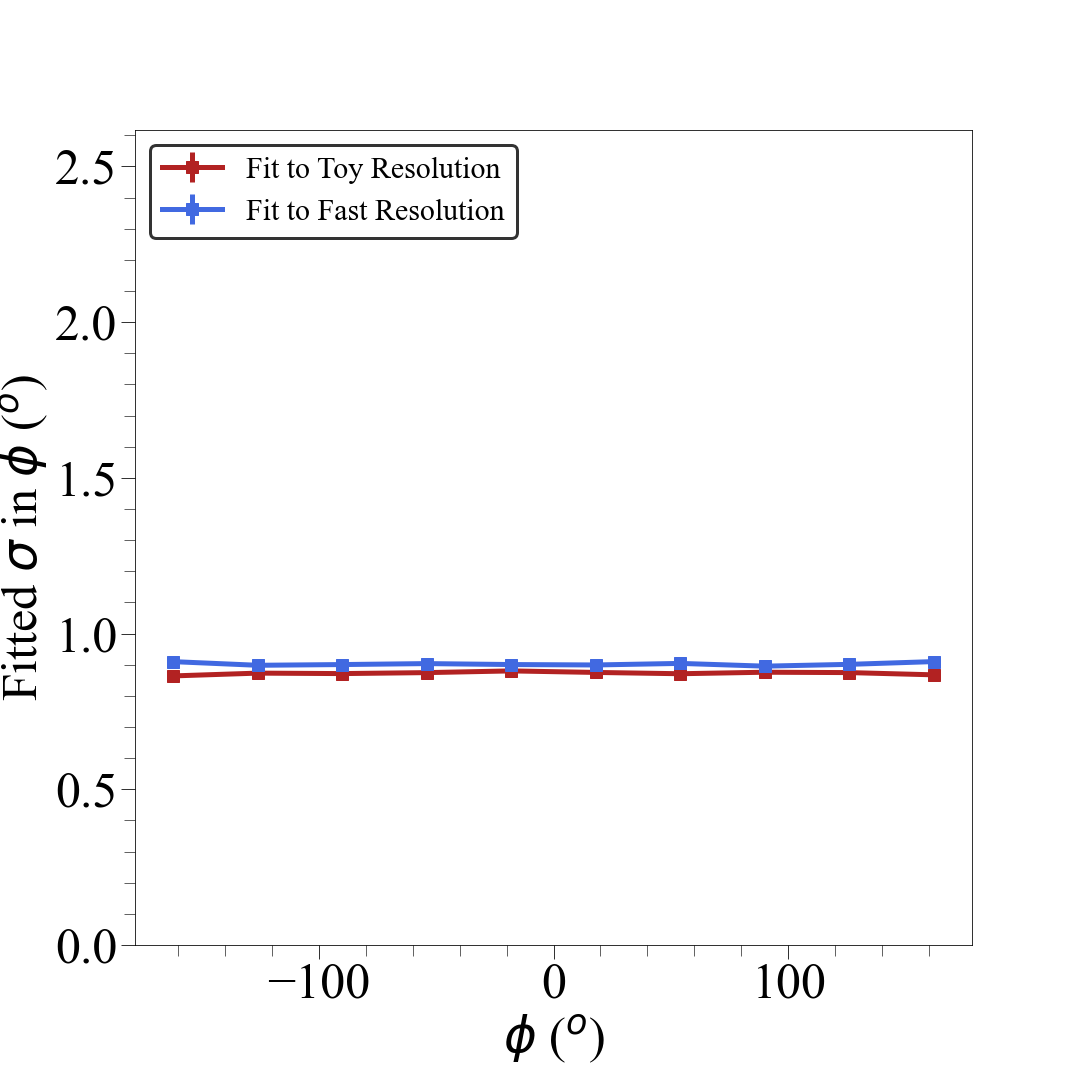}}
  \caption[]{ \label{fig:dt_fitAll} The standard deviation extracted by fitting a crystal ball function to the resolution of momentum and a normal distribution to the resolution of $\theta$ and $\phi$ in bins of momentum in the first row, $\theta$ in the second, $\phi$ in the third. The red curves were made by fitting the Toy simulation resolution. The blue curves were made by fitting the Fast simulation resolution produced by the DT.}
 \end{center}
\end{figure}

Finally we plot the fitted $\sigma$ of the resolution for p, $\theta$, $\phi$ as a function of these components for Fast and Toy simulations. As shown in Figure \ref{fig:dt_fitAll}, these match well.

Overall, the method presented here is shown to accurately map the resolution of three variables and their dependencies on these three variables, whilst also providing smooth random resolution distributions. On top of its reliability, it is also capable of processing order $10^5\mbox{s}^{-1}$ events per second which is orders of magnitude faster than full detector simulations, while being reasonably fast (150$s$) and consistent to train. Once the DTs are trained they can be saved requiring around 150 MB each for a total of 50x150 MB.

\section{Example Reaction}\label{sec_reaction}

To give an example of how the ML simulation works for full reaction kinematics, we use 
the two pion photoproduction reaction $\gamma + p \longrightarrow p' + \pi^{+} +\pi^{-}$.
Five million events were generated for a $\gamma$ beam energy of 10 GeV, with the three final state particles decaying into flat phase space.

We used the same Toy detector described in Section \ref{sec_toy} for each of the three particles, training a model for each particle separately.
We then used the ML acceptance simulation to determine whether a particle was 
successfully detected or not, if it was we generated its reconstructed momentum components. 
In addition, the particles were also passed through the parameterised Toy detector to get the 
true response for comparison.

\begin{figure}[hbt!]
 \begin{center}
   \raisebox{0.5mm}{\includegraphics[trim = 2 2 2 2, clip, width=0.99\linewidth]{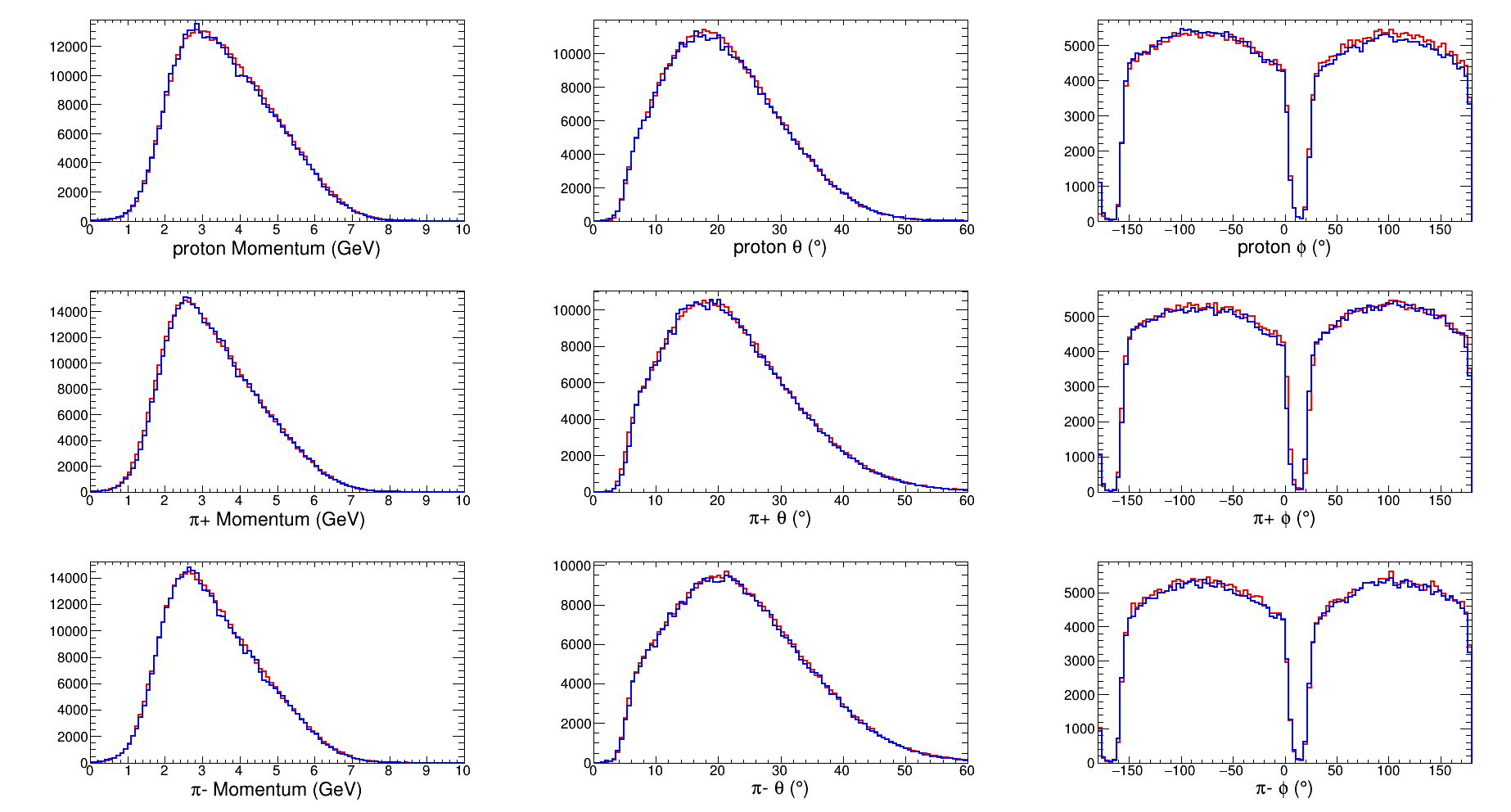}}
  \caption[]{ \label{fig:react_partDist} Accepted and reconstructed momentum components for the Fast (blue) and Toy (red) simulations. Top is for generated protons, middle $\pi^{+}$ and bottom $\pi^{-}$. The red line shows the real toy simulation, while blue is the resulting machine learned distributions.}
 \end{center}
\end{figure}

\begin{figure}[hbt!]
 \begin{center}
   \raisebox{0.5mm}{\includegraphics[trim = 2 2 2 2, clip, width=0.99\linewidth]{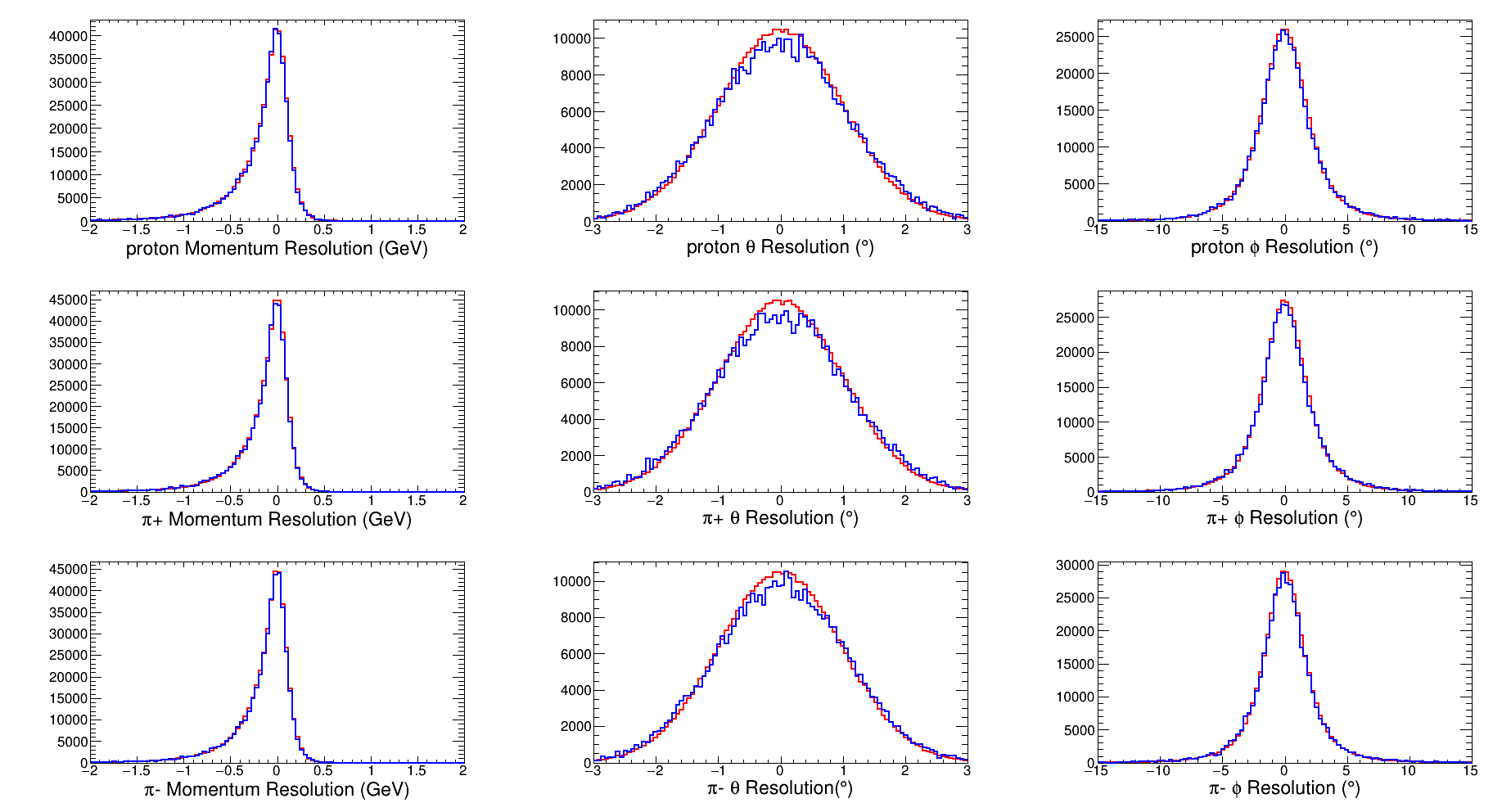}}
  \caption[]{ \label{fig:react_partRes} Resolutions for the momentum components for the Fast (blue) and Toy (red) simulations. Top is for generated protons, middle $\pi^{+}$ and bottom $\pi^{-}$.}
 \end{center}
\end{figure}

In  Fig. \ref{fig:react_partDist} we show the momentum components for each particle for true Toy and Fast ML simulations, while in  Fig. \ref{fig:react_partRes} we show the resolutions. Overall the agreement is excellent. The only slight deviation is with the particle $\theta$ resolutions which are slightly broader in the machine learned model. We also note that the resolution distributions from this model are less smooth than for the true detector distributions. This is an artifact of the resolution algorithm as discussed in Section \ref{sec_recon}. We do not expect this to give significant distortions to full simulations as the resolution distribution is summed event-by-event with randomly generated particle momentum. This is demonstrated in the momentum distributions of Fig. \ref{fig:react_partDist} where the blue lines are smoother.

In general, we wish to reconstruct distributions based on more than one particle at a time.
This actually is  a benefit to the method as it provides further mixing of random 
variables and we might expect these distributions to be smoother than those for the individual momentum components.

\begin{figure}[hbt!]
 \begin{center}
   \raisebox{0.5mm}{\includegraphics[trim = 2 2 2 2, clip, width=0.95\linewidth]{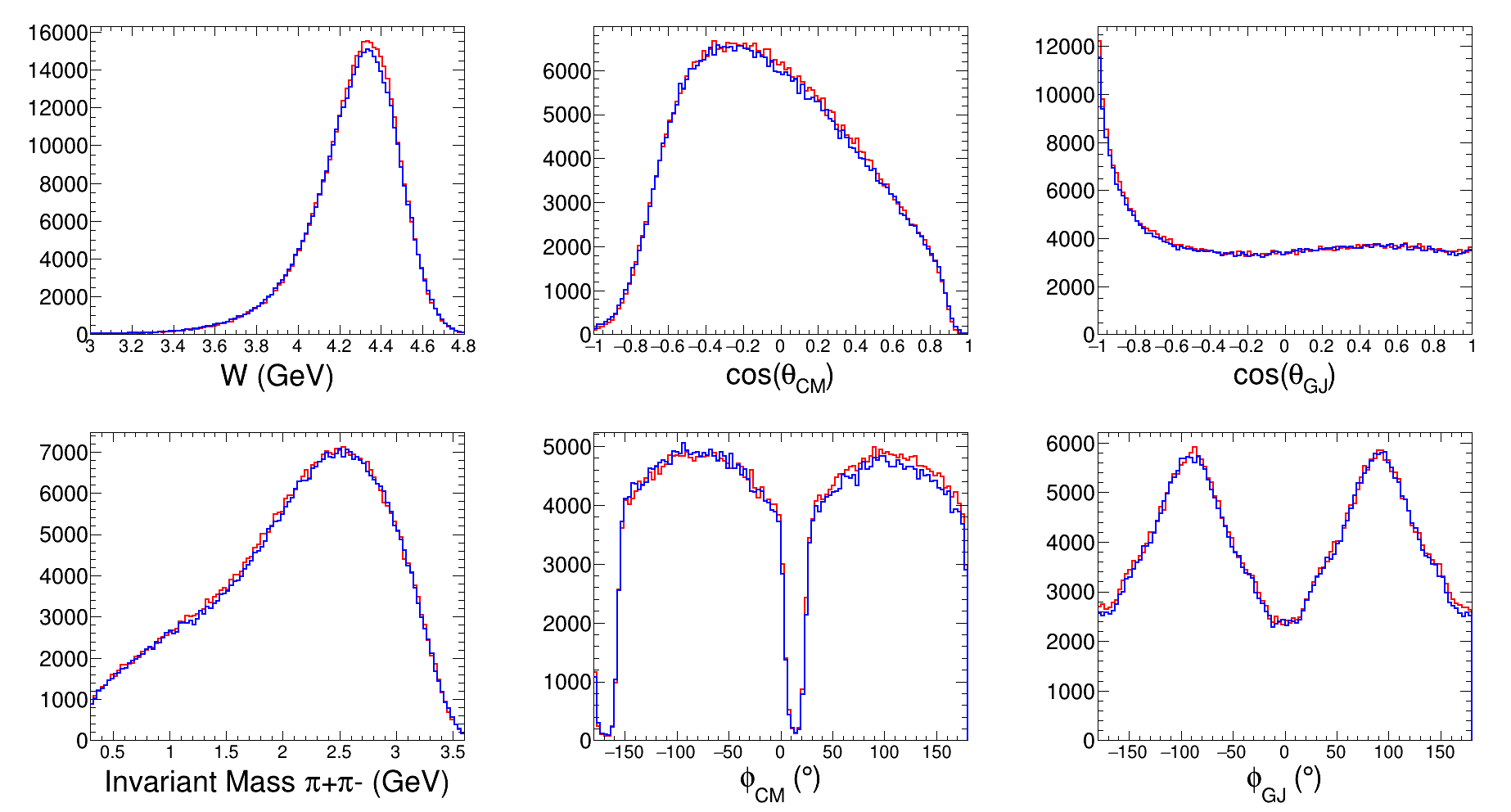}}
  \caption[]{ \label{fig:react_physDist} Accepted and reconstructed physics variables for the Fast (blue) and Toy (red) simulations of the 2 pion photoproduction reaction. The distributions show: the invariant mass of the three final state particles, W; the invariant mass of the two pions, $M(2\pi)$; the production angles in the centre-of-mass system ($\cos(\theta_{CM}),\phi_{CM}$); and the decay angles of the two pions.}
 \end{center}
\end{figure}

For our reaction we reconstructed distributions for the invariant mass of the three final state particles, W; the invariant mass of the two pions, $M(2\pi)$; the production angles in the centre-of-mass system ($\cos(\theta_{CM}),\phi_{CM}$); and the decay angles of the two pions  in their combined rest frame (Gottfried-Jackson frame) ($\cos(\theta_{GJ}),\phi_{GJ}$). In Fig. \ref{fig:react_physDist} we show the comparison of the distributions in each of these variables for the 5 million simulated events. Given the model was trained on 1 million events it is not guaranteed that the ML simulation would
provide the same smoothness as the parameterised models, but the results show very similar levels of fluctuations for the two simulations. 

\begin{figure}[hbt!]
 \begin{center}
   \raisebox{0.5mm}{\includegraphics[trim = 2 2 2 2, clip, width=0.95\linewidth]{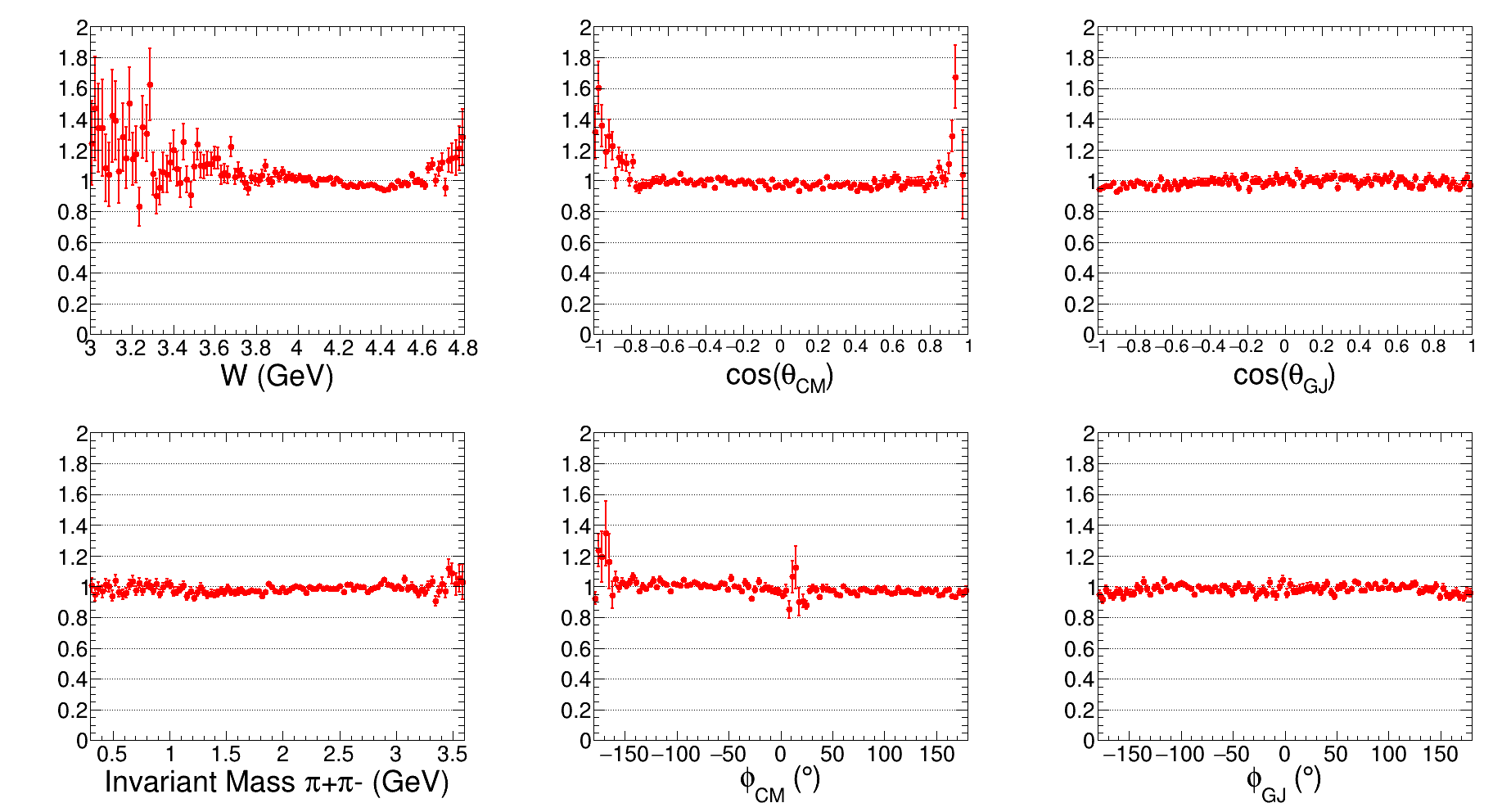}}
  \caption[]{ \label{fig:react_physRatio} Ratio of Toy to Fast simulations for the 2 pion photoproduction reaction. The distributions show: the invariant mass of the three final state particles, W; the invariant mass of the two pions, $M(2\pi)$; the production angles in the centre-of-mass system ($\cos(\theta_{CM}),\phi_{CM}$); and the decay angles of the two pions.}
 \end{center}
\end{figure}

In Fig. \ref{fig:react_physRatio} we show the ratio of the the two models. These are all reasonably flat and close to 1 for all variables. Slight deviations are visible where the counts in the variable distributions are very small, this reflects slight deviations in the resolution distributions modelled. The weighted average ratio in this test was found to be just over 0.99, giving an overall normalisation good to the $1\%$ level. 

\section{Discussion}\label{sec_dis}

The availability of large volumes of simulated data can enhance the analysis of 
complicated physics reactions. The application of machine learning algorithms to the production of this data is very natural due to their speed and generality coupled with readily available software for different techniques and algorithms. Here we have investigated a method for fast simulation which splits the process into two
distinct parts: particle acceptance modelling and three-vector reconstruction. This
allowed us to produce simulated particles at a rate of at least 10 kHz.

For modelling the acceptance we found that density ratio estimation via probabilistic classification provides a neat solution when applied to multidimensional problems. 
In principle, any appropriate algorithm may be used for the classification. In our studies 
we found deep neural networks (DNN), consisting of at least 500 neurons, were best suited to reproduce the multidimensional correlations while preserving regions of zero acceptance which Boosted Decision Trees were not quite able to reproduce.

One downside to DNN is their inherent randomness and the training often arrived at solutions
which did not give an accurate description of our toy model. However, by making multiple training passes an adequate solution was readily found. In addition, we found we could apply
a reweighting procedure to help correct for any small deviations and provide more stable
final models. Boosted Decision Trees were well suited to this due to their relatively small
training and prediction times.

We point out that modelling acceptance in this simple manner must incorporate many factors, such as particle identification, which may reduce the apparent acceptance. In this procedure,
particle identification would be accounted for by assuming the PID is already applied before the event is classed as accepted. Correlations between particles that may be induced by,
for example, a data-acquisition trigger, would also not be modelled accurately. Such effects would require an additional classification stage with many momentum vectors as input at once allowing the correlations between particles to be modelled. 
We did not test the performance of such a many particle classification in the context of this work, but in principle a similar approach could work.

For modelling the reconstruction of the particle momentum we used a Decision Tree to mimic a nearest neighbours algorithm, as we found this to be significantly faster than a kNN algorithm, with similar results. We used 1 million events with the particle momentum components and 5 random numbers as input and the differences between truth and reconstructed momentum components as 3 outputs. Effectively, given a new event, the algorithm finds the nearest in the training sample, with some randomisation from the random inputs, and provides
its reconstruction difference for each momentum component. We found that by using multiple of such decision trees we were able to move towards a smoother distribution for given momentum components. By using such a nearest neighbours type algorithm the correlations in the training data were automatically retained, at the expense of having a somewhat jagged resolution function, limited by the number of training events. This algorithm benefits from being very fast, with a single tree trained on 1 million events able to make predictions at around 1 MHz. We note that to save a single such Decision Tree requires 150MB, so if using many decision trees, for safe memory usage, it is best to read the decision tree model one at a time and apply to a masked set of events from the full data set.

Here we have limited the investigation to reproducing simulated momentum vectors, however 
in practise there may be many more variables that could be generated in a similar manner using additional decision tree regressor with inputs still based on particle three momentum, but with outputs giving directly measured quantities such as time-of-flight, flight path length, energy deposits, etc. Simple tests showed that adding more outputs to the reconstruction algorithm did not affect the quality of the predictions. From these lower level variables particle identification algorithms would be applied as for a full simulation. Of course, if this approach is followed, particle identification should not be used as part of the acceptance modelling.

Applying this approach to full detector simulations requires some further considerations.
In particular, what to use as training data for which we see two main approaches: (I) use a particle gun type of event generator to simulate just single particle events of known species
covering the momentum range and particle type produced in all possible reactions; (II) perform full simulations for a particular final state, separate and create a model for each particle.
Method (I) has the advantage of generality, while (II) would benefit from focusing the model training on the most important regions of phase space. In either case computationally it is most efficient to process data in columns of each sample particle species, and train or predict one particle at a time, rather than one event at a time. This has the benefit of only requiring one model at a time while the data being processed can be stored contiguously in memory. 

Fast simulated data produced in this manner can have many uses, although we would expect that
any measurements requiring accurate normalisation would use the most realistic simulation
available. Indeed, concerns were raised on the use of GANs for replacing full Monte-Carlo pipelines by \cite{Konstantin22}, pointing out that the statistical significance of the fully GAN produced simulated data will not be more than the original training set, regardless of how many more events are produced. We note our approach does not try and replace the full pipeline with one model, but factorises the event generator part from models (for resolution and acceptance) for each particle, which should allow production of larger volumes of statistically significant datasets. This is more akin to the "Replacing parts of the Monte Carlo pipeline with GANs" discussion in Section 7.1 of that paper. Of course, the individually trained models are themselves still open to statistical uncertainties related to the number of training events, which will result in systematic deviations from the true model when used to predict. Therefore, applications of our approach must be carefully considered. 

Cheap, approximate, simulated data would be useful for performing toy studies
on parameter extraction. That is generating events with particular values for parameters of 
interest, passing them through the fast simulation and seeing how close the results are
to the true parameter values. This would provide a good approximation to distortions induced by acceptances, resolutions or fitting, even if there are some small deviations in the trained and true detector model

Simulated data may be used to provide templates for performing signal and background yield
extractions. Background may consist of many different channels for which it can require very
large generated event samples to leak sufficiently into the signal discrimination variable.
Using this fast method to simulate the backgrounds would allow such templates to have
smaller statistical fluctuations in the resulting probability distribution function, making the yield extraction fits more reliable.

Another possibility would be to use it as a preprocessing step to full simulations that have low efficiency in some regions of phase space. The acceptance model proposed here could be used to determine whether to perform a full simulation for each event, via accept and reject, and subsequently using the acceptance weight to correct the distributions post simulation.

\include{summary}

\section*{Acknowledgements}
{
This work has been funded by the U.K. Science and Technology Facilities Council under grants ST/P004458/1 and ST/V00106X/1.}

\bibliographystyle{apsrev4-2.bst}
\bibliography{references}

\appendix

\section{Appendix: Evaluating the Reconstruction}\label{appendix_a}

\begin{figure}[hbt!]
 \begin{center}
   \raisebox{0.5mm}{\includegraphics[trim = 2 2 2 2, clip, width=0.5\linewidth]{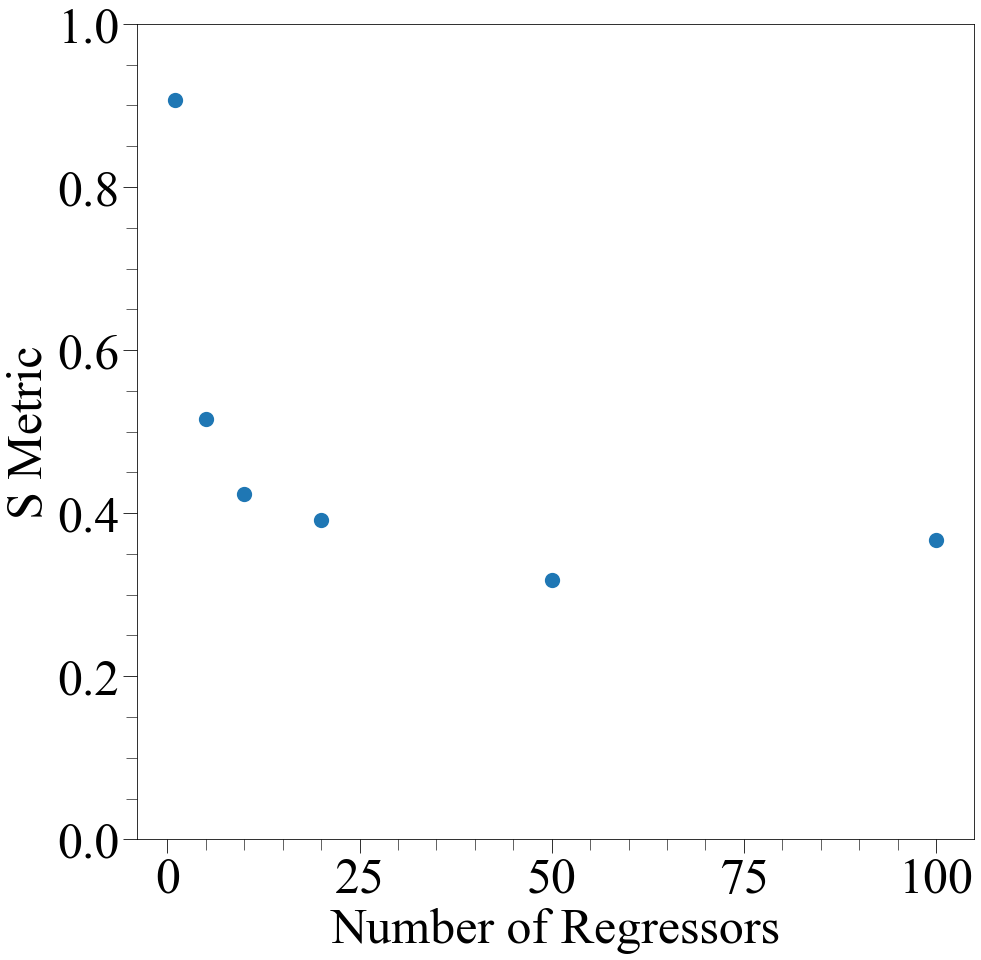}
   \includegraphics[trim = 2 2 2 2, clip, width=0.5\linewidth]{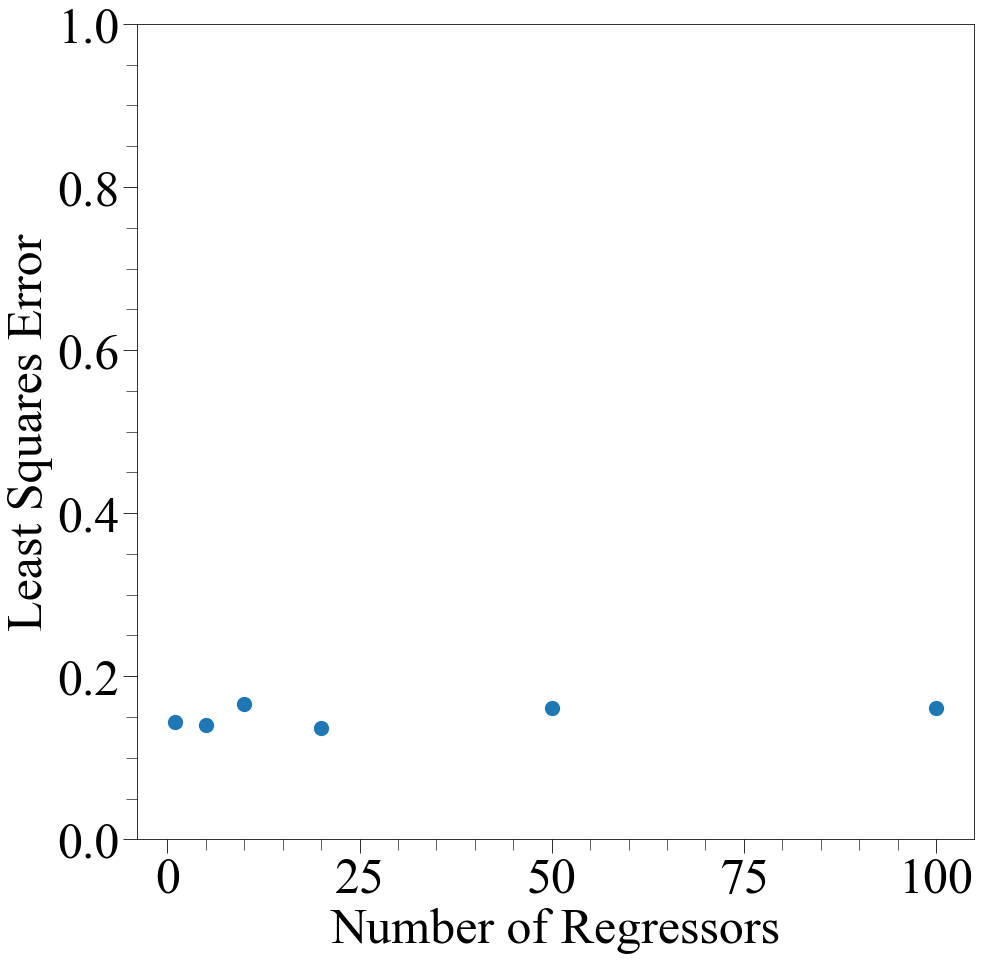}}
  \caption[]{ \label{fig:metrics_nReg} From left to right, the $S$ metric and $LSE$ as a function of number of DTs with 1 random input and 0 random outputs. The $S$ metric improves up until it plateaus at 50 DTs whilst there is little variation in the $LSE$.}
 \end{center}
\end{figure}

\begin{figure}[hbt!]
 \begin{center}
   \raisebox{0.5mm}{\includegraphics[trim = 2 2 2 2, clip, width=0.5\linewidth]{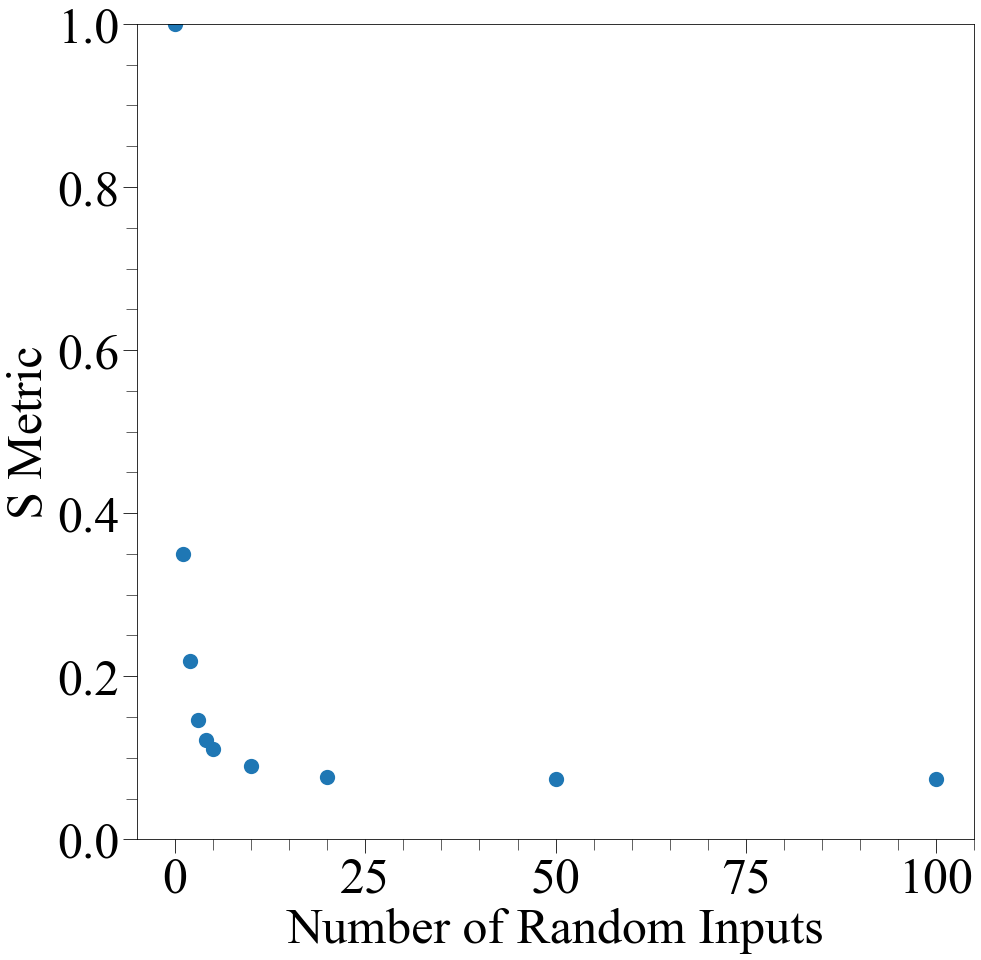}
   \includegraphics[trim = 2 2 2 2, clip, width=0.5\linewidth]{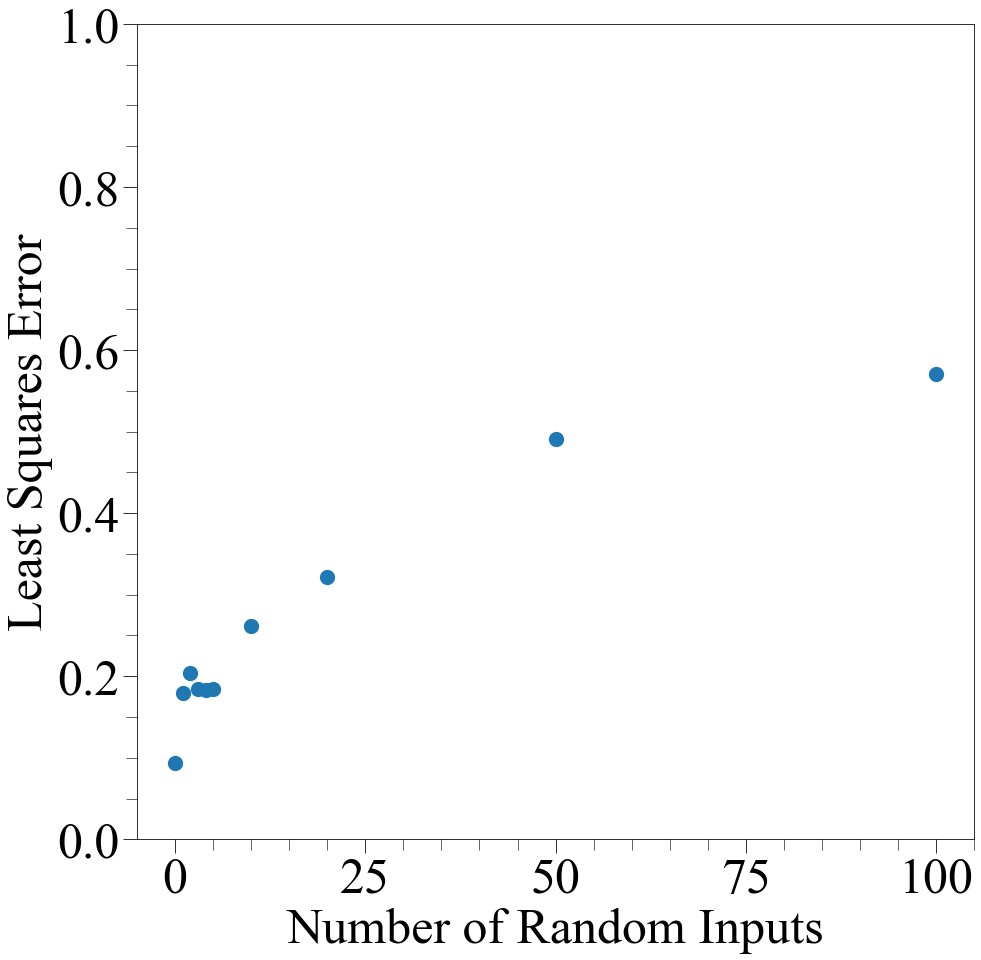}}
  \caption[]{ \label{fig:metrics_nRIn} From left to right, the $S$ metric and $LSE$ as a function of number of random inputs with 50 DTs and 0 random outputs. The $S$ value improved as more random inputs are added, whereas the $LSE$ got worse. This may lead to a balancing game between improving the smoothness of our output compared to the quality of the predictions. }
 \end{center}
\end{figure}

\begin{figure}[hbt!]
 \begin{center}
   \raisebox{0.5mm}{\includegraphics[trim = 2 2 2 2, clip, width=0.5\linewidth]{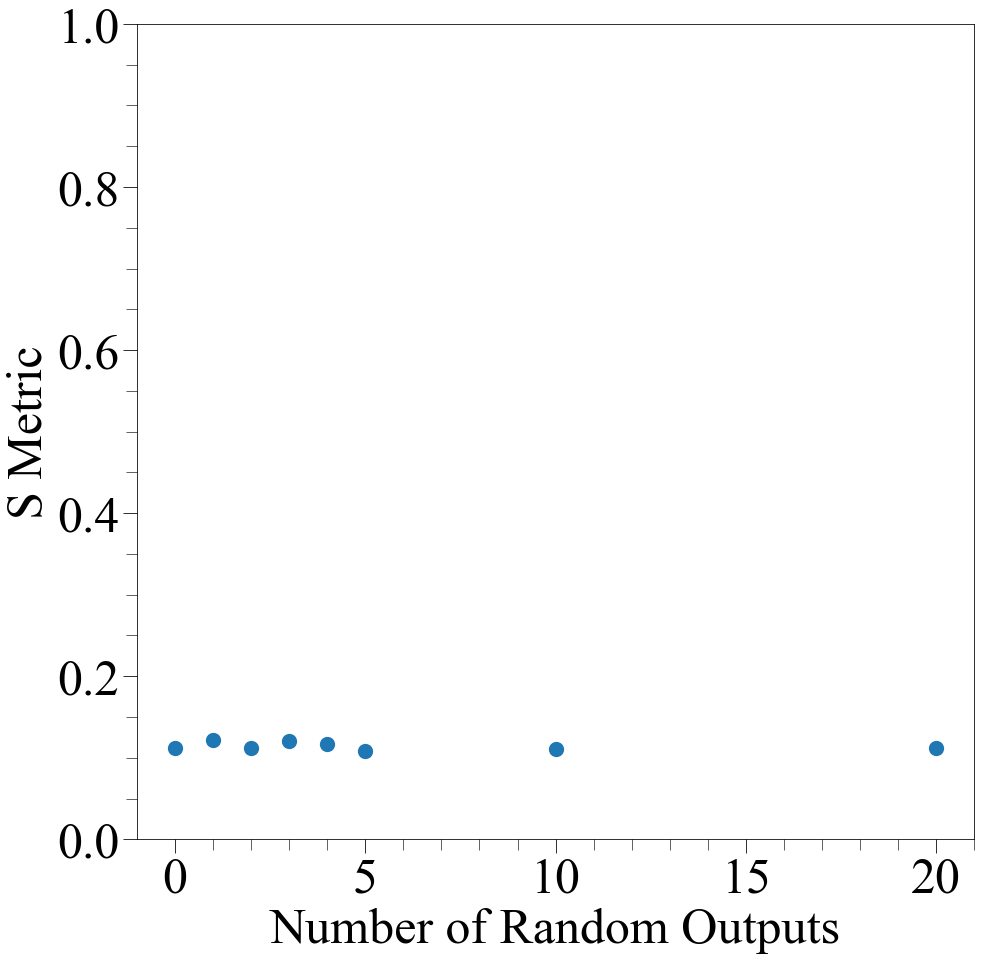}
   \includegraphics[trim = 2 2 2 2, clip, width=0.5\linewidth]{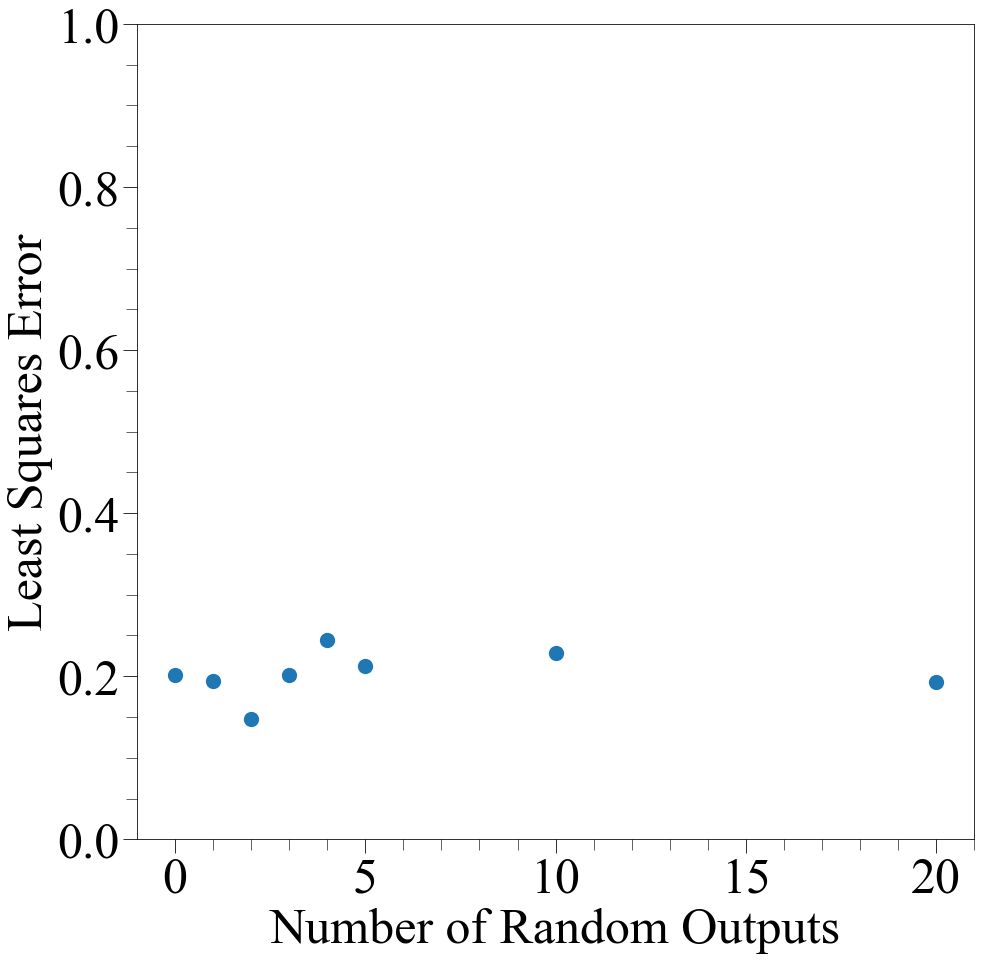}}
  \caption[]{ \label{fig:metrics_nROut} From left to right, the $S$ metric and $LSE$ as a function of number of random outputs with 50 DTs and 5 random inputs. Adding additional random outputs seems to have little impact on either of the two metrics.}
 \end{center}
\end{figure}

\begin{figure}[hbt!]
 \begin{center}
   \raisebox{0.5mm}{\includegraphics[trim = 2 2 2 2, clip, width=0.5\linewidth]{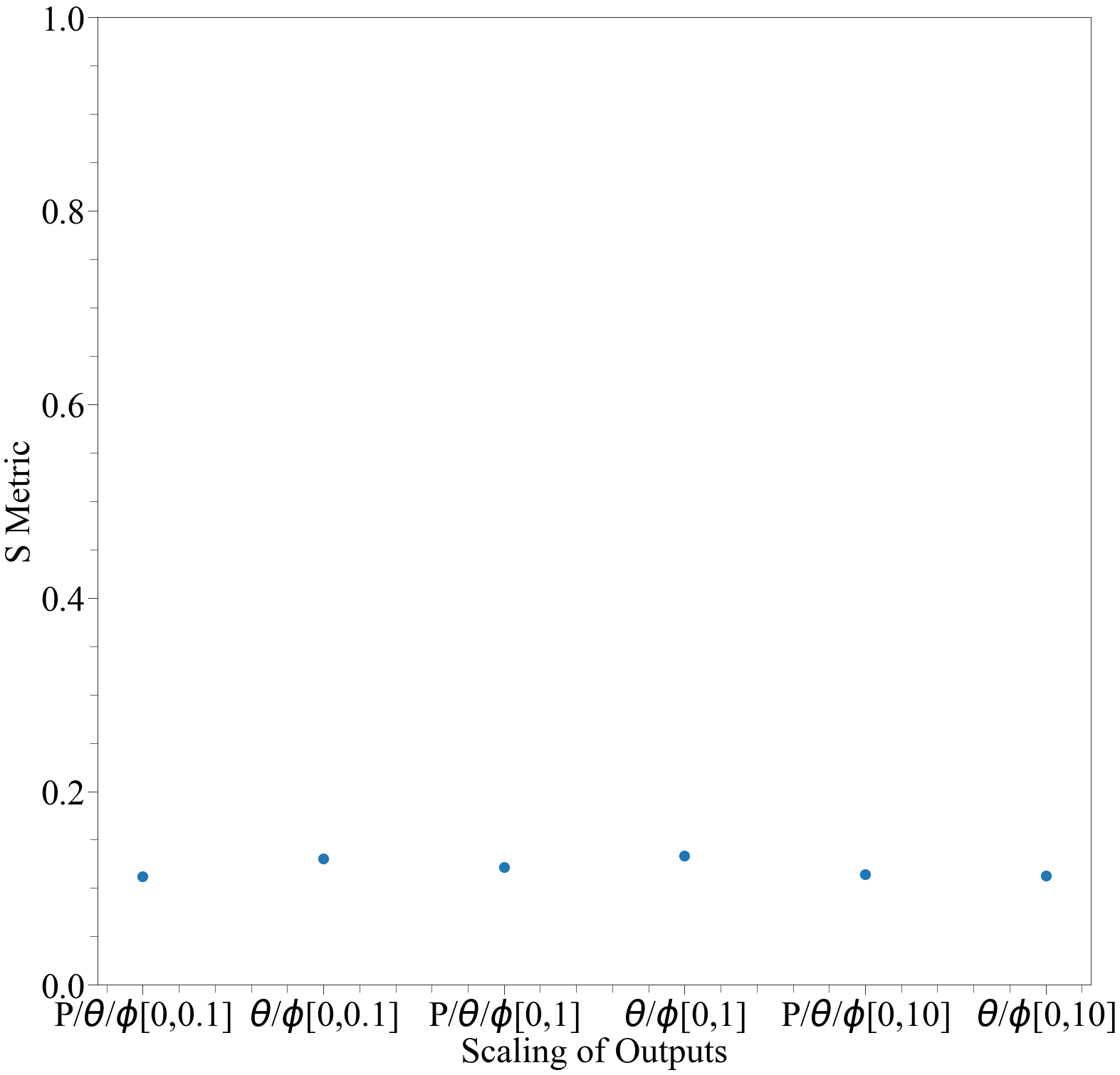}
   \includegraphics[trim = 2 2 2 2, clip, width=0.5\linewidth]{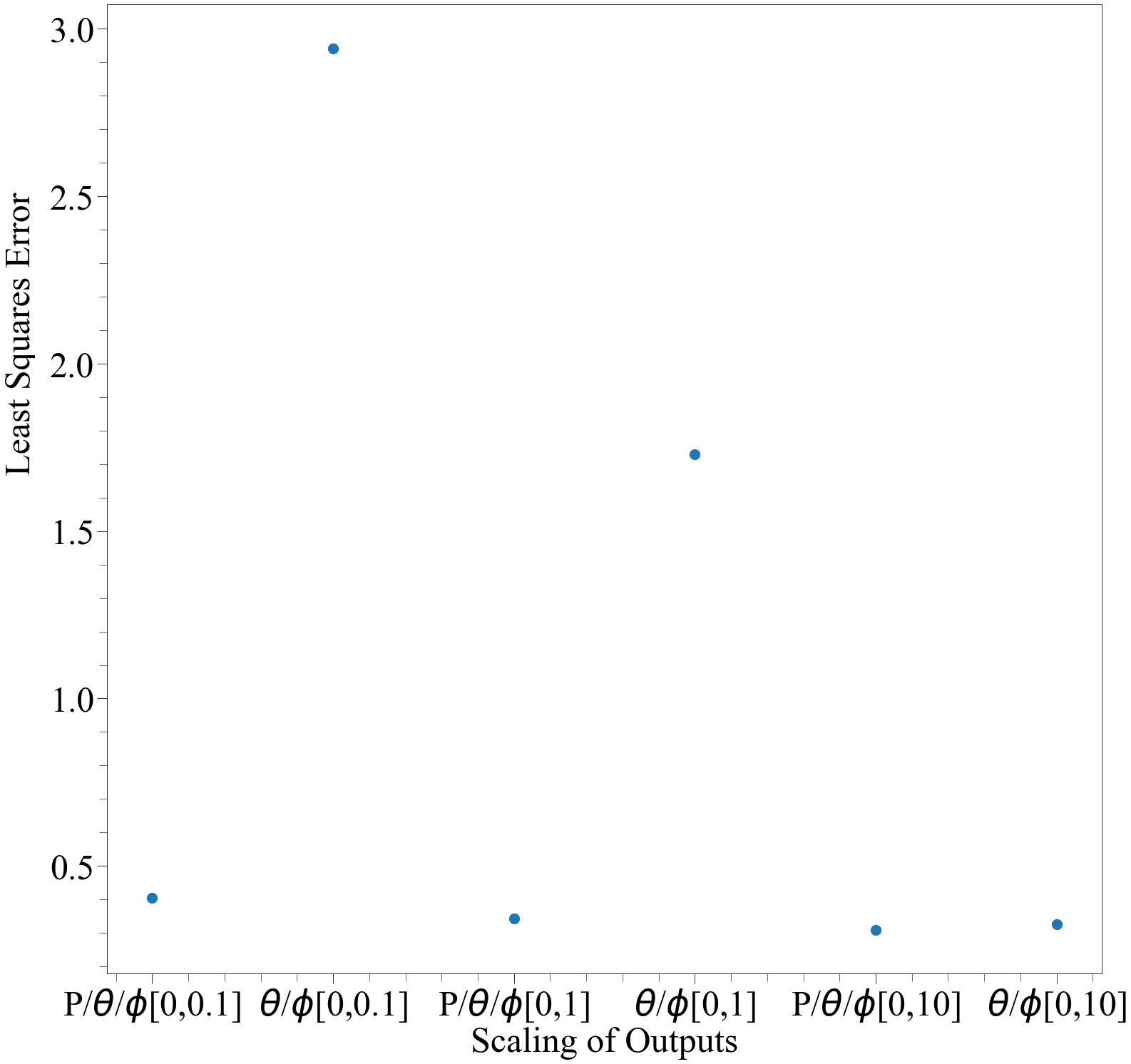}}
  \caption[]{ \label{fig:metrics_nROutScale} From left to right, the $S$ metric and $LSE$ as a function of the scale of non random outputs for 20 DTs trained with 4 random inputs ranging from 0 to 1 and 0 random outputs. A comparison is made between scaling both $\theta$ and $\phi$ but not momentum, and scaling all three variables. When only $\theta$ and $\phi$ are scaled to lower ranges than momentum, the $LSE$ increases, which is not the case when scaling all three variables. The $S$ value however is not affected by the scale of the non random outputs.}
 \end{center}
\end{figure}

\begin{figure}[hbt!]
 \begin{center}
   \raisebox{0.5mm}{\includegraphics[trim = 2 2 2 2, clip, width=0.5\linewidth]{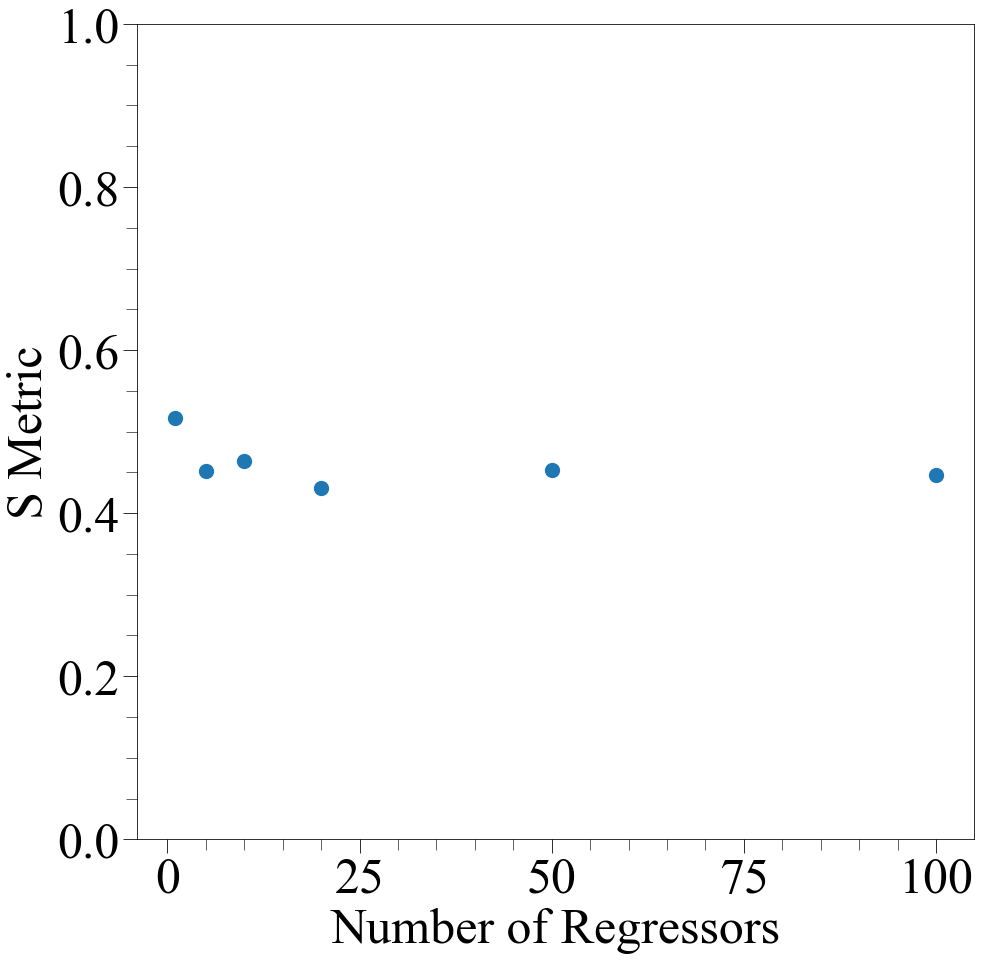}
   \includegraphics[trim = 2 2 2 2, clip, width=0.5\linewidth]{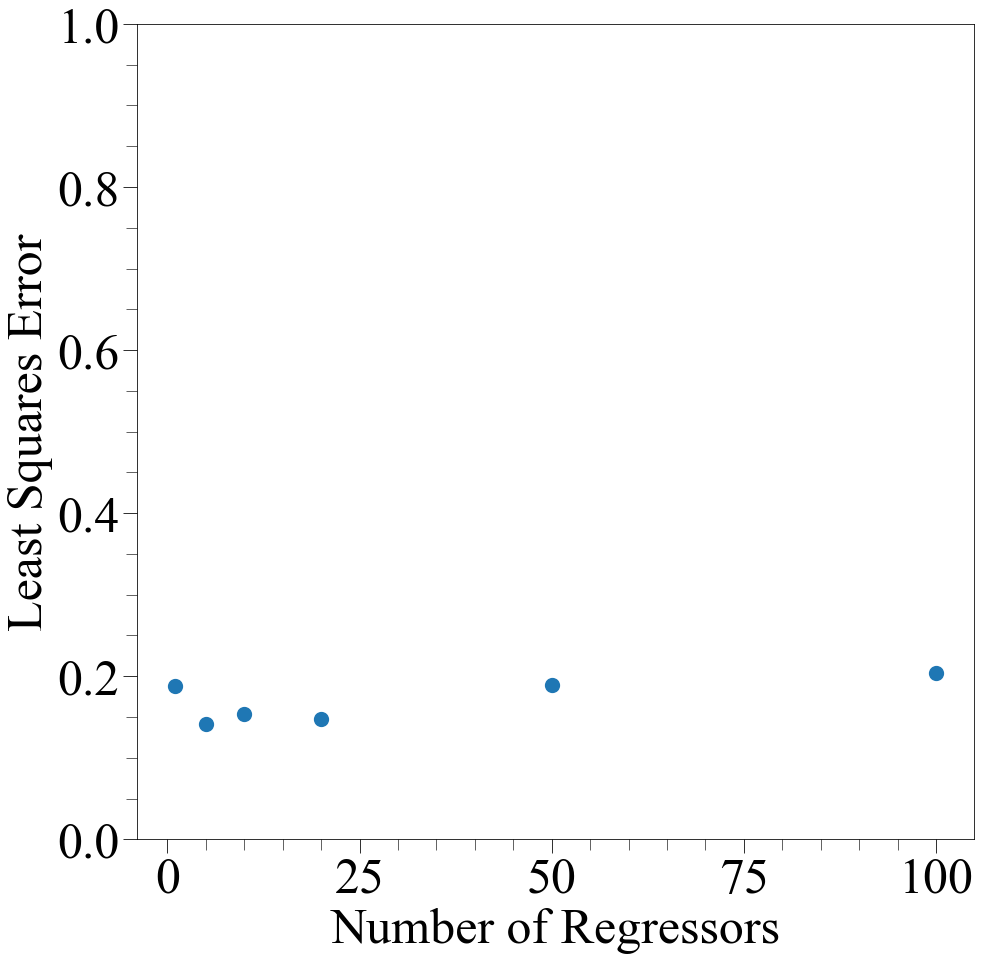}}
  \caption[]{ \label{fig:metrics_nReg_kNN} From left to right, the $S$ metric and $LSE$ as a function of number of kNNs with 1 random input ranging from 0 to 10 and 0 random outputs. Varying the number of kNNs has little impact on either metric.}
 \end{center}
\end{figure}

\begin{figure}[hbt!]
 \begin{center}
   \raisebox{0.5mm}{\includegraphics[trim = 2 2 2 2, clip, width=0.5\linewidth]{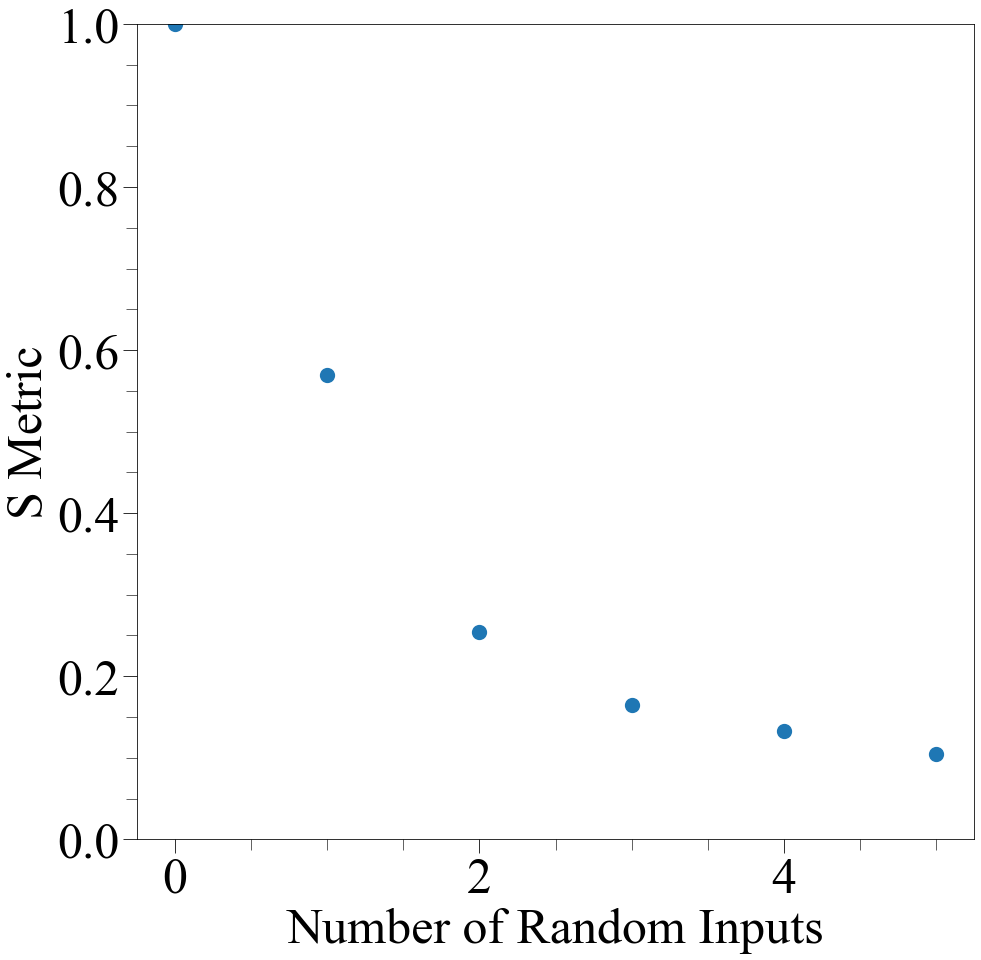}
   \includegraphics[trim = 2 2 2 2, clip, width=0.5\linewidth]{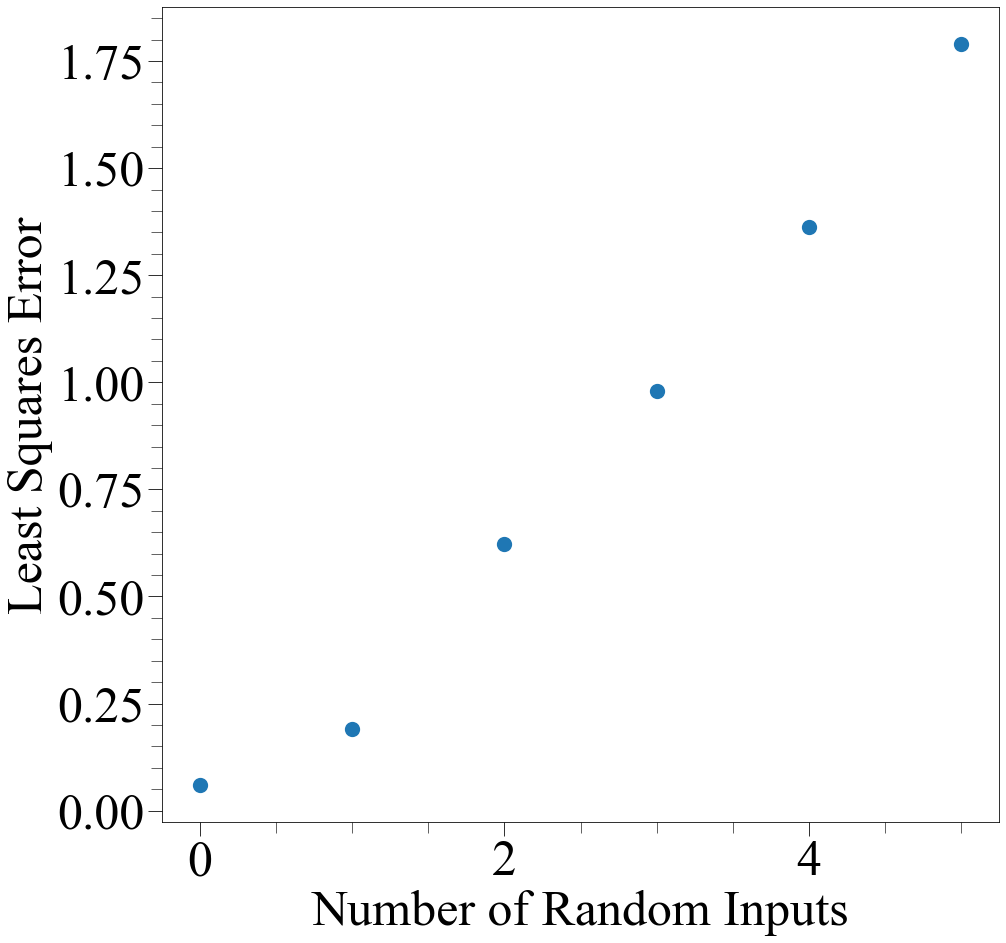}}
  \caption[]{ \label{fig:metrics_nRIn_kNN} From left to right, the $S$ metric and $LSE$ as a function of number of random inputs ranging from 0 to 1 with 1 kNN and 0 random outputs. As for the DT, the $S$ value improved as more random inputs are added, whereas the $LSE$ got worse.}
 \end{center}
\end{figure}

\begin{figure}[hbt!]
 \begin{center}
   \raisebox{0.5mm}{\includegraphics[trim = 2 2 2 2, clip, width=0.5\linewidth]{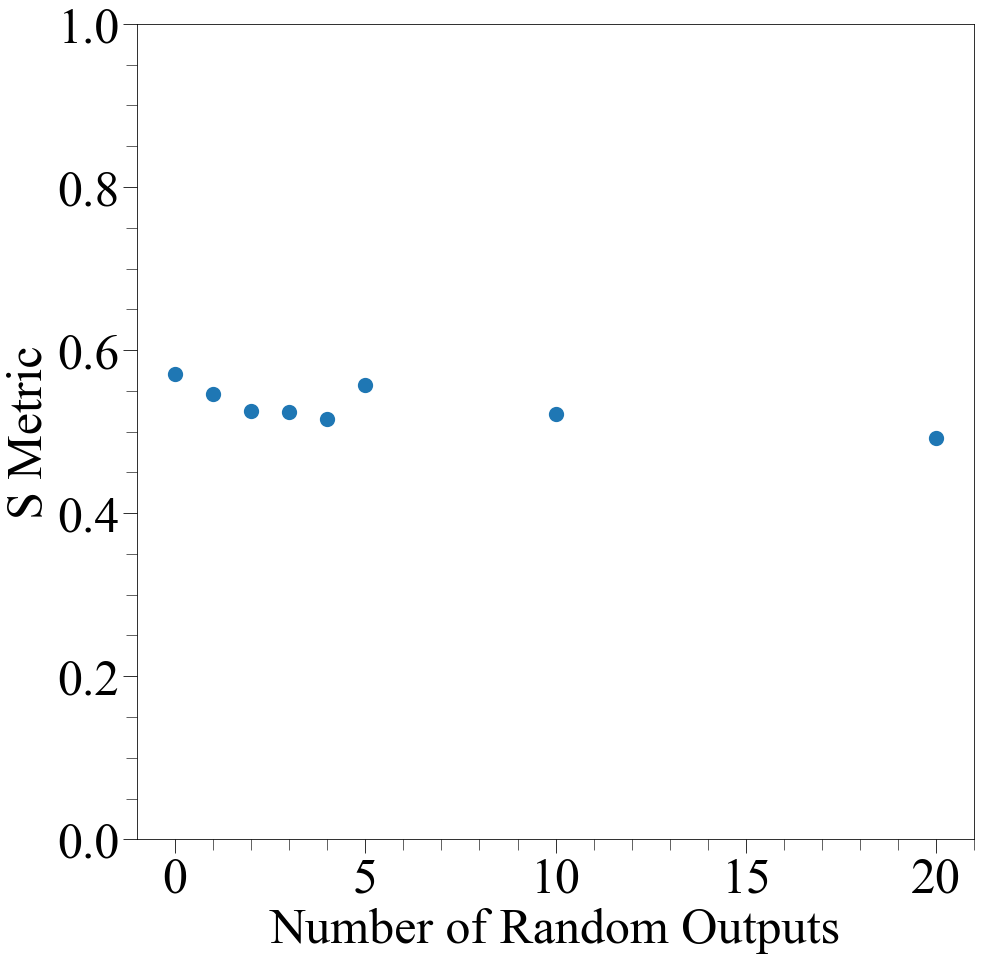}
   \includegraphics[trim = 2 2 2 2, clip, width=0.5\linewidth]{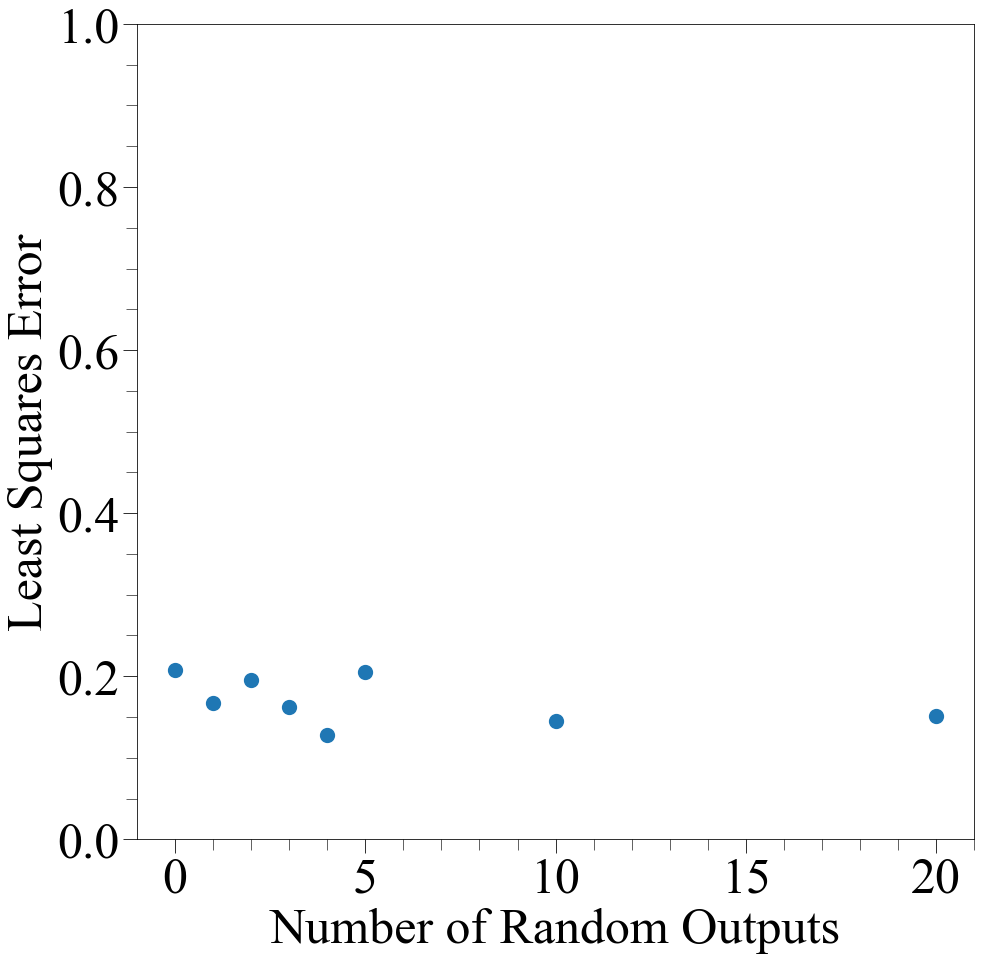}}
  \caption[]{ \label{fig:metrics_nROut_kNN} From left to right, the $S$ metric and $LSE$ as a function of number of random outputs with 1 kNN and 1 random input ranging from 0 to 10. As for the DT, varying the number of random outputs does not significantly impact either metrics }
 \end{center}
\end{figure}

\begin{figure}[hbt!]
 \begin{center}
   \raisebox{0.5mm}{\includegraphics[trim = 2 2 2 2, clip, width=0.5\linewidth]{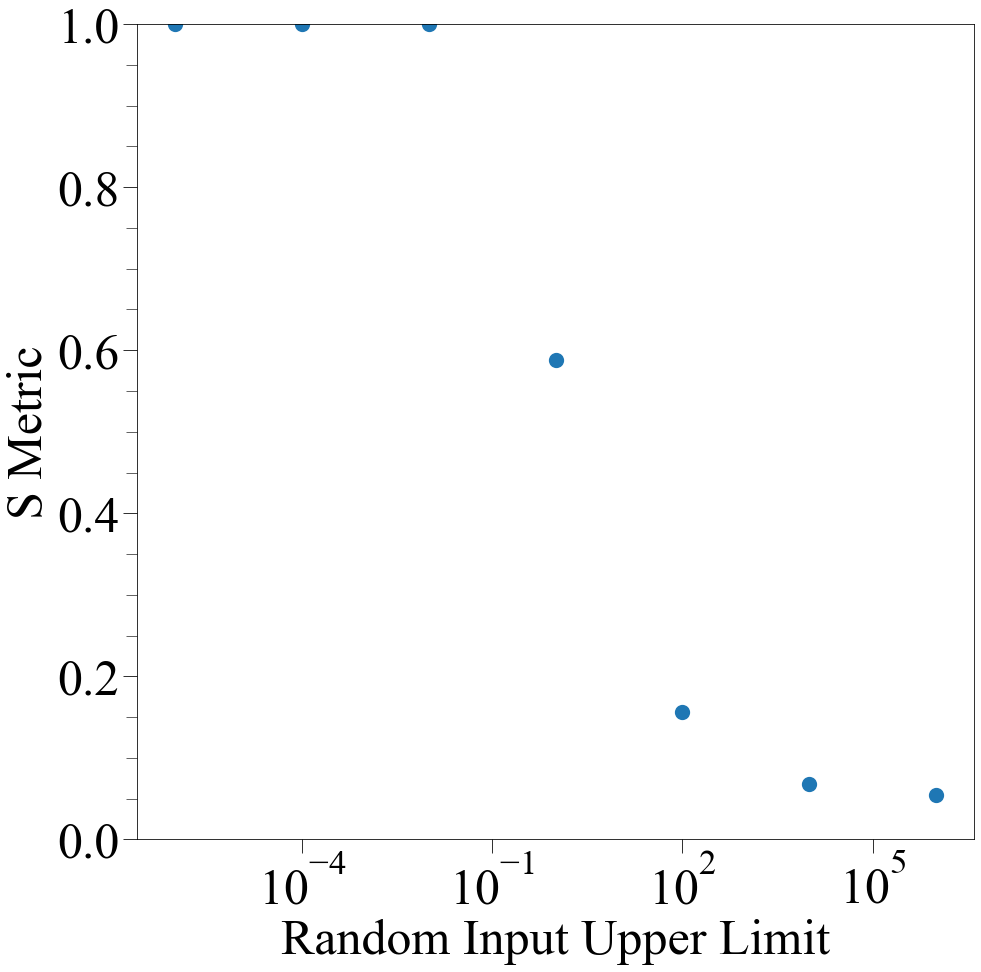}
   \includegraphics[trim = 2 2 2 2, clip, width=0.5\linewidth]{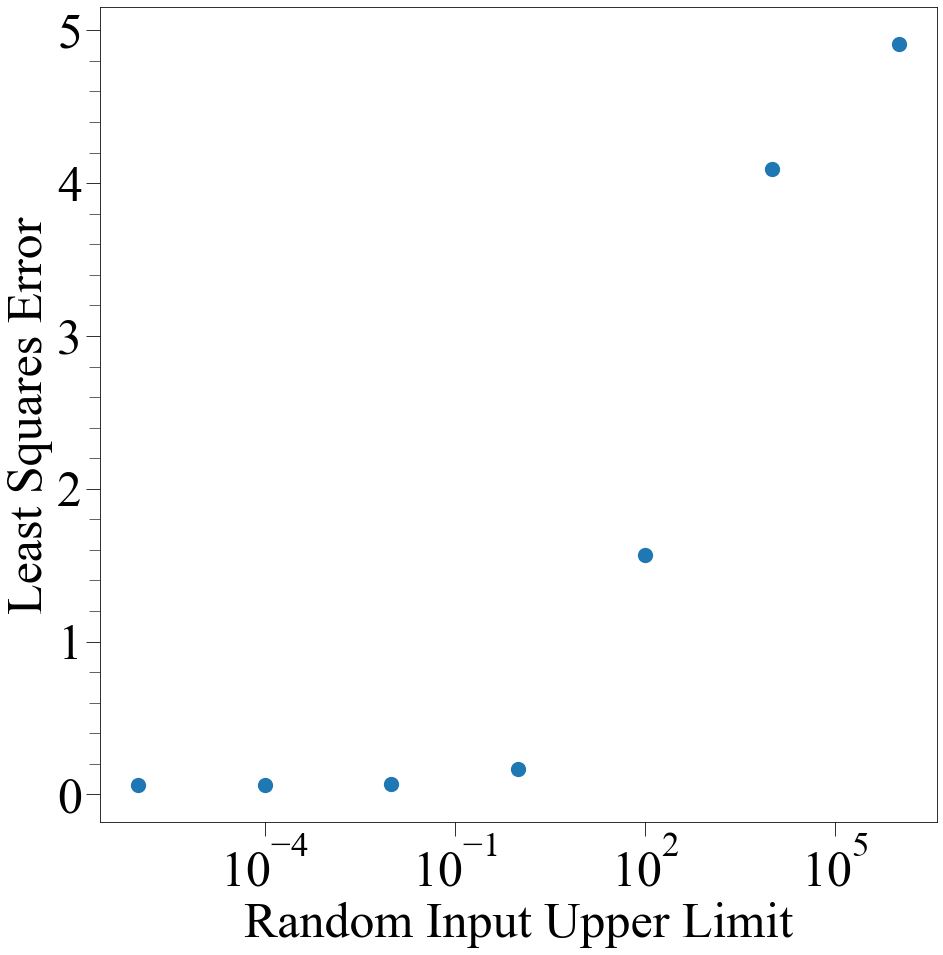}}
  \caption[]{ \label{fig:metrics_nRINLim_kNN} From left to right, the $S$ metric and $LSE$ as a function of the upper limit on the range of the random input with 1 kNN and 1 random input and 0 random outputs. Similarly to varying the number of random inputs, the $S$ value improved as the scale is increased, whereas the $LSE$ got worse.}
 \end{center}
\end{figure}

\end{document}